\documentclass[a4paper,12pt]{amsart}
\usepackage[foot]{amsaddr}

\theoremstyle{definition}

\theoremstyle{remark}

\usepackage[margin=2.7cm]{geometry}
\usepackage{systeme}
\usepackage{graphicx}
\usepackage[]{units}
\usepackage{color}
\usepackage{mathrsfs}
\usepackage{bm}
\usepackage[ansinew]{inputenc}
\usepackage{float}
\usepackage{lineno}
\usepackage{bm}  
\usepackage{amsfonts} 
\usepackage{esint}
\usepackage{amsmath}
\usepackage{amssymb}
\usepackage{framed} 
\usepackage{multicol} 
\usepackage{tabularx}
\usepackage{tabulary}
\usepackage{amsfonts}
\usepackage{booktabs}
\usepackage{siunitx}
\usepackage{caption}
\usepackage{float}
\usepackage{url} 
\usepackage{bm}
\usepackage{nomencl} 
\usepackage{breakurl} 
\usepackage[breaklinks]{hyperref} 
\usepackage{algorithm,algorithmic}
\numberwithin{equation}{section}
\usepackage{amssymb}
\providecommand{\norm}[1]{\lVert#1\rVert}

\newcommand{\ltwonorm}[1]{\langle #1 \rangle_{L_2(\Omega)}}

\DeclareMathAlphabet\mathsfcal{OMS}{cmsy}{b}{n}

\usepackage{subcaption}  
\usepackage{xpatch}
\usepackage{nomencl} 
\renewcommand*{\nompreamble}{\footnotesize}
\makenomenclature
\RequirePackage{ifthen}
\let\oldref\ref      
\renewcommand{\ref}[1]{(\oldref{#1})}

\renewcommand*\nompreamble{\begin{multicols}{2}}
\renewcommand*\nompostamble{\end{multicols}}
\newcolumntype{Y}{>{\centering\arraybackslash}X}  

\xpatchcmd{\thenomenclature}{%
  \section*{\nomname}
}{
}{\typeout{Success}}{\typeout{Failure}}

\usepackage{ifthen}
\renewcommand{\nomgroup}[1]{%
  \ifthenelse{\equal{#1}{A}}{\item[\textbf{Abbreviations}]}{%
    \ifthenelse{\equal{#1}{G}}{\item[\textbf{Symbols}]}{%
      \ifthenelse{\equal{#1}{C}}{\item[\textbf{Abbreviations}]}{%
        \ifthenelse{\equal{#1}{S}}{\item[\textbf{Subscripts}]}{%
          \ifthenelse{\equal{#1}{Z}}{\item[\textbf{Mathematical Symbols}]}{}
        }
      }
    }
  }
}

\begin{document}
\renewcommand\texteuro{-} 
\nomenclature{$\text{ROM}$}{Reduced Order Model}
\nomenclature{$\text{POD}$}{Proper Orthogonal Decomposition}

\nomenclature[G]{$\bm{u}$}{velocity field}
\nomenclature[G]{$\bm{f}$}{Dirichlet boundary condition for velocity}
\nomenclature[G]{$g$}{Dirichlet boundary condition for temperature}
\nomenclature[G]{$\bm{k}$}{initial condition for velocity}
\nomenclature[G]{$l$}{initial condition for temperature}
\nomenclature[G]{${p}$}{pressure field}
\nomenclature[G]{${\theta}$}{temperature field}
\nomenclature[G]{${\nu}$}{dimensionless kinematic viscosity}
\nomenclature[G]{${\alpha_{dif}}$}{thermal diffusivity}
\nomenclature[G]{${N_u^h}$}{number of unknowns for velocity at full-order level}
\nomenclature[G]{${N_p^h}$}{number of unknowns for pressure at full-order level}
\nomenclature[G]{${N_{sup}^h}$}{number of unknowns for supremizer at full-order level}
\nomenclature[G]{${N_{sup}^s}$}{number of unknowns for supremizer at reduced level before the truncation}
\nomenclature[G]{${N_{sup}^r}$}{number of unknowns for supremizer at reduced order level after the truncation}
\nomenclature[G]{${N_{\theta}^h}$}{number of unknowns for temperature at full-order level}
\nomenclature[G]{${N_u^s}$}{number of uknowns for velocity at reduced level before the truncation}
\nomenclature[G]{${N_p^s}$}{number of unknowns for pressure at reduced order level before the truncation}
\nomenclature[G]{${N_{\theta}^s}$}{number of unknowns for temperature at reduced order level before the truncation}
\nomenclature[G]{${N_u^r}$}{number of unknowns for velocity at reduced order level after the truncation}
\nomenclature[G]{${N_p^r}$}{number of unknowns for pressure at reduced order level after the truncation}
\nomenclature[G]{${N_{\theta}^r}$}{number of unknowns for temperature at reduced order level after the truncation}
\nomenclature[G]{$C$}{correlation matrix}
\nomenclature[G]{$\bm{W}$}{eigenvector matrix}
\nomenclature[G]{$N_t$}{number of time instances}
\nomenclature[G]{$Q$}{space-time domain}
\nomenclature[G]{$T$}{final time}
\nomenclature[G]{$\mathcal{P}$}{parameter space}
\nomenclature[G]{$\mathcal{K}$}{training set space}
\nomenclature[G]{${\Omega}$}{bounded domain}
\nomenclature[G]{${\Gamma}$}{boundary of $\Omega$}
\nomenclature[G]{${\epsilon}_{L^2}$}{$L^2$ norm error}
\nomenclature[G]{${\Theta_{IF}}$}{boundary of $\Omega$ and initial condition for the temperature internal field}
\nomenclature[G]{$\bm{n}$}{outward normal vector}
\nomenclature[G]{$\bm{\varphi_i}$}{i-th POD basis function for velocity}
\nomenclature[G]{$\bm{\eta_i}$}{i-th POD basis function for supremizers}
\nomenclature[G]{${\psi_i}$}{i-th POD basis function for pressure}
\nomenclature[G]{${\chi_i}$}{i-th POD basis function for temperature}
\nomenclature[G]{$\bm{M}$}{ROM mass matrix} 
\nomenclature[G]{$\bm{Q(u)}$}{ROM convection matrix}
\nomenclature[G]{$\bm{L}$}{ROM diffusion matrix}
\nomenclature[G]{$\bm{K}$}{ROM mass matrix for the heat equation}
\nomenclature[G]{$\bm{G}$}{ROM convection matrix for the heat equation}
\nomenclature[G]{$\bm{N}$}{ROM diffusion matrix for the heat equation}   
\nomenclature[G]{${N_{\mu}}$}{number of parameters in the training set $\mathcal{K}$}
\nomenclature[G]{$\otimes$}{tensor product}
\nomenclature[G]{$\bm{\nabla}\cdot$}{divergence operator}
\nomenclature[G]{$\bm{\nabla}\times$}{curl operator}
\nomenclature[G]{$\bm{\nabla}$}{gradient operator}
\nomenclature[G]{$\bm{\nabla}^s$}{symmetric gradient operator}
\nomenclature[G]{$\bm{\alpha}$}{reduced vector of unknowns for velocity}
\nomenclature[G]{$\bm{b}$}{reduced vector of unknowns for pressure}
\nomenclature[G]{$\bm{c}$}{reduced vector of unknowns for temperature}
\nomenclature[G]{$\Delta$}{laplacian operator}
\nomenclature[G]{$\beta$}{inf-sup stability constant}
\nomenclature[G]{$\left\lVert \cdot\right\rVert$}{norm in $L^2(\Omega)$}
\nomenclature[G]{$\langle \cdot , \cdot \rangle$}{inner product in $L^2(\Omega)$}
\nomenclature[G]{$E_{N_{u,p,{\theta},{sup}}^r}$}{Energy of the snapshots}


\title[Parametric ROM for Unsteady Heat Transfer Problems]{Parametric POD-Galerkin Model Order Reduction for Unsteady-State Heat Transfer Problems}

\author{Sokratia Georgaka\textsuperscript{1,*}}
\address{\textsuperscript{1}Imperial College London, Department of Mechanical Engineering, London, SW7 2BX, UK.}
\thanks{\textsuperscript{*}Corresponding Author.}
\thanks{\textsuperscript{**}Second Corresponding Author.}
\email{s.georgaka16@imperial.ac.uk}

\author{Giovanni Stabile\textsuperscript{2,**}}
\address{\textsuperscript{1}SISSA, International School for Advanced Studies, Mathematics Area, mathLab Trieste, Italy.}
\email{gstabile@sissa.it}

\author{Gianluigi Rozza\textsuperscript{2}}
\email{grozza@sissa.it}

\author{Michael J Bluck \textsuperscript{1}}
\email{m.bluck@imperial.ac.uk}

\keywords{}

\date{}

\dedicatory{}

%



\begin{abstract} 
A parametric reduced order model based on proper orthogonal decomposition with Galerkin projection has been developed and applied for the modeling of heat transport in T-junction pipes which are widely found in nuclear power reactor cooling systems. Thermal mixing of different temperature coolants in T-junction pipes leads to temperature fluctuations and this could potentially cause thermal fatigue in the pipe walls. The novelty of this paper is the development of a parametric ROM considering the three dimensional, incompressible, unsteady Navier-Stokes equations coupled with the heat transport equation in a finite volume regime. Two different parametric cases are presented in this paper: parametrization of the inlet temperatures and parametrization of the kinematic viscosity. Different training spaces are considered and the results are compared against the full order model. The first test case results to a computational speed-up factor of $374$ while the second test case to one of $211$.  
\end{abstract}


\maketitle

%
%
%
%
\section{Introduction}\label{sec:intro}


Partial differential equations (PDEs) describe a variety of physical systems occurring in nature and in engineering. PDEs are complex and generally nonlinear and their numerical solution requires considerable computational effort. For example, fluid flow, a phenomenon very common in many engineering fields, is governed by the Navier-Stokes equations and accurate numerical solutions provide vital insight into complex physical processes. Analytical solutions of these equations is impossible in almost all circumstances. For this reason, computational fluid dynamics (CFD) has seen progressive development since the 1970s and is now capable of solving many practical problems in fluid flow and heat transfer. With the continued development of improved algorithms and increasing computational power, CFD is now used in various engineering fields such as aerospace, nuclear, civil, mechanical as well as non-engineering fields such us neuroscience and meteorology etc. 

Despite its popularity and applicability, the computational burden for simulating realistic large scale and many query systems is still very high, even with the use of supercomputers. A good example of the challenges involved can be found in nuclear applications, where turbulence, multiphase flow and heat transfer phenomena occur in complex geometries; a fairly accurate CFD simulation of a single instance of an accident case scenario could take months or more to be performed. To address these challenges, Systems Codes (SC), such as RELAP, CATHARE, etc and sub-channel codes (COBRA, etc), constitute phenomenological reduced order methods based on considerable limiting physical assumptions. These codes, that were developed in the 1950s, rely on major physical and geometrical simplifications, such as averaging over the flow cross section leading to essentially 1D simulations. These simplifications can save great amounts of computational time. However, the compromise is that they rely exclusively on experimental and phenomenological correlations to take account of heat transfer and turbulence and the like. In particular, these assumptions are particularly inadequate for 3D flows. In the recent years although these codes have been improved allowing some limited 3D capability, the accuracy is still inadequate and their application is very limited. The same applies in the field of neutronics for the study of reactor dynamics. Geometrical and physical simplifications are made to the governing equations in order to obtain a computationally affordable model. These simplifications include 1D geometries, homogenous core dynamics, uniform axial fluxes, etc. The challenge then, is to bridge the considerable gap between high fidelity full-order models (eg CFD and its variants) and these over-simplistic reduced order models (systems and sub-channel codes).

Modern reduced order models (ROMs) \cite{quarteroniRB2016, hesthaven2015certified, Grepl2007} have been proposed as an alternative way of approximating full-order systems (such as those arising in conventional CFD) in a more sophisticated and reliable way. Unlike phenomenological methods, modern ROMs potentially retain the high fidelity of the full order model (FOM) while exhibiting performance akin to phenomenological methods. Reduced order modeling is a highly promising area, which is currently flourishing in the science and engineering community. 
 
An essential tool in the development of ROMs is the Proper orthogonal decomposition (POD) or Karhunen - Lo$\grave{e}$ve decomposition. Originally conceived as a data analysis method for finding an optimal lower-dimensional orthonormal basis in a least-squares sense, POD can be used as a model order reduction method for multidimensional dynamical systems, using data from high fidelity simulations (in this case CFD) or from experiments. POD can be seen as a modal decomposition technique, which provides modes ranked according to their energy. In fluid dynamics, POD has been successfully applied in two main areas: Firstly in the search for an optimal basis in a lower dimensional space and secondly in the identification of hidden patterns (in terms of size, shape, location) in complex datasets. Amongst other related methods, POD is usually considered the most efficient method for capturing the dominant structures of large scale systems. Lumley \cite{lumley} was the first to apply POD in the study of turbulent flow, using spatial velocity correlations. Classical statistical methods which rely on averaging quantities consider turbulence as a complex chaotic phenomenon with little or no underlying structure. On the contrary, coherent structures exist and turbulent flow is composed of organised motions and it is the superposition of these that presents the apparent complexity. To identify large eddy structures, Bakewell and Lumley \cite{Bakewell1967} applied POD to experimental data taken in the study of the boundary layer of homogeneous turbulent pipe flow. The authors came to an important conclusion regarding the formation of shear turbulent flow, that it is created and sustained not only in the wall region but also in the viscous sub-layer. They also showed that in the wall region, the creation and evolution of counter-rotating eddy pairs is governed by the non-linear mechanism of vortex stretching. Payne and Lumley \cite{payne} studied cylinder wake flows using POD. As the dominant mode, they observed a counter-rotating eddy pair, although they mentioned that for more accurate results, more data and grid points are needed. A detailed review on identification of coherent structures in turbulent flows can be found in \cite{berkooz}. The theory of Lumley had proven very successful but the necessary processing of large datasets of experimental and numerical data became a limitation. To overcome this, Sirovich \cite{Sirovich1987} introduced the snapshot POD (as opposed to the direct POD) method as an efficient way of identifying the dominant modes of large scale systems, when the spatial dimension is larger than the temporal dimension. Snapshots are instantaneous solutions obtained by a high-fidelity solver (eg CFD) or from experimental data on which POD is performed for the calculation of the reduced basis. Rempfer and Fasel in \cite{rempfer}, performed simulations on a flat plate boundary layer to prove that, in the case of flow fields which present symmetry along a coordinate, POD can describe spatially evolving structures. Baltzer \textit{et al} \cite{baltzer} used snapshot POD for identification of coherent structures in a turbulent boundary layer, where the evolution of large-scale motions appears. Bernero and Fiedler \cite{bernero} applied snapshot POD to Particle Image Velocimetry (PIV) data obtained from a jet in a counterflow, to show that even in such chaotic structures, a combination of PIV and snapshot POD could reveal a few dominant patterns. A related application of POD methods is in data reconstruction: Thanh \textit{et al} in \cite{willcoxtransonic}, showed that POD is an efficient method for reconstructing flow fields in aerodynamics when data is missing.  
 
The use of POD in the construction of reduced order models is a more recent development. Hall \textit{et al} \cite{hall}, applied snapshot POD in transonic and subsonic unsteady aerodynamic flows, in a study of an isolated airfoil and a cascade of flat plate airfoils. The authors obtained accurate ROMs with meaningful results, and suggested that ROMs could be suitable in active control applications. So-called POD-Galerkin ROMs have been widely used in optimal control problems, design optimisation, data reconstruction and many query systems. Ravindran \cite{ravindran}, developed a POD-Galerkin ROM for optimal control of fluid flows in a channel flow problem. The results showed accurate short-time ROM behaviour and high computational savings. These two characteristics are essential for real-time control applications. Bourguet and Braza \cite{Bourguet2007}used a POD-Gelerkin ROM in the study of 2D transonic, compressible, unsteady flows around a NACA0012 airfoil, where two dominant flow structures were identified: the von Karman instability and buffeting. The resulting ROM is in an excellent agreement with the dynamics of the high fidelity model. An observation from this work is that the non-linear terms arising in the calculation of the ROM are (relatively) computationally expensive. Examples of very effective reduced order models based on finite volume FOMs of the Navier-Stokes equations are demonstrated in the pioneering work of \cite{Stabile2018,Lorenzi2016,Haasdonk2008}. Regarding parametric PDEs, which are the main interest of this study, in \cite{ballarinmono} Ballarin \textit{et al} proposed a monolithic model order reduction approach based on POD-Galerkin for parametrized fluid-structure interaction problems. Also in \cite{Ballarin2014}, stable POD-Galerkin for the parametrized, incompressible, steady Navier-Stokes equations is presented. Stabile \textit{et al} in \cite{Stabile2018} presented a POD-Galerkin ROM for the parametrized, incompressible, unsteady Navier-Stokes equations. POD-Galerkin model reduction for parametric PDEs can be also found in the study of haemodynamics, in the work of Ballarin \textit{et al} \cite{BALLARIN2016609}.

In regard to non-isothermal problems, a first attempt to develop a POD-Galerkin ROM for modeling the temperature field in a rapid thermal processing chamber is described in \cite{aling1997nonlinear}, where the authors considered a 2D steady state problem. In \cite{Alonso2009}, Alonso \textit{et al.} presented a ROM for studying heat transfer in a backwards facing step flow, using a combination of POD and a genetic algorithm. A heat transfer POD-Galerkin ROM is presented in \cite{Raghupathy2009}, where the 1D conduction heat equation has been used. A POD study for the heat conduction equation is also presented in \cite{Wang2012} and in \cite{Han2014}. The problem of natural circulation is studied in \cite{li2013fast} where a FOM of the coupled Navier-Stokes and energy equations are used to develop a ROM. However, the resulting POD-Galerkin ROM only considers perturbations of the (two-dimensional) temperature field, and assumes the flow field remains fixed. These restrict the study to small perturbation temperature control applications. A POD-Galerkin methodology for groundwater flow problems driven by spatially distributed stochastic forcing terms is presented in \cite{PASETTO20111450} where the authors considered collecting the POD snapshots in the probability space. Their proposed method results in a reduced order Monte Carlo framework (ROMC). Another reduced order modeling technique, other than the POD-Galerkin, can be found in the study of uncertainty propagation in porous media \cite{MULLER20111527} where the authors applied the Karhunen - Lo$\grave{e}$ve (KL) decomposition (or POD) and polynomial chaos with sparse Smolyak quadrature for the flow problem. In \cite{LI20112489}, the POD method has been applied to 2D solute transport problems. In \cite{BuStaRo2018}, the authors proposed a POD-Galerkin ROM for the Navier-Stokes weakly coupled heat transport equations based on a hybrid finite element - finite volume method.           

In the work presented in this article, a POD-Galerkin method is developed for the parametric 3D unsteady Navier-Stokes equations minimally coupled with the heat transport equation. The parametrization is applied on two test cases: first on the boundary conditions, considering the temperature inlets, and in the second case, on a physical parameter, considering the kinematic viscosity. The open-source finite volume solver OpenFOAM \cite{jasak1996error} is used to generate the FOM solutions which are then used as a training space for the ROM. In this paper, the work of \cite{Stabile2018} is extended, taking into account the heat transport equation. To the best of the authors knowledge, a parametric POD-Galerkin ROM for modeling problems which are governed by the full set of the parametric 3D Navier-Stokes equations and the heat transport equation, including transient, diffusive and convective terms is introduced in this paper for the first time.

The work is organised as follows: in \autoref{sec:math_form} the mathematical formulation is presented and in \autoref{sec:ROM} the reduced order methodology is introduced and discussed. In \autoref{sec:num_exp} the proposed ROM is used to model thermal-mixing in a T-junction pipe, applied to two different parametric cases: the inlet temperatures and the kinematic viscosity. Finally in \autoref{sec:conclusions} conclusions and perspectives are drawn, highlighting the directives for future improvements and developments.

%
%
%
%
\section{Mathematical Framework for the Full Order Model}\label{sec:math_form}

The full order model (FOM) is governed by the incompressible, transient parametrized Navier-Stokes equations along with the parametrized heat transport equation. In a Eulerian framework and domain $Q = \Omega \times [0,T_s] \subset \mathbb{R}^d\times\mathbb{R}^+$ with $d=2,3$, these equations can be expressed as follows:  
\begin{equation}\label{eq:navstokes}
\begin{cases}
\frac{\partial \bm{u}}{\partial t}+ \bm{\nabla} \cdot (\bm{u} \otimes \bm{u})- \bm{\nabla} \cdot 2 \nu(\bm\mu) \bm{\nabla^s} \bm{u}=-\bm{\nabla}p &\mbox{ in } Q,\\
\bm{\nabla} \cdot \bm{u}=\bm{0} &\mbox{ in } Q,\\
\frac{\partial \theta}{\partial t} + \nabla \cdot (\bm{u} \theta ) - \alpha_{dif}\Delta \theta = 0 &\mbox{ in } Q,\\
\bm{u} (\bm{x},\bm\mu,t) = \bm{f}(\bm{x}) &\mbox{ on } \Gamma_{In} \times [0,T_s],\\
\theta (\bm{x},\bm\mu,t) = g(\bm{x},\bm\mu) &\mbox{ on } \Gamma_{In} \times [0,T_s],\\
\bm{u} (\bm{x},\bm\mu,t) = \bm{0} &\mbox{ on } \Gamma_{w} \times [0,T_s],\\ 
(\nu(\bm\mu))\nabla \bm{u} - p\bm{I})\bm{n} = \bm{0} &\mbox{ on } \Gamma_{o} \times [0,T_s],\\ 
\bm{u}(0,\bm{x})=\bm{k}(\bm{x}) &\mbox{ in } \theta_{s_0},\\      
\theta(0,\bm{x})=l(\bm{x}) &\mbox{ in } \theta_{s_0},\\        
\end{cases}
\end{equation}
where $\bm{u}$ is the fluid velocity, $p$ the normalized pressure, $\theta$ is the fluid temperature, $\alpha_{dif}$ is the thermal diffusivity, $\nu(\bm\mu)$ is the kinematic viscosity and $\bm\mu$ is the vector of parameters. $T_s$ represents the time of the simulation, $\Gamma = \Gamma_{In} \cup \Gamma_{w} \cup \Gamma_{o}$ is the boundary of $\Omega$ and it consists of three different parts $\Gamma_{In}$, $\Gamma_{o}$ and $\Gamma_w$ that indicate, respectively, inlet, outlet and physical wall boundaries. The functions $\bm{f}(\bm{x},\bm\mu)$ and $g(\bm{x},\bm\mu)$ represent the boundary conditions for the non-homogeneous boundaries. $\bm{k}(\bm{x})$ and $l(\bm{x})$ denote the initial conditions for the velocity and the temperature at $t=0$. Time independence of the boundary conditions $\bm{f}$ and $\bm{g}$ is also assumed. In this work, the parametric dependency of interest is on the temperature inlet boundary conditions as well as on the kinematic viscosity. A nomenclature with all the symbols can be found at the end of this paper.

%
\subsection{Full Order Approximation via Finite Volume}

\begin{figure}[]
\centering
\includegraphics[width=0.49\textwidth]{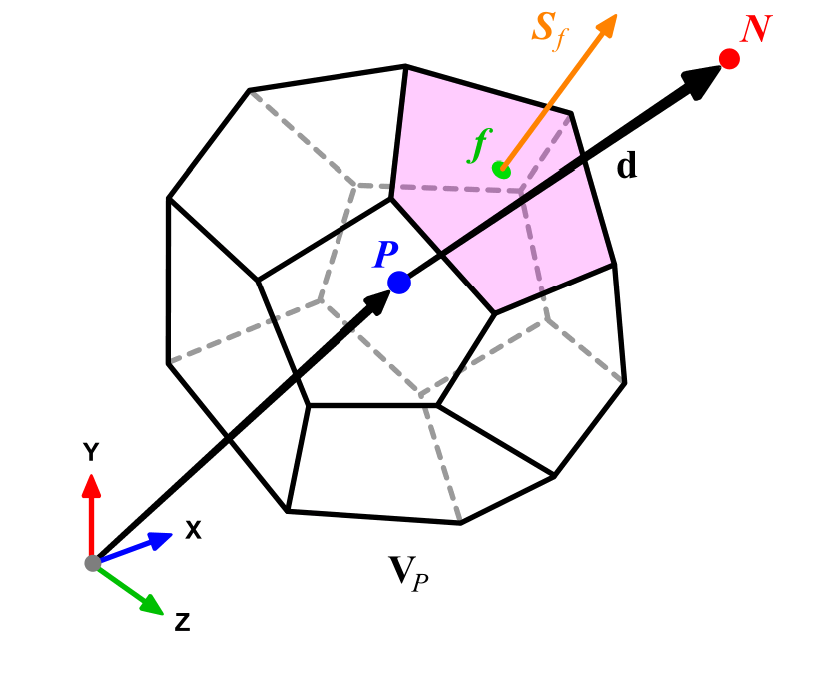}
\caption{Example of a polyhedral control volume, $V_P$, around a control volume centroid $P$ with volume $V$. The picture is taken from \cite{joel}}\label{fig:volume}
\end{figure}

The full order system which is represented by the partial differential equations (\ref{eq:navstokes}), is transformed into a system of discrete algrebraic equations which can be then solved with any iterative or direct numerical method. The system is discretized in a finite volume regime using the open source C++ library OpenFOAM \cite{OF}. These transport equations include temporal derivatives as well as conventive and diffusive terms and each of these terms is treated in a different way. The first step towards discretization of the spatial terms, is the division of the computational domain into arbitrarily small control volumes (cells) such the one depicted in figure (\ref{fig:volume}). The transient term is discretized in time by splitting the total time interval of the simulation into a number of time steps. In the finite volume regime, the integral form of the equations is discretized over a control volume and therefore the quantities of interest are conserved (mass, momentum etc).             

Considering a general transported quantity $\phi$, the transport equation can be written as:

\begin{equation}\label{eq:transport}
\int_{V_P}\frac{\partial \phi}{\partial t}dV + \int_{V_P}\nabla\cdot(\bm{u}\phi)dV - \int_{V_P}\nabla\cdot(\Gamma_{\phi}\nabla\phi)dV = 0,  
\end{equation}

where the source term $S_{\phi}$ has been set to zero. Therefore we are not considering any external sources. The first term in equation (\ref{eq:transport}) represents the temporal term, the second is the convective and the third the diffusive term.  Using Gauss theorem the volume integrals in equation (\ref{eq:transport}) are transformed into surface integrals:

\begin{equation}\label{eq:transportsurf}
\frac{\partial}{\partial t}\int_{V_p}\phi dV + \oiint_{\partial V_P} d\bm{S}\cdot(\bm{u}\phi) - \oiint_{\partial V_P} d\bm{S}\cdot(\Gamma_{\phi}\nabla\phi) = 0,
\end{equation}

where $\partial V_P$ represents a closed surface which bounds the control volume $V_P$ and $\bm{n}d\bm{S}=d\bm{S}$. Since the known quantities are in the centre of the cell, interpolation of the cell centred values to the cell faces is needed. Taking each term in equation (\ref{eq:transportsurf}) separately and starting with the convective term:

\begin{equation}\label{eq:convectiveapprox}
\oiint_{\partial V_P} d\bm{S}\cdot(\bm{u}\phi) = \sum_{f}\int_{f} d\bm{S}\cdot(\bm{u}\phi)_f \approx \sum_{f}\bm{S}_{f}\cdot(\bm{u}\phi)_{f},
\end{equation}

where the integral has been approximated with a second order accurate midpoint rule. Similarly, the diffusive term reads:

\begin{equation}\label{eq:diffusiveapprox}
\oiint_{\partial V_P} d\bm{S}\cdot(\Gamma_{\phi}\nabla\phi) = \sum_{f}\int_{f} d\bm{S}\cdot(\Gamma_{\phi}\nabla\phi) \approx \sum_{f}\bm{S}_{f}\cdot(\Gamma_{\phi}\nabla\phi)_{f} .
\end{equation}

Replacing the terms in equation (\ref{eq:transportsurf}) with the approximated ones, (\ref{eq:convectiveapprox}) and (\ref{eq:diffusiveapprox}), we obtain the following equation: 

\begin{equation}\label{eq:transportflux}
\frac{\partial}{\partial t}\int_{V_p}\phi dV + \sum_{f}\bm{S}_{f}\cdot(\bm{u}\phi)_{f} - \sum_{f}\bm{S}_{f}\cdot(\Gamma_{\phi}\nabla\phi)_{f} = 0 ,
\end{equation}

where the second and third terms correspond to the convective and diffusive fluxes, respectively. The convective fluxes can be interpolated using linear interpolation or, in some cases, upwind or second order linear upwind schemes. The diffusive terms are usually discretized by a central difference scheme which is second order accurate. For non-orthogonal meshes, the interpolation of the diffusive fluxes takes also into account a non-orthogonal correction. The temporal discretization can be performed using any temporal discretization scheme such as the Euler implicit, Crank-Nicolson, backward differencing, forward Euler etc. For more information the reader could refer to \cite{jasak1996error}.

\section{Reduced Order Model Framework}\label{sec:ROM}

The main idea of reduced order modeling is to find a spatial basis $\bm{\phi}(\bm{x})$, which spans a subspace $\mathcal{S}$, to express the full order state vector (velocity, pressure, temperature etc) as $\bm{u}(\bm{x},\bm\mu,t) \approx\bm{u}_s=\sum_{i=1}^{N_{u}^s}\alpha_i(t,\bm\mu)\bm{\phi}_i(\bm{x})$, where $\bm{u}_s$ denotes the reduced field, $\alpha_i(t,\bm\mu)$ are some temporal coefficients which depend on the parameter vector $\bm\mu$ and $N_{u}^s$ is the cardinality of the POD space for the velocity. The same principle is applied for temperature and pressure. The basis can be generated using a plethora of methods, for example POD, Reduced Basis with a greedy approach, Proper Generalized Decomposition (PGD) etc. In this work, the reduced basis is calculated using the snapshot POD method.  For the assembly of the snapshot matrix, an equispaced grid (Cartesian grid) both in the time and in the parameter space has been utilized. This method results to a global snapshot matrix which combines together snapshots for every time-step and for every parametric value. This method has given satisfactory results for the scope of our paper, given that cases in the laminar region have been considered. For parametrized problems, other sampling techniques include the greedy-POD method \cite{Haasdonk2013}, the goal-oriented POD-greedy sampling \cite{hoang2015efficient} or a two-field greedy sampling strategy\cite{hoangfast2015,hoang2018fast}. For more details about the Reduced Basis and PGD methods, the reader could refer to \cite{Rozza2008,ChinestaEnc2017,Kalashnikova_ROMcomprohtua,quarteroniRB2016,Chinesta2011,Dumon20111387}.   

\subsection{Proper Orthogonal Decomposition}  

In the snapshot POD, state vector solutions are gathered using a high fidelity solver. Considering, for example, the velocity snapshots, $\bm{u}_s$, are then placed into an $N^h_u\times N_{u}^s$ snapshot matrix, $\bm{U}$, where $N^h_u$ is the number of degrees of freedom (grid points$\times$ number of components) and $N_{u}^s$ is the number of snapshots. Since we are dealing with parametric model order reduction, the total number of snapshots is not equal to the number of time instances only. The size of the parameter space should also be taken into account. The FOM is solved for each $\bm\mu^k \in \mathcal{K}=\{ \mu^1, \dots, \mu^{N_{\mu}}\} \subset \mathcal{P}$ where $\mathcal{K}$ is a finite dimensional training set of samples chosen inside the parameter space $\mathcal{P}$ and for each time instance $t^k \in \{t^1,\dots,t^{N_t}\} \subset [0,T]$. Therefore, the total number of snapshots, $N_{u}^s$, is equal to $N_{\mu} \cdot N_t$. One of the attributes of the POD basis is the minimization of the error between the velocity snapshots and their projection onto the POD basis. In the $L^2$-norm, this statement leads to the following least-squares problem:

\begin{gather}\label{eq:pod_energy}
\mathcal{V} = \mbox{arg min} \frac{1}{N_{u}^s}\sum_{i=1}^{N_{u}^s}||\bm{u}_i(\bm{x},\bm{\mu},t) - \sum_{i=1}^{N_{u}^s}(\bm{u}_i(\bm{x},\bm{\mu},t),\bm{\phi}_i(\bm{x}))\bm{\phi}_i(\bm{x})||^2_{L^2} .
\end{gather}

Using the property $||\bm A\bm x-\bm b||^2_{L^2} = (\bm A\bm x-\bm b)^T(\bm A\bm x-\bm b)$, the above problem (\ref{eq:pod_energy}) can be written as:

\begin{equation}
\bm{C}\bm{W} = \bm{W}\bm{\lambda} ,  
\end{equation} 

where $\bm{C}\in\mathbb{R}^{N_{u}^s\times N_{u}^s}$ is the correlation matrix, $\bm{W}\in\mathbb{R}^{N_{u}^s\times N_{u}^s}$ a matrix for the eigenvectors and $\bm{\lambda}\in\mathbb{R}^{N_{u}^s\times N_{u}^s}$ is a diagonal matrix which contains the eigenvalues. Since the correlation matrix is positive and semi-definite, it can be written as follows:

\begin{equation}
C_{ij} = \ltwonorm{\bm{u}_i,\bm{u}_j} \mbox{\hspace{0.5cm} for } i,j = 1,\dots, N_{u}^s . 
\end{equation} 

To take into advantage the $L^2$-norm optimality of the POD method, the 'most-energetic' modes should be retained. Thus, the original spatial POD basis, $\mathcal{V}=$span$[\bm{\phi}_1,\bm{\phi}_2,...\bm{\phi}_{N_u^s}]$, is truncated using the following energy retained criterion:

\begin{equation}
E_{N^r_u}=\frac{\sum_{i=1}^{N^r_u}\lambda_{i}}{\sum_{j=1}^{N_{u}^s}\lambda_{j}} ,
\end{equation}

where $\lambda_{i}$ are the eigenvalues and $N^r_u$ is the number of the most energetic modes which are retained. Therefore, the truncated POD space, $\hat{\mathcal{V}} =$span$[\bm{\phi}_1,\bm{\phi}_2,...\bm{\phi}_{N_u^r}]$ $\subset\mathcal{V}$, has a new cardinality $N^r_u$. 

The orthogonal POD basis functions, $\bm{\phi}(\bm{x})$, are calculated and normalized as:

\begin{eqnarray}\label{eqn2}
\bm{\phi}_j &=&   \frac{1}{\sqrt{\lambda}_{i} N^r_u}\sum_{j=1}^{N^r_u}\bm{u}_j \bm{W}_{ij}, \\
\ltwonorm{\bm{\phi}_i,\bm{\phi}_j} &=& \delta_{ij}  \hspace{0.5cm}\forall\mbox{ } i,j = 1,\dots,N^r_u.
\end{eqnarray}

The same approximation is applied for the pressure and temperature fields. However, as pressure and temperature are scalar fields, the basis functions which are denoted as $\psi(\bm{x})$ $\in \mathbb{R}^{N_p ^h}$ and $\chi(\bm{x})$  $\in \mathbb{R}^{N_{\theta}}$ respectively, are now scalar functions. For each field, different temporal coefficients are considered, denoted as $b(t,\bm\mu)$ and $c(t,\bm\mu)$ respectively. Thus, the POD decomposition of the velocity, pressure and temperature reads:

\begin{eqnarray}
\bm{u}(\bm{x},\bm\mu,t) \approx \bm{u}_r &=& \sum_{i=1}^{N_{u}^r}\alpha_i(\bm\mu,t)\bm\phi_i(\bm{x}), \\
p(\bm{x},\bm\mu,t) \approx p_r &=& \sum_{i=1}^{N_p^r}b_i(\bm\mu,t)\psi_i(\bm{x}), \\
\theta(\bm{x},\bm\mu,t) \approx \theta_r &=&  \sum_{i=1}^{N_{\theta}^r}c_i(\bm\mu,t)\chi_i(\bm{x}) ,
\end{eqnarray}

where $\bm{u}_r$, $p_r$ and $\theta_r$ are the reduced fields. For more information about the reconstruction and stabilization of the pressure field the reader could refer to the equation (\ref{eq:sup_problem}). 

\subsection{Galerkin Projection}

The reduced order model can be obtained by projection techniques including Galerkin or Petrov-Galerkin projection of the full order Navier-Stokes/temperature equations (\ref{eq:navstokes}) onto the POD spatial basis $\bm{\phi}(\bm{x})$, $\psi(\bm{x})$ and $\chi(\bm{x})$. The projection leads to an ordinary differential equation (ODE) for the evolution of the temporal coefficients $\alpha(t,\bm\mu)$, $b(t,\bm\mu)$ and $c(t,\bm\mu)$ respectively. In this work, Galerkin projection is utilized but the reader could read \cite{Fang2013,Amsallem2012,Carlberg2010} for more information regarding the Petrov-Galerkin method.

 
Taking the projection of the Navier-Stokes equations onto the POD bases $\bm{\phi}(\bm{x})$ and $\psi(\bm{x})$ and exploiting the orthogonality, we obtain the following ODEs:

\begin{eqnarray}
\sum_{j=1}^{N_u^r}M_{ij}\frac{\partial \alpha_j}{\partial t} &=& \sum_{j=1}^{N_u^r}\sum_{k=1}^{N_u^r}Q_{ijk}\alpha_{j}\alpha_{k} + \nu\sum_{i=1}^{N_u^r}L_{ij}\alpha_{i} - \sum_{i=1}^{N_p^r}P_{ij}b_{i}, \\
\sum_{j=1}^{N_p^r}R_{ij}\alpha_{j} &=& 0 ,
\end{eqnarray}

where the reduced quadratic and linear terms, $Q_{ijk}$, $M_{ij}$, $L_{ij}$ and $K_{ij}$ are represented by the following matrices:

\begin{align}
M_{{ij}} & =  \ltwonorm{\bm{\phi}_i, \bm{\phi}_j}, \\
Q_{{ijk}} & =  \ltwonorm{\nabla\cdot(\bm{\phi}_i\otimes\bm{\phi}_j),\bm{\phi}_k}, \\
L_{{ij}} & =   \ltwonorm{\nu\Delta\bm{\phi}_i,\bm{\phi}_j},\\
P_{{ij}} & =   \ltwonorm{\nabla\psi_i,\bm{\phi}_j},\\
R_{{ij}} & =   \ltwonorm{\nabla\cdot\bm\phi_i,{\psi}_j}.
\end{align}

For computational efficiency reasons, the non-linear convective term which is represented by a third order tensor $Q_{{ijk}}$ evaluated as $(\bm{Q}(\alpha)\alpha)=\bm{\alpha^T}\bm{Q}_{{i\bullet \bullet}}\bm{\alpha}$. 
\\
\\ 
The projected initial conditions read:

\begin{equation}
\alpha_i(0) = (\bm{u}(\bm{x},\bm\mu,0),\bm{\phi}_i) .
\end{equation}

For the projection of the heat transport equation, we follow the same procedure, considering now the projection of the heat equation onto the POD bases $\chi(\bm{x})$ which, after some manipulation of the terms becomes as follow: 

\begin{equation}\label{eqn9}
\sum_{j=1}^{N_{\theta}^r}K_{ij}\frac{\partial c_j}{\partial t} = \sum_{j=1}^{N_u^r}\sum_{k=1}^{N_{\theta}^r}G_{ijk}\alpha_{j}c_{k} + \alpha_{dif}\sum_{j=1}^{N_{\theta}^r}N_{ij}c_{j} ,
\end{equation}

where the reduced quadratic and linear terms, $G_{ijk}$, $K_{ij}$ and $N_{ij}$ are defined as: 

\begin{align}\label{eqn8}
K_{{ij}}  & =  \ltwonorm{\chi_i, \chi_j}, \\
G_{{ijk}} & =  \ltwonorm{\nabla\cdot(\bm{\phi}_i\chi_j),\chi_k}, \\
N_{{ij}}  & =  \ltwonorm{\alpha_{dif}\Delta\chi_i,\chi_j} .
\end{align}

The initial conditions for the temperature are also projected onto the POD basis as $c_i(0) = (\theta(\bm{x},\bm\mu,0),\chi_i)$.

To summarize all the above, the reduced order model is governed by the following set of ODEs, which are then discretized in time using any temporal discretization scheme.

\begin{equation}\label{eq:romequations}
\begin{cases}
\bm{M}\bm{\dot{\alpha}} = \bm{\alpha}^T\bm{Q}\bm{\alpha} + \nu\bm{L}\bm{\alpha}  - \bm{P}\bm b, \\
\bm{K}\bm{\dot{c}} = \bm{\alpha}^T\bm{G}\bm{c} + \alpha_{dif}\bm{N}\bm{c}, \\
\bm{O}^T\bm{\alpha} = 0 ,      
\end{cases}
\end{equation}  

where $\bm{O_{ij}} = \ltwonorm{\nabla\cdot\bm{\phi}_i,\bm{\phi}_j}$ is the reduced matrix associated with the continuity equation $\nabla\cdot\bm{u} = 0$.

\subsection{Pressure Field Reconstruction and Stabilization using the Supremizer Enrichment Method}
The projection of the pressure gradient, ($\nabla p$), onto the POD basis can be derived using Green's theorem as follows:

\begin{equation}
\ltwonorm{\bm\phi, \nabla p} = \int_{\Omega} \bm\phi\cdot\nabla p d\Omega = -\int_{\Omega}\nabla\cdot\bm\phi p d\Omega + \int_{\partial{\Omega}} p\bm\phi\cdot\bm{n}dS = \int_{\partial{\Omega}}p\bm\phi\cdot\bm{n}dS .
\end{equation}

In ROMs, the contribution of the pressure field is not always taken into account. The volume integral term is taken equal to zero since, for incompressible flows, the velocity basis functions are computed using divergence free snapshots. Therefore, the pressure term depends only on the boundary $\Gamma$. In the case where enclosed flows ($\bm\phi\cdot\bm{n} = 0$ on $\partial\Omega$) or flows with inlet-outlet conditions with the outlet being far away from the obstacle are considered, the pressure term vanishes completely \cite{Ma2002,deane1991}. However, as indicated in \cite{noack_papas_monkewitz_2005}, the pressure term can not always be neglected, especially when unstable shear layers are considered or when pressure drop calculations are important, such as pressure drop in pipes. To solve this issue, many different solutions have been proposed. In \cite{akhtar2009stability} a method of taking the divergence of the Navier-Stokes momentum equation to obtain a Poisson equation for pressure which is projected onto a POD basis is proposed. In \cite{Stabile2017}, the Poisson equation method is adapted to a finite volume context. Bergmann \textit{et al} in \cite{Bergmann2009}, suggested a global POD basis for both the pressure and the velocity fields and decomposed the fields using the same temporal coefficients. In \cite{Stabile2018} in a finite volume and in \cite{Rozza2007,Ballarin2014} in a finite element context, a supremizer enrichment method has been proposed. This approach is also followed on this paper for modeling the pressure field in the ROM. 

The idea is that the velocity POD space is enriched with velocity supremizer snapshots where these additional basis functions are chosen in a way to satisfy the inf-sup (Ladyzhenskaya-Brezzi-Babuska) condition \cite{BREZZI199027,boffi_mixed}:

\begin{equation}
\inf_{q_h \in \mathcal{Q}} \sup_{\bm{v_h} \in \mathcal{V}} \frac{\langle \nabla \cdot \bm{v_h}, q_h \rangle}{\norm{\nabla \bm{v_h}}\norm{q_h}} \ge \beta > 0.
\end{equation}

where $\beta$ is a constant which does not depend on the discretization parameter $h$. The size of the enriched velocity POD spaces is now a subset of $\mathbb{R}^{N_u ^h \times (N_u ^s+N_{sup}^s)}$ where $N_{sup}^s$ is the size of the supremizer basis functions. The supremizer enrichment is given by solving the following equations for each pressure basis function:        
\begin{equation}\label{eq:sup_problem}
\begin{cases}
\Delta \bm{s_i} = - \bm{\nabla} p_i &\mbox{ in } \Omega \\
\bm{s_i}=\bm{0} &\mbox{ on } \partial\Omega,
\end{cases} 
\end{equation}
where $\bm{s_i}$ denotes the supremizer solution. For a more detailed description of the above method, the reader could see \cite{Stabile2018,Ballarin2014}. 

\subsection{Boundary Conditions and Snapshot Homogenization}

One of the key aspects of the present work is the development of reduced order methods with parametrized boundary conditions. For this reason particular attention is devoted to this aspect. To enforce Dirichlet boundary conditions in the reduced order model we employ a similar approach as the one employed in \cite{Stabile2018}. This method was firstly proposed in \cite{NME:NME537} for boundary conditions that can be parametrized by a single multiplicative coefficient, as in the present case, and generalized for every type of function in \cite{Gunzburger2007}.
  
A lifting function is used to homogenize the snapshots so that they become independent of the boundary conditions. At the reduced order level, it is possible to specify the new boundary values and these values are then added back. The homogenized velocity value is written as:

\begin{equation}
\bm{u}'(\bm{x},\bm\mu,t) = \bm{u}(\bm{x},\bm\mu,t) - \sum_{j=1}^{N_{BC}}u_{D_j}(\bm\mu,t)\bm{\phi}_{c_j} ,
\end{equation} 

where $\bm{\phi}_{c_k}$ are divergence free control functions which are equal to the number of the parametrized boundaries, and $N_{BC}$ is the number of parametrized boundary conditions. The coefficients $u_{D_j}$ are determined is such a way to make the snapshots homogeneous after the subtraction of the chosen control function multiplied by the coefficient itself. Since we chose to have a number of control functions which is equal to the number of parametrized boundaries and that each control function assumes a uniform and unitary value at the boundary to which it refers and uniform null values on the other parametrized boundaries, the coefficient $u_{D_j}$ will assume the value that the snapshots have at the boundary. This process is described in algorithm~(\ref{alg1:liftingvelocity}).

The POD is applied to the homogeneous snapshots and the boundary value is added back so that:
 
\begin{equation}
\bm{u}(\bm{x},\bm\mu,t) = \sum_{j=1}^{N_{BC}}u_{D_j}(\bm\mu,t)\bm{\phi}_{c_j} + \sum_{i=1}^{N_{u}^s}\alpha_i(t,\bm\mu)\bm{\phi}_i(\bm{x}).
\end{equation} 

The values of the lifting functions are obtained by dividing the Dirichlet boundary in different parts $\Gamma_D = \bigcup_{i=1}^{N_{BC}} \Gamma_{D_i}$, one for each parametrized boundary condition. Then a full order problem is solved for each boundary condition following algorithm \ref{alg:lifting}. In the case of a problem with a non-linear dependency with respect to the boundary conditions, the full order problem should be solved with values of the boundaries as close as possible to those that one would like to test during the online stage. Also, in case of a non-zero forcing term, the forcing term should also be considered in the evaluation of the lifting functions.

\begin{algorithm}[t]
\caption{The algorithm for the generation of the velocity lifting functions}
\label{alg1:liftingvelocity}
\hspace*{\algorithmicindent} \textbf{Input:} $N_{BC}$, $\Gamma_D = \bigcup_{i=1}^{N_{BC}} \Gamma_{D_i}$, $N_{u}^s$=Total number of snapshots\\
\hspace*{\algorithmicindent} \textbf{Output:} $\{\bm{\phi}_{c_i}\}_{i=1}^{N_{BC}}$
\begin{algorithmic}[1]
\FOR{$i=1$ to $N_{BC}$}
\FOR{$j=1$ to $N_{BC}$}
\STATE if $i=j$ then $\bm{u}|_{\Gamma_{D_j}} = 1$; else $\bm{u}|_{\Gamma_{D_j}} = 0$
\ENDFOR
\FOR{$l=1$ to $N_{u}^s$}
\STATE Solve the full order problem and store the solution $\to \bm{u}_{il}$
\ENDFOR
\STATE $\bm{\phi}_{c_i} = \frac{1}{N_{u}^s} \sum_{l=1}^{N_{u}^s} \bm{u}_{il}$ 
\ENDFOR
\end{algorithmic}
\end{algorithm}
 
For the heat transport equation a similar approach is followed. Unlike with the velocity case, where a 'no-slip' condition is specified on the walls, in heat transfer problems, a homogeneous Neumann boundary condition is usually assigned (adiabatic walls). Usually, together with the boundary conditions, an initial condition for the internal field ($IF$) is also prescribed. A modification of the algorithm (\ref{alg:lifting}) is proposed here where also the initial value of the internal field is removed from the snapshots. In this way, one could parametrize the internal field initial condition as well. Therefore, apart from the lifting functions that are obtained for every Dirichlet boundary condition, the domain is now divided into $N_{BC}+1$  different parts $\Omega_R = \bigcup_{i=1}^{N_{BC}} \Gamma_{D_i}\bigcup\Theta_{IF}$ where the extra lifting function accounts for the initial internal field. The algorithm (\ref{alg:lifting}) is modified as follows:  

\begin{algorithm}[ht!]
\caption{The algorithm for the generation of the temperature lifting functions}
\label{alg:lifting}
\hspace*{\algorithmicindent} \textbf{Input:} $N_{BC}+1$, $\Omega_{R} = \bigcup_{i=1}^{N_{BC}} \Gamma_{D_{i}}\bigcup\Theta_{IF}$, $N_{\theta}^s$= Total number of snapshots\\
\hspace*{\algorithmicindent} \textbf{Output:} $\{{\chi}_{c_i}\}_{i=1}^{N_{BC+1}}$
\begin{algorithmic}[1]
\FOR{$i=1$ to $N_{BC}+1$}
\FOR{$j=1$ to $N_{BC}+1$}
\STATE if $i=j$ then $\theta|_{\Gamma_{R_j}} = 1$; else $\theta|_{\Gamma_{R_j}} = 0$
\ENDFOR
\FOR{$l=1$ to $N_{\theta}^s$}
\STATE Solve the full order problem and store the solution $\to \theta_{il}$
\ENDFOR
\STATE ${\chi}_{c_i} = \frac{1}{N_{\theta}^s} \sum_{l=1}^{N_{\theta}^s} \theta_{il}$ 
\ENDFOR
\end{algorithmic}
\end{algorithm}

During the calculation of the lifting functions, the adiabatic walls and the outlet still have homogeneous Neumann conditions as in the FOM.


The boundary condition independent temperature is written as:
 
\begin{equation}
\theta'(\bm{x},\bm\mu,t) = \theta(\bm{x},\bm\mu,t) - \sum_{j=1}^{N_{BC+1}}\theta_{R_j}(\bm\mu,t)\chi_{c_j}(\bm{x}) .
\end{equation} 

The POD is then applied to the temperature snapshots and, at the reduced order level, the boundary values, as well as the internal field initial value, are added back to the temperature equation:

\begin{equation}
\theta(\bm{x},\bm\mu,t) = \sum_{j=1}^{N_{BC}+1}\theta_{R_k}(\bm{x},\bm\mu,t)\chi_{c_j}(\bm{x}) + \sum_{i=1}^{N_{\theta}^s}c_i(t,\bm\mu)\chi_i(\bm{x}).
\end{equation}

\section{Numerical Experiments}\label{sec:num_exp}
In this section the proposed method is applied to a test case which consists of the well-studied non-isothermal mixing in a T-junction pipe. Two parametric cases are considered here: parametrization of the inlet temperature boundary conditions and parametrization of the kinematic viscosity.

\subsection{Non-isothermal Mixing in T-junction - Parametrization of the Temperature Inlet Boundary Conditions}\label{sec:tempinlet}
  
\begin{figure}
\centering
\includegraphics[width=\textwidth]{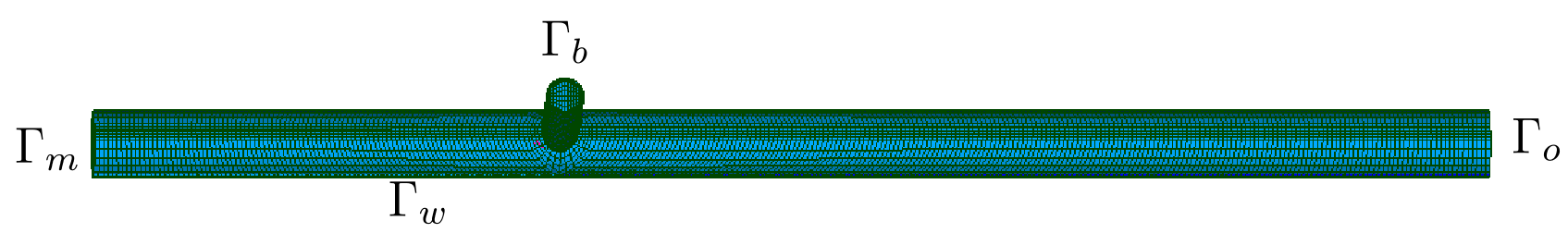} 
\caption{Sketch of the T-junction 3D mesh.}\label{fig:mesh_tjunction}
\end{figure}

\begin{figure*}[!tbp]
\centering
\includegraphics[width=0.55\textwidth]{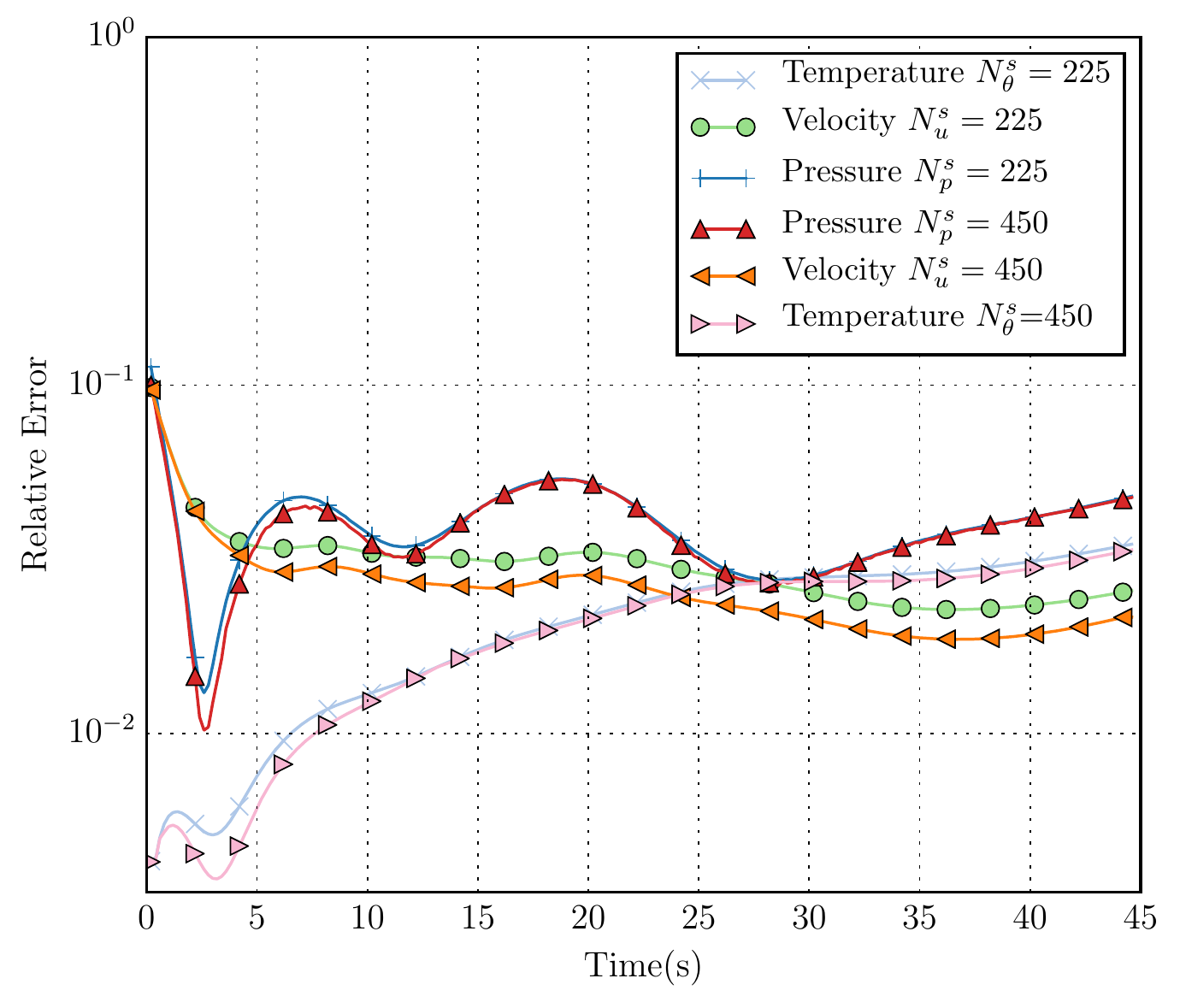}
\caption{$\epsilon_{L^2}(t)$ error ($\epsilon_{L^2}(t) = \frac{||X_{FOM}(t)-X_{ROM}(t)||_{L^2 (\Omega)}}{||X_{FOM}(t)||_{L^2 (\Omega)}}$) for two sampling frequencies for the snapshot collection, per $0.2$s where $N_u^s=N_{\theta}^s=N_p^s = 225$ and per $0.1s$ where $N_u^s=N_{\theta}^s=N_p^s = 450$s.  The ROM is run on $\theta_m=50^{\circ}$C and $\theta_b=70^{\circ}$C.}\label{fig:convergence}
\end{figure*}

\begin{figure}[!tbp]
  \centering
  \begin{minipage}[b]{0.45\textwidth}
    \includegraphics[width=\textwidth]{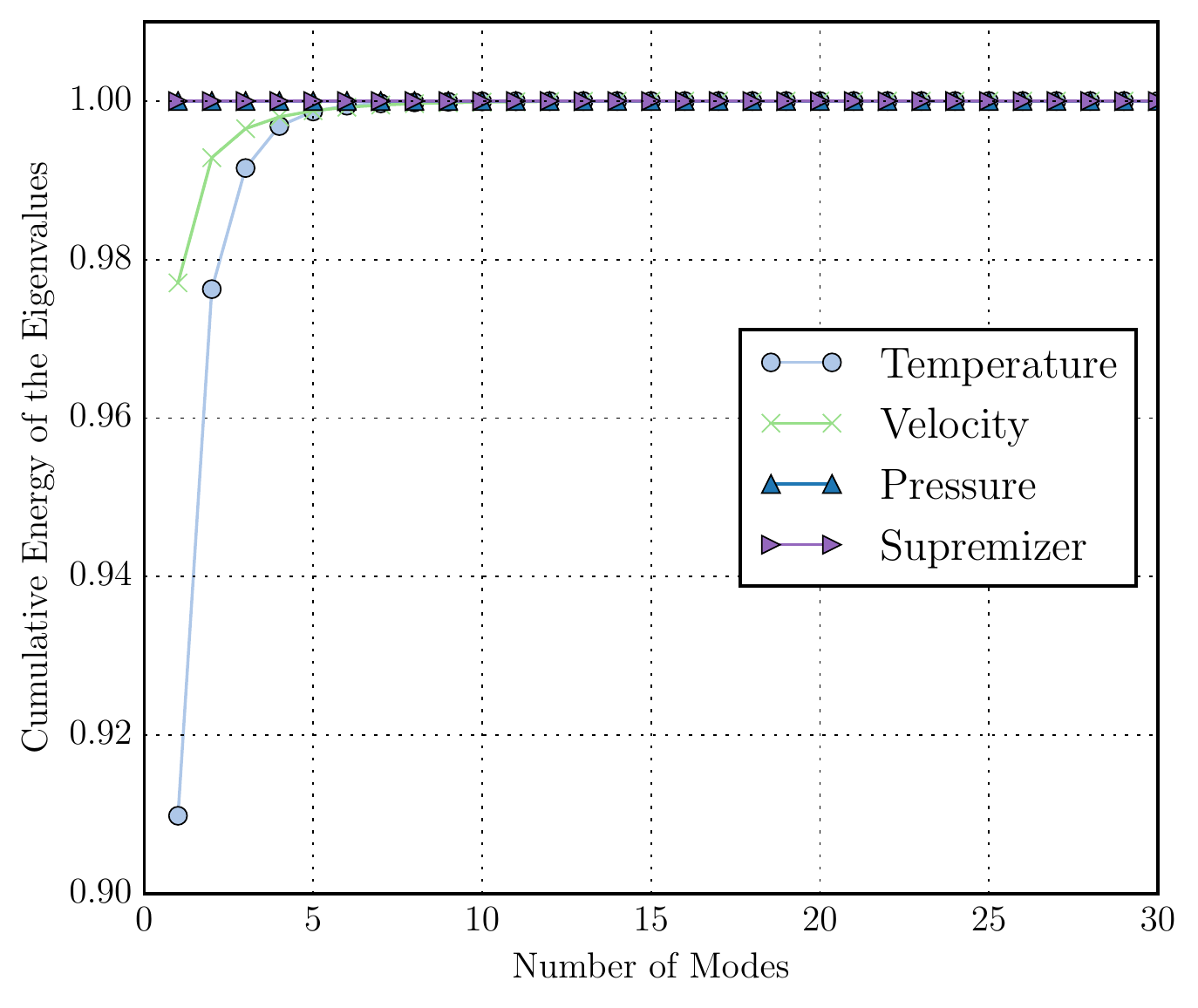}
    \caption{Cumulative energy of the eigenvalues for temperature, velocity, pressure and supremizer fields for the training case with temperature inlet boundary conditions  $\theta_m=50^{\circ}$C and $\theta_b=70^{\circ}$C.}\label{fig:eigenvalues}
  \end{minipage}
  \hfill
  \begin{minipage}[b]{0.45\textwidth} 
    \includegraphics[width=\textwidth]{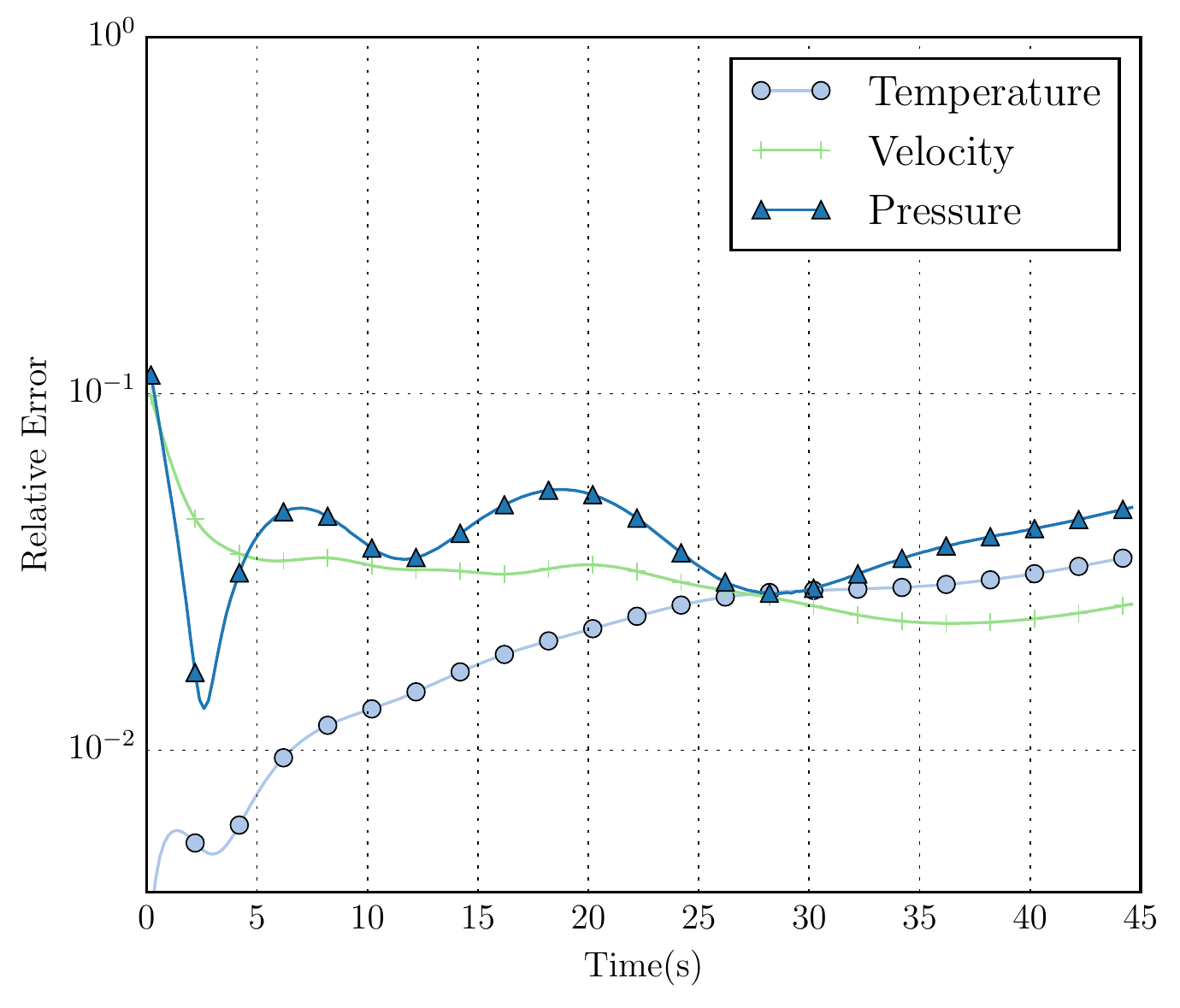}
    \caption{$\epsilon_{L^2}(t)$ error ($\epsilon_{L^2}(t) = \frac{||X_{FOM}(t)-X_{ROM}(t)||_{L^2(\Omega)}}{||X_{FOM}(t)||_{L^2(\Omega)}}$) plots for temperature, velocity and pressure fields obtained on the test case for temperature inlet boundary conditions $\theta_m=60^{\circ}$C and $\theta_b=80^{\circ}$C.}\label{fig:l2error}
  \end{minipage}
\end{figure}

\begin{figure*}[!tbp]
\begin{minipage}{1\textwidth}
\centering
\includegraphics[width=0.4\textwidth]{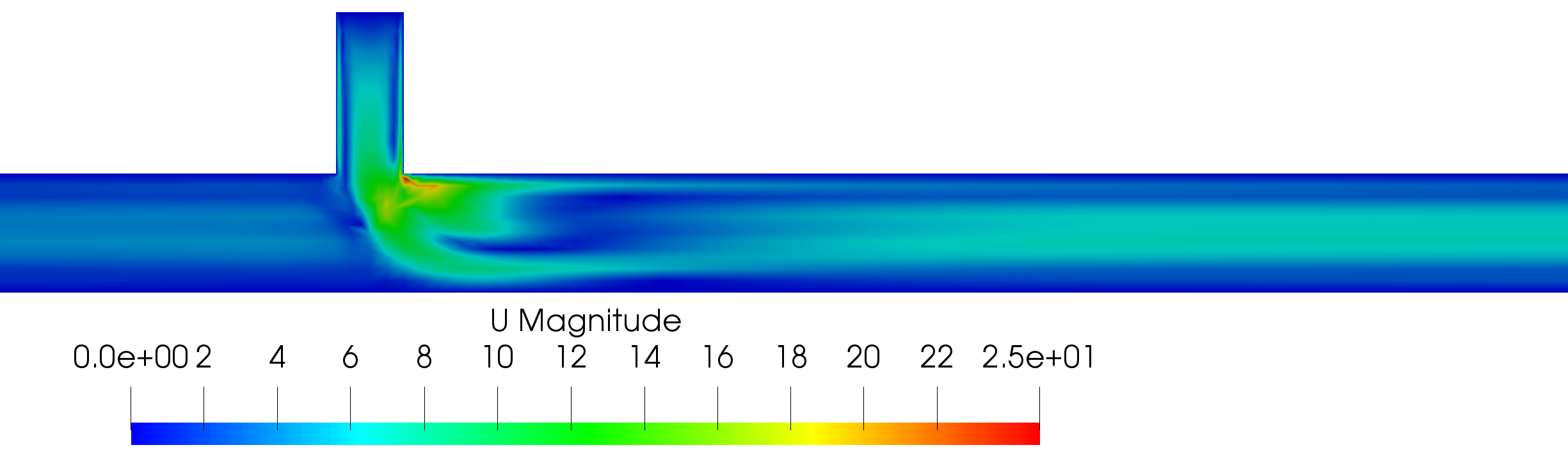}
\includegraphics[width=0.4\textwidth]{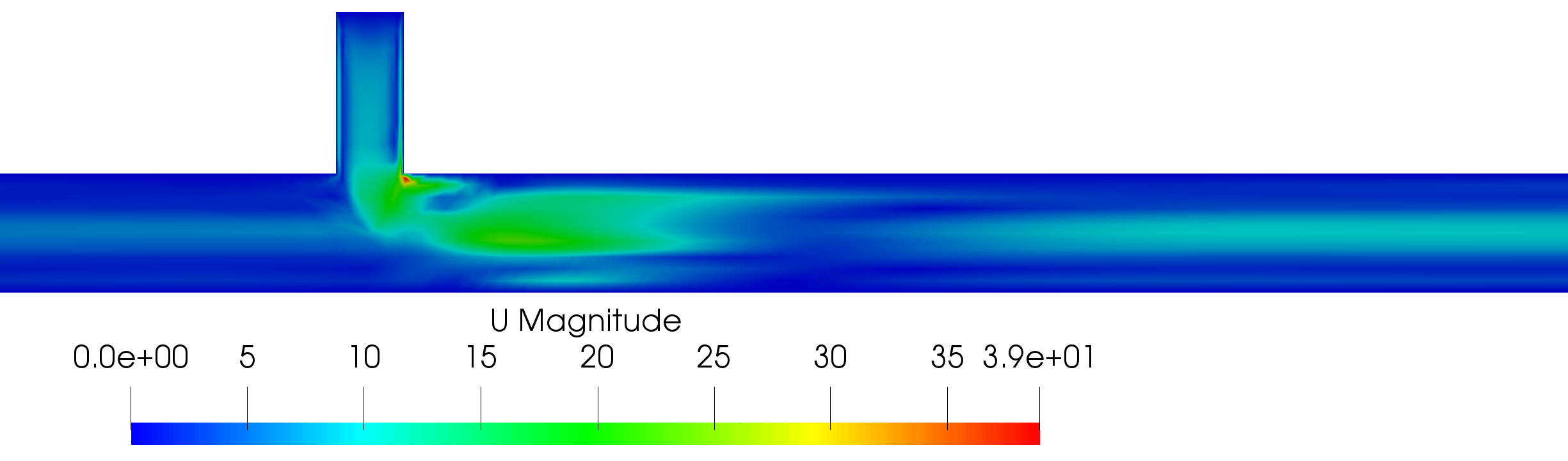}
\includegraphics[width=0.4\textwidth]{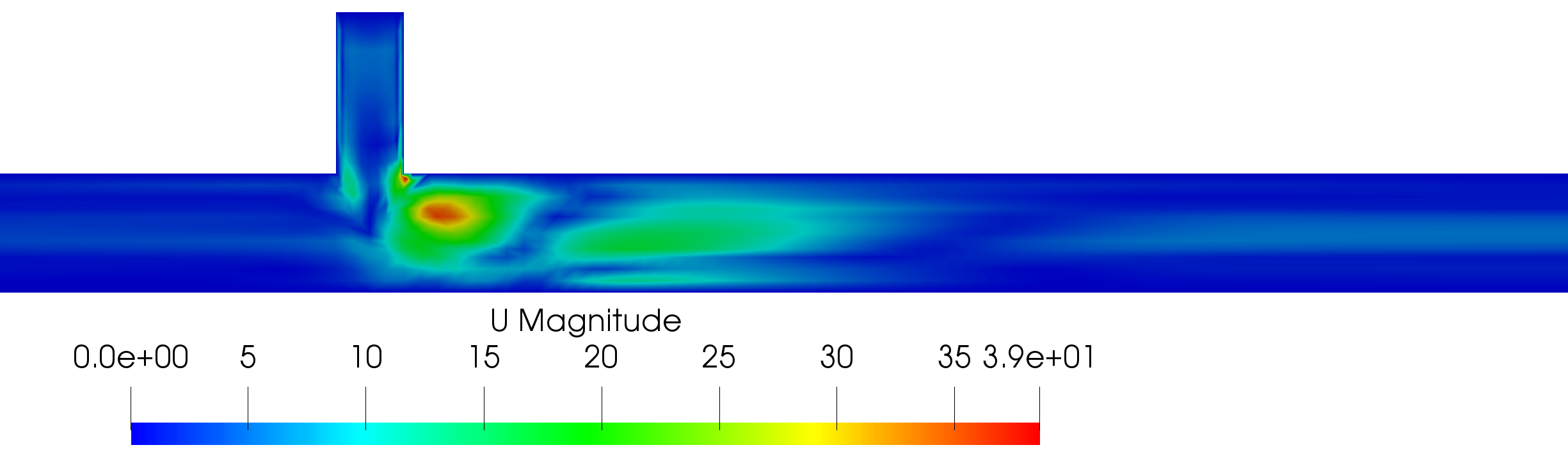}
\includegraphics[width=0.4\textwidth]{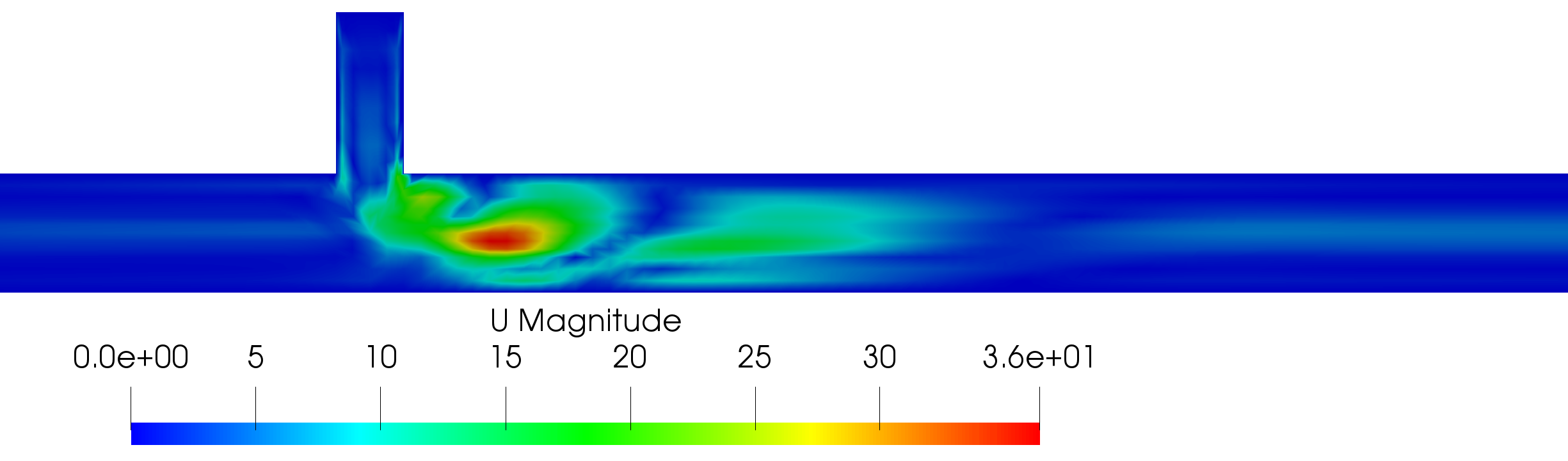}
\includegraphics[width=0.4\textwidth]{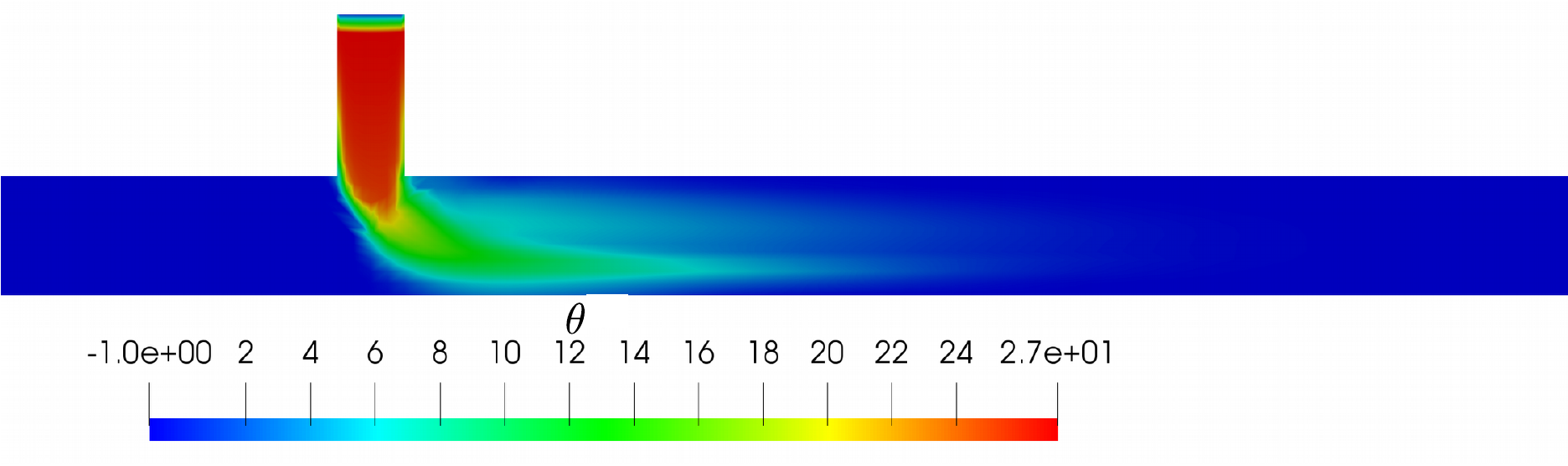}
\includegraphics[width=0.4\textwidth]{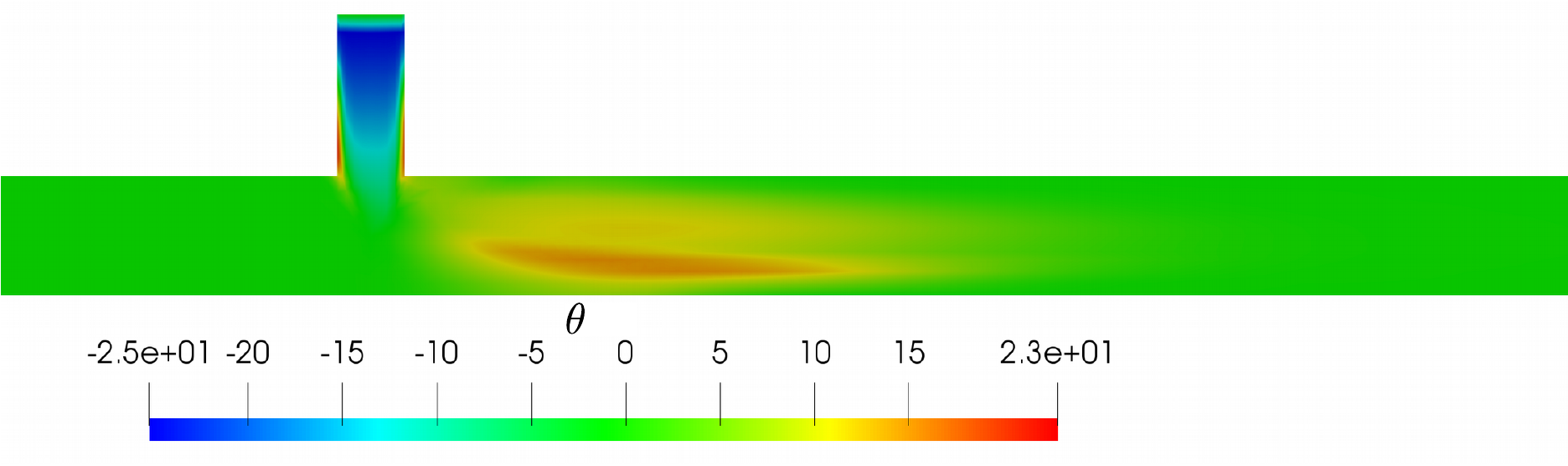}
\includegraphics[width=0.4\textwidth]{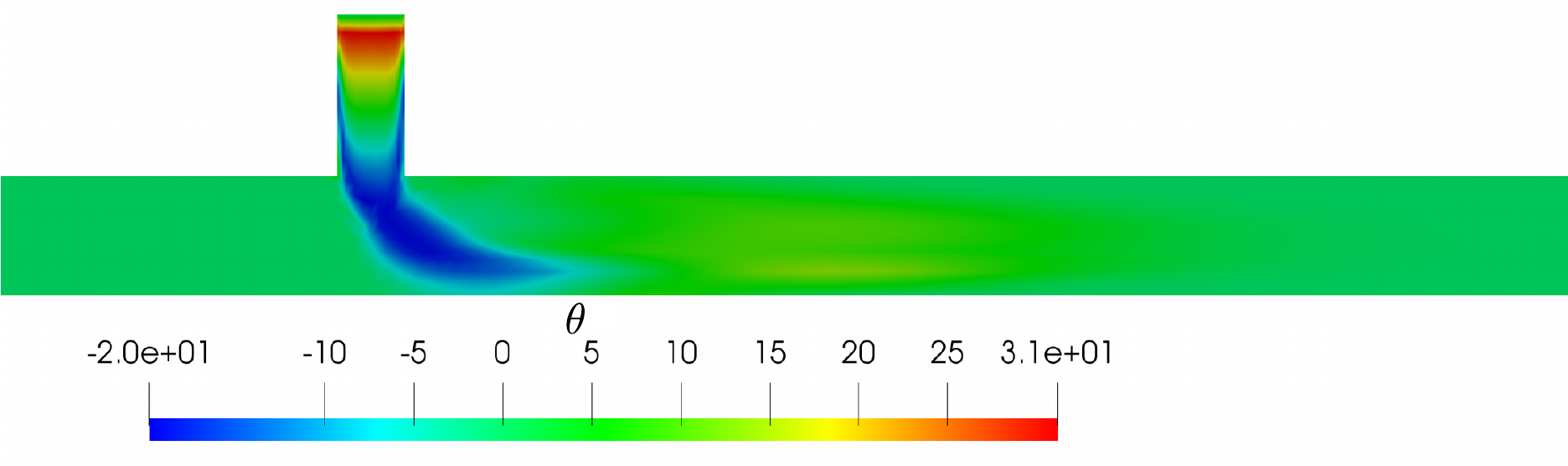}
\includegraphics[width=0.4\textwidth]{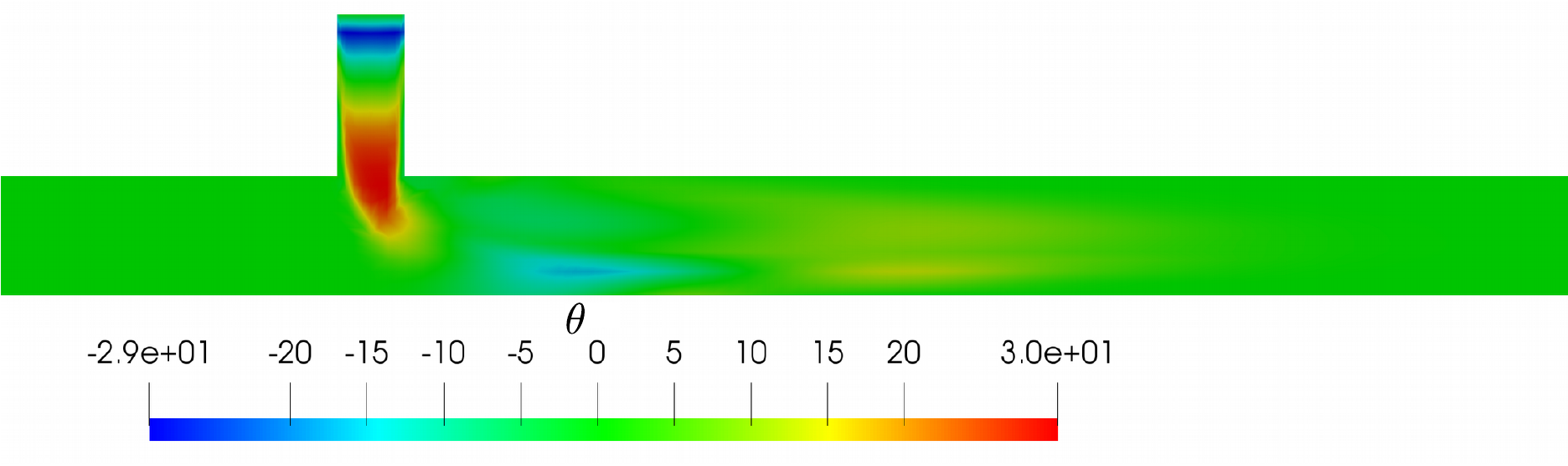}
\includegraphics[width=0.4\textwidth]{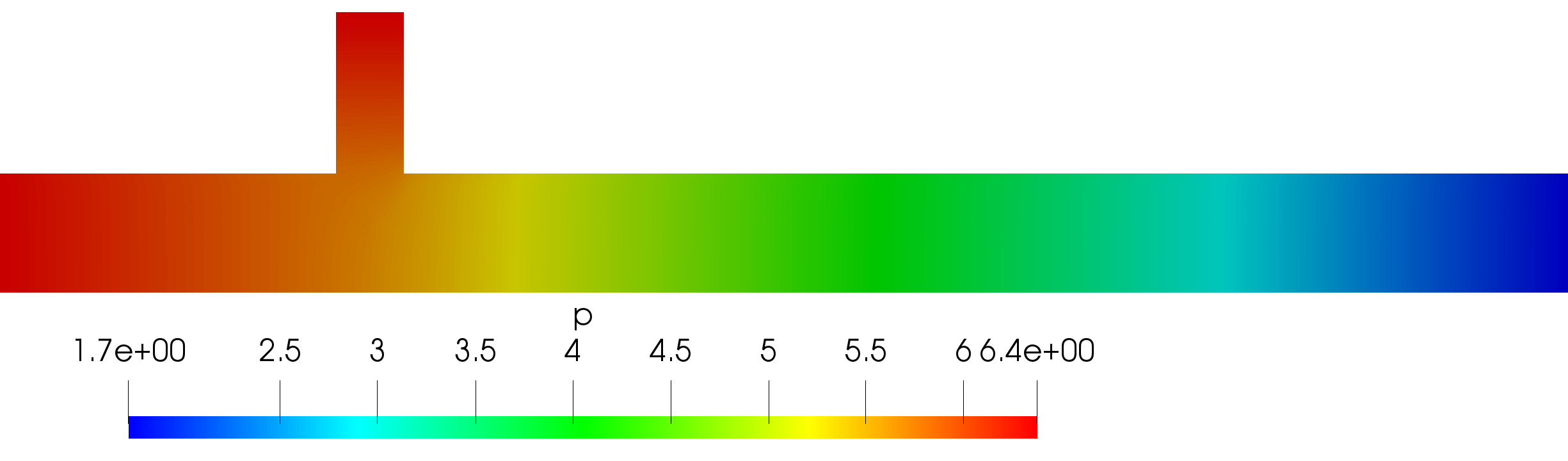}
\includegraphics[width=0.4\textwidth]{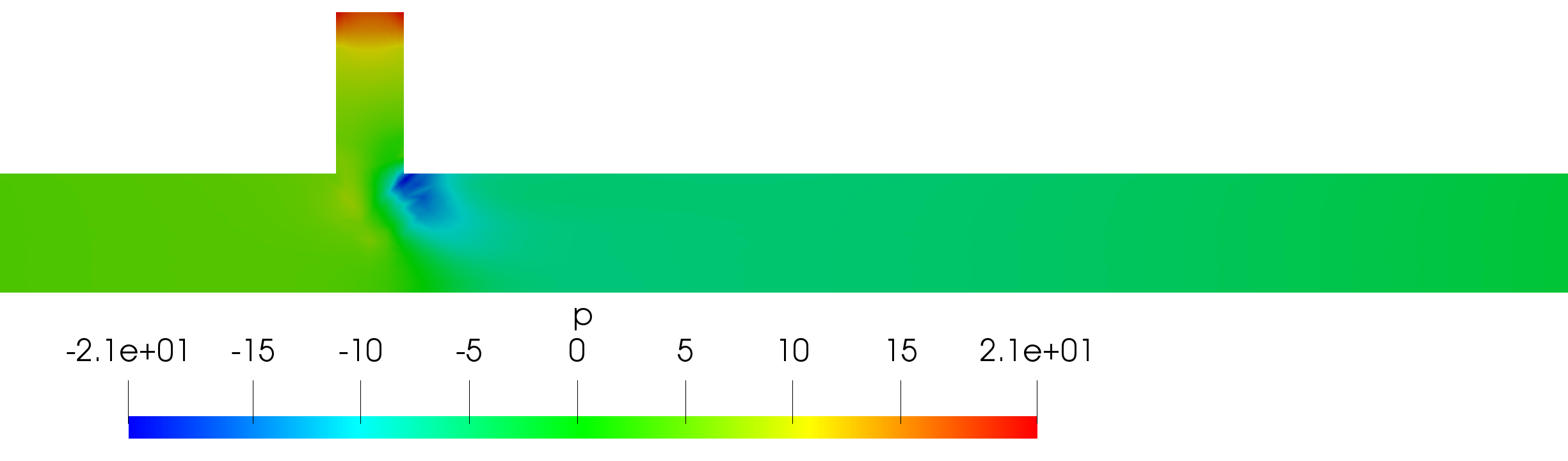}
\includegraphics[width=0.4\textwidth]{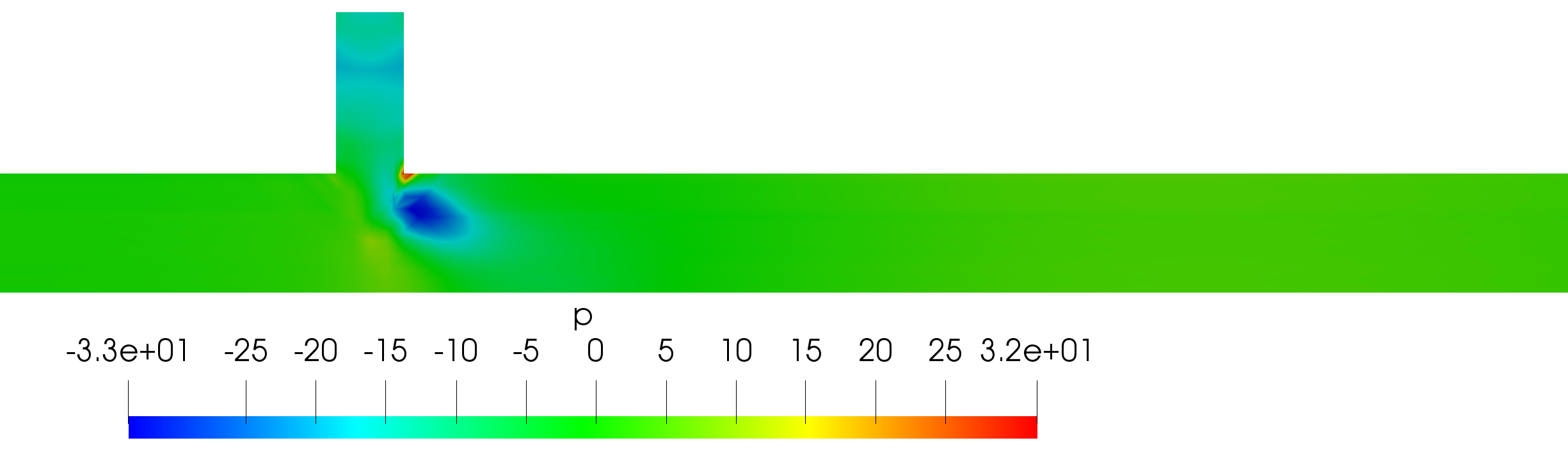}
\includegraphics[width=0.4\textwidth]{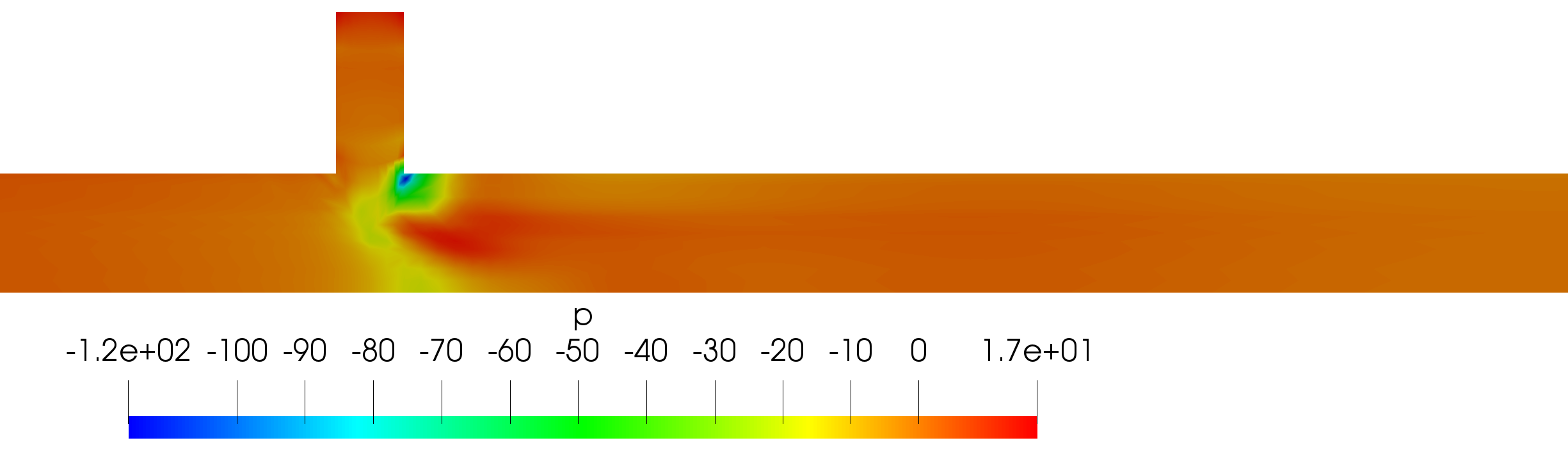}
\end{minipage} 
\caption{First four basis functions for velocity (first two rows),temperature (rows three and four) and pressure (last two rows) corresponding to $\theta_{m}=60^{\circ}$C and $\theta_{b}=80^{\circ}$C.}\label{fig:pod_modes} 
\end{figure*}

The test case consists of a 3D T-junction shaped pipe with main pipe hydraulic diameter $D_m=140$mm and branch pipe hydraulic diameter $D_b=80$mm and lengths of $L_m=3$m and $L_b=0.44$m respectively. The branch pipe is placed at the position of $0.33*L_m$. Streams of cold and hot water enter the system from the horizontal and the branch pipe and mix together in the T-junction region. The thermal diffusivity is taken as $0.160\times 10^{-6}$m/s$^2$ under atmospheric pressure. A summary of the physical parameters is shown on table (\ref{fig:parameters}). The computational domain which consists of 34490 elements, is divided into three boundary parts plus one part for the initial condition of the internal field, $\Omega_{R} = \Gamma_{m} \bigcup \Gamma_{b} \bigcup \Gamma_{o} \bigcup\Theta_{IF}$, as shown in figure (\ref{fig:mesh_tjunction}). The initial conditions are as shown in table(\ref{fig:boundary_con}). The FOM simulation is performed in OpenFOAM using a modified IcoFoam solver, which accounts also for the temperature transport equation. IcoFoam \cite{jasak1996error} is a transient solver which uses the PISO algorithm \cite{ISSA198640} to solve the incompressible Navier-Stokes equations. The spatial discretization of the convenctive terms is achieved using a combination of a second order central-differencing and upwind schemes. The diffusive terms are discretized using second order central-differencing corrected schemes. For more information about OpenFOAM numerical schemes, the reader could refer to \cite{jasak1996error}. For the temporal discretization, a first order Euler backward implicit scheme is used. The simulation is performed for $T=45s$ with timestep $\Delta T = 5\times10^{-3}$s and the snapshots are collected every $0.2$s using an equispaced grid method in time. Therefore, the dimension of the correlation matrix is $225\times225$ and $N_u^s=N_{\theta}^s=N_p^s = 225$. A convergence test as the number of snapshots increases has been performed. The frequency with which the snapshots are collected has been doubled, thus the snapshots are collected every $0.1$s. The dimension of the correlation matrix is now $450\times450$ and $N_u^s=N_{\theta}^s=N_p^s = 450$. Figure (\ref{fig:convergence}) shows the comparison between the two different sampling frequencies, showing that the relative error between the FOM and the ROM converges slightly better as the number of snapshots increases. However, since the differences in the convergence are very small, for computational saving reasons, the first sampling frequency (per $0.2$s) will be used for the generation of the results. Table (\ref{fig:errorstatistics}) shows the minimum, maximum and average $\epsilon_{L^2}$ error for each sampling frequency. Figure ($\ref{fig:eigenvalues}$) shows the cumulative energy of the eigenvalues for velocity, temperature, pressure and supremizer fields. In order to retain the $99.9 \% $ of the system's energy, $5$ modes for velocity, $5$ for temperature and $3$ for the pressure and supremizer are selected. This truncation reduces the original POD space from $N^{s}_u = ^{s}_p = N^{s}_{\theta} = N^{s}_{sup}= 225$ to $N_u^r=5$, $N_{\theta}^r = 5$  ,$N_p^r = 3$  and $N_{sup}^r=3$. Figure (\ref{fig:pod_modes}) shows the first $4$ POD modes and it is clear from Figure~\ref{fig:eigenvalues}) that the first mode captures most of the energy of the system.
   
The ROM computations are performed in the ITHACA-FV C\texttt{++} library \cite{RoSta17} to simulate a ROM with the same conditions as the FOM. To provide some quantitative results, the $\epsilon_{L^2}$ error is calculated as

\begin{equation}
\epsilon_{L^2}(t) = \frac{||X_{FOM}(t)-X_{ROM}(t)||_{L^2 (\Omega)}}{||X_{FOM}(t)||_{L^2 (\Omega)}}\% ,
\end{equation}

where $X_{FOM}$ is the value of a particular field in the FOM model and $X_{ROM}$ the one that is calculated using the ROM.
 
\begin{figure*}[!tbp]
\centering  
\includegraphics[width=0.55\textwidth]{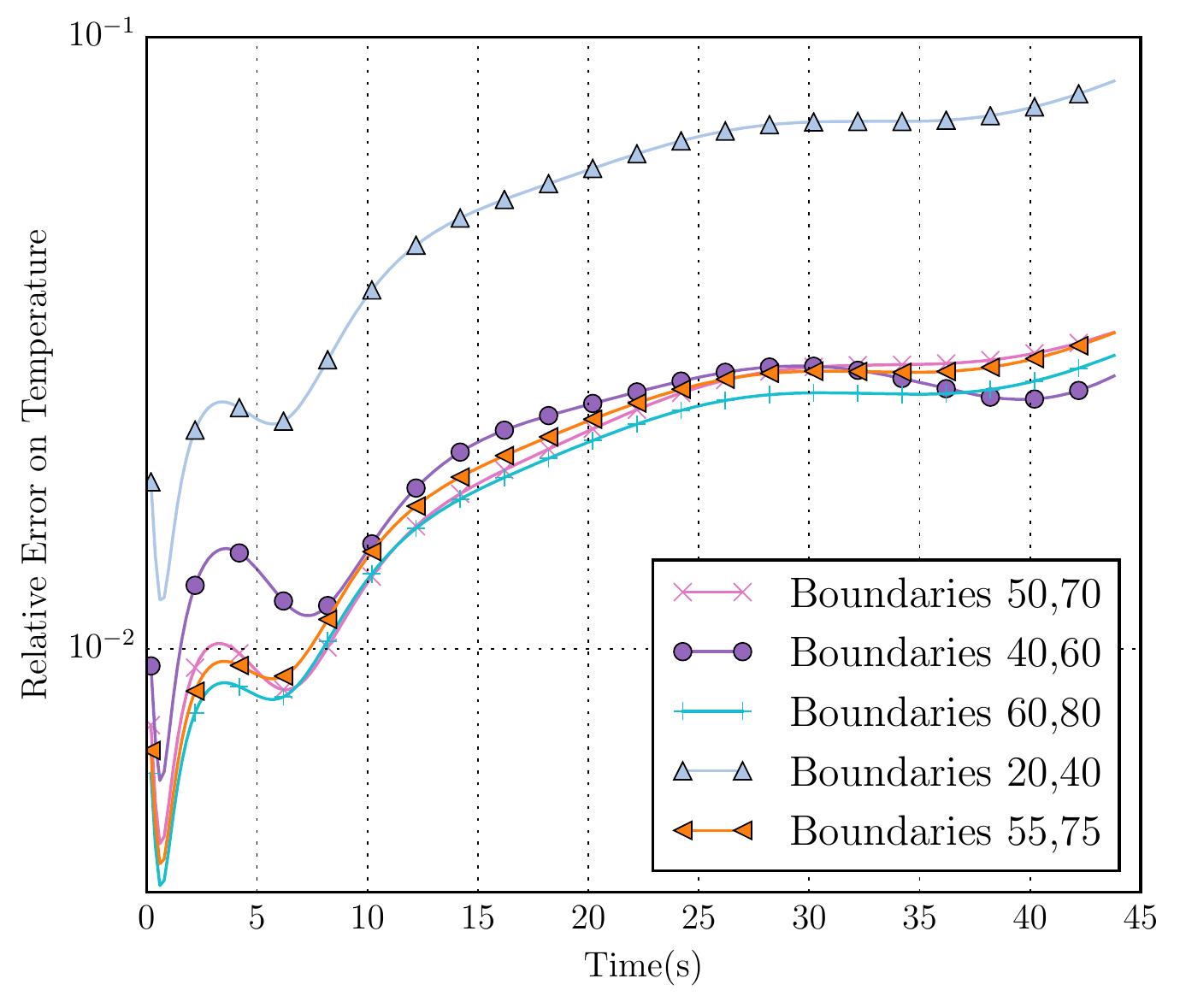}
\caption{$\epsilon_{L^2}(t)$ error ($\epsilon_{L^2}(t) = \frac{||X_{FOM}(t)-X_{ROM}(t)||_{L^2(\Omega)}}{||X_{FOM}(t)||_{L^2(\Omega)}}$) for different temperature inlet conditions. The ROM is trained on $\theta_m=50^{\circ}$C and $\theta_b=70^{\circ}$C and then is used to predict the temperature, velocity and pressure fields on four different test cases with sets of temperature inlets.}\label{fig:error_dif_runs}
\end{figure*}

\begin{figure*}[!tbp]
\begin{minipage}{1\textwidth}
\centering
\includegraphics[width=0.32\textwidth]{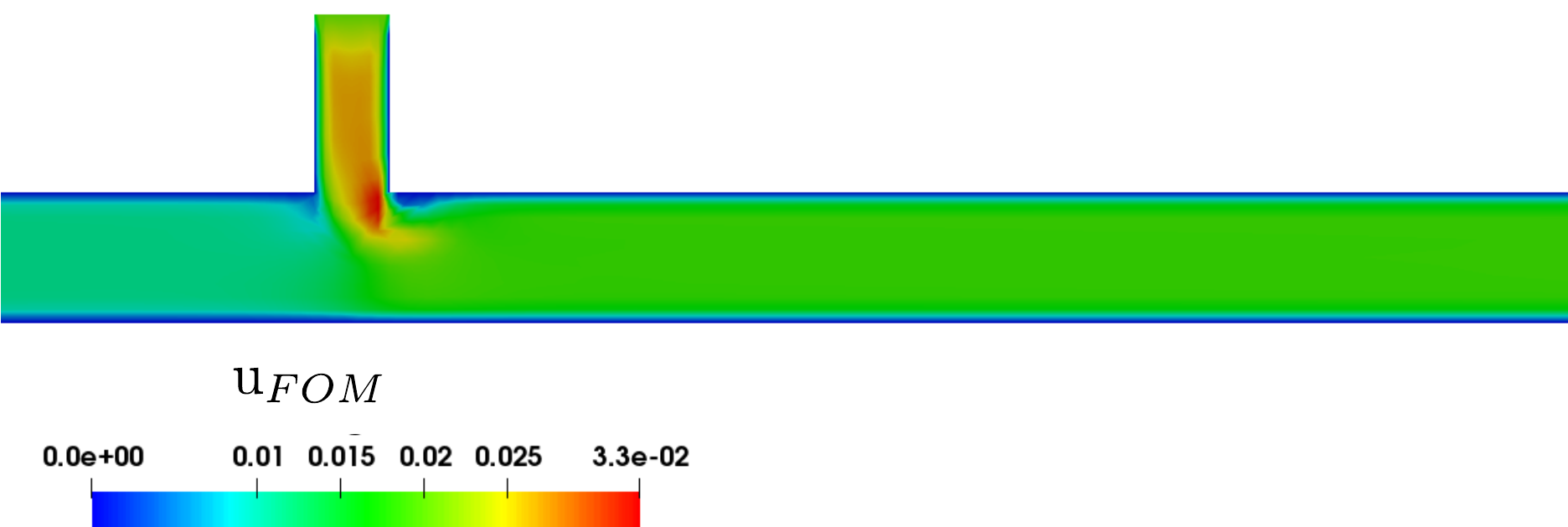}
\includegraphics[width=0.32\textwidth]{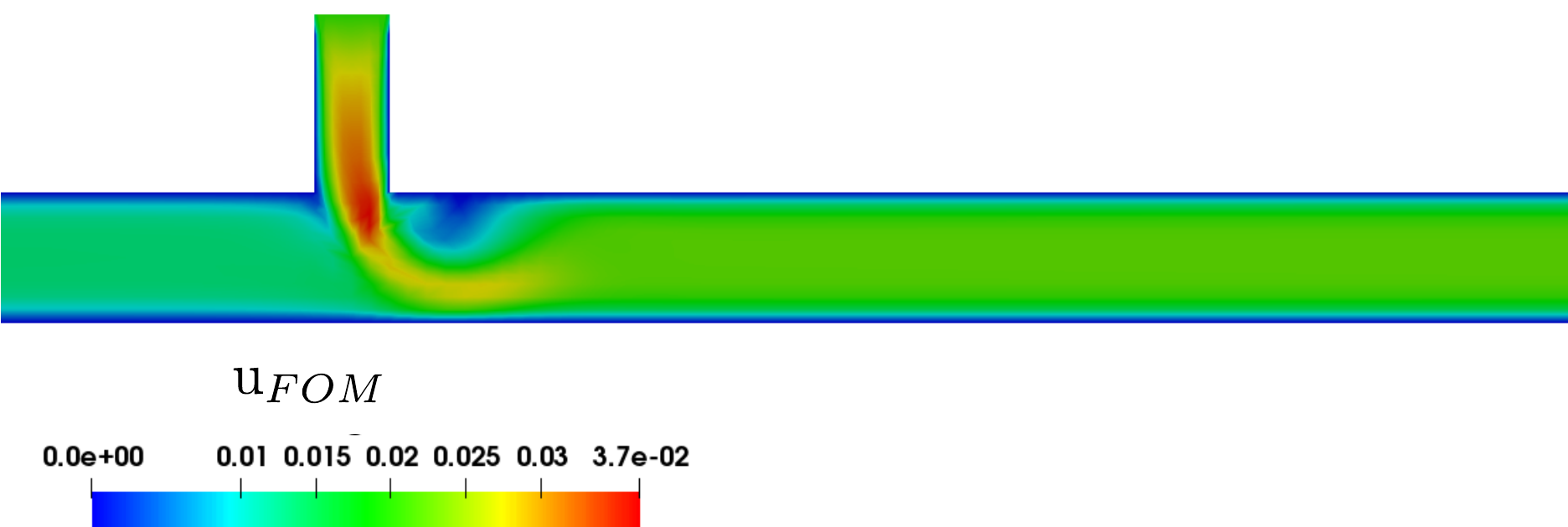}
\includegraphics[width=0.32\textwidth]{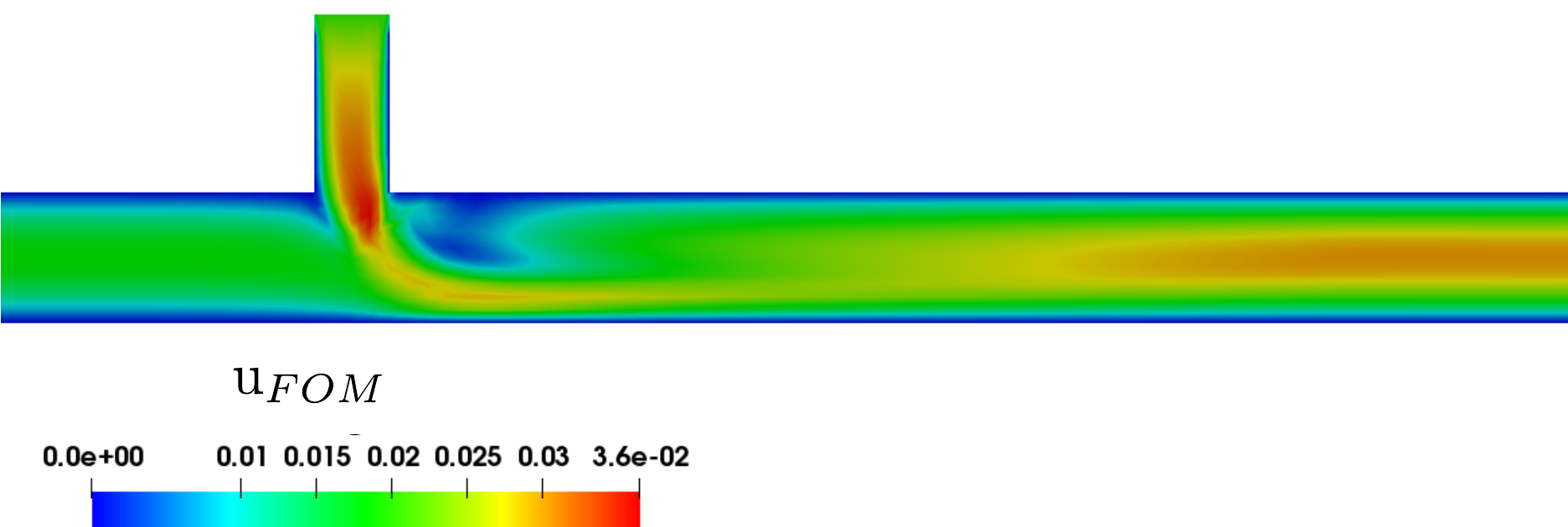}   
\includegraphics[width=0.32\textwidth]{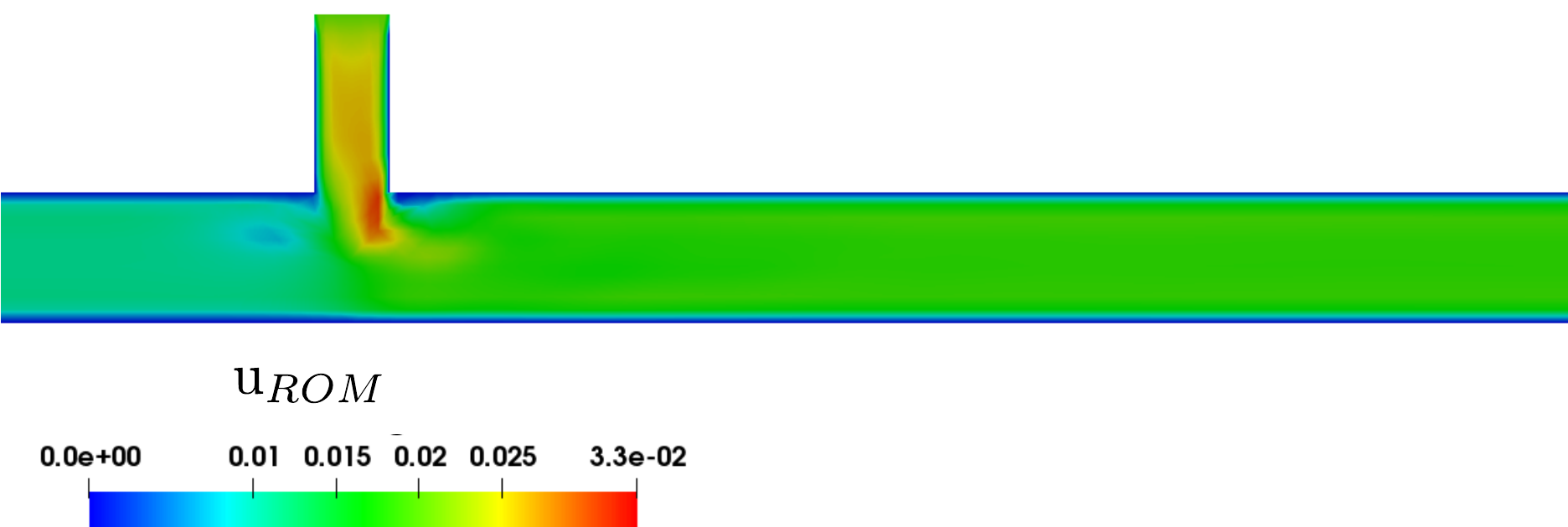}
\includegraphics[width=0.32\textwidth]{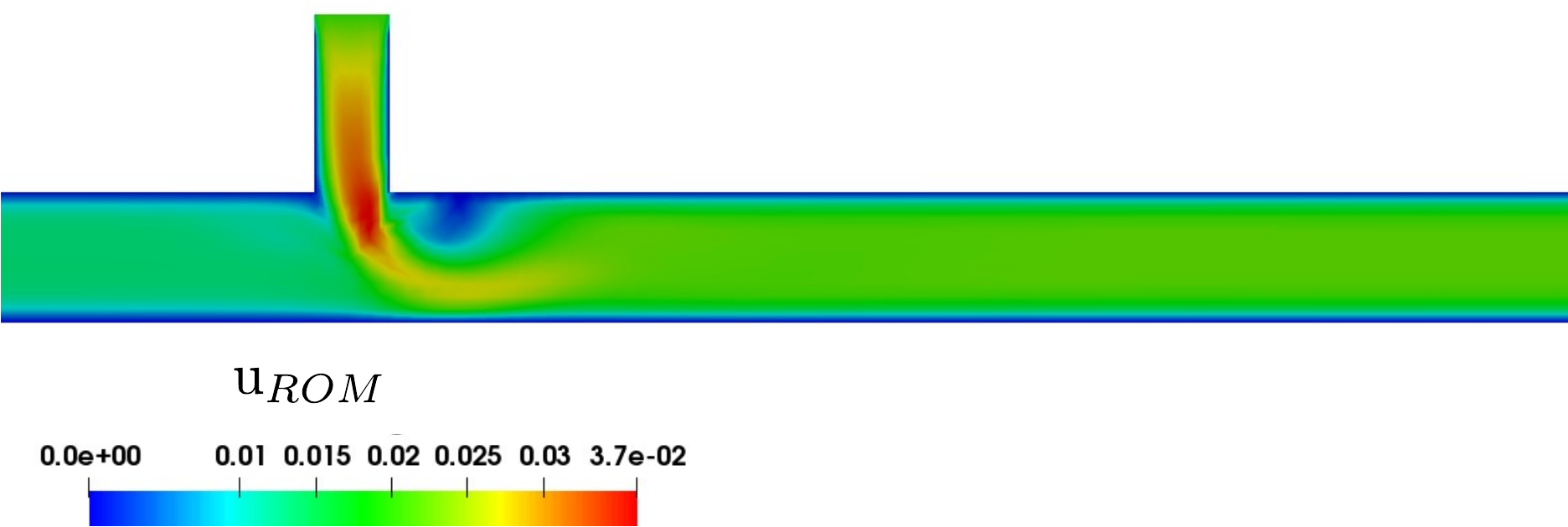}
\includegraphics[width=0.32\textwidth]{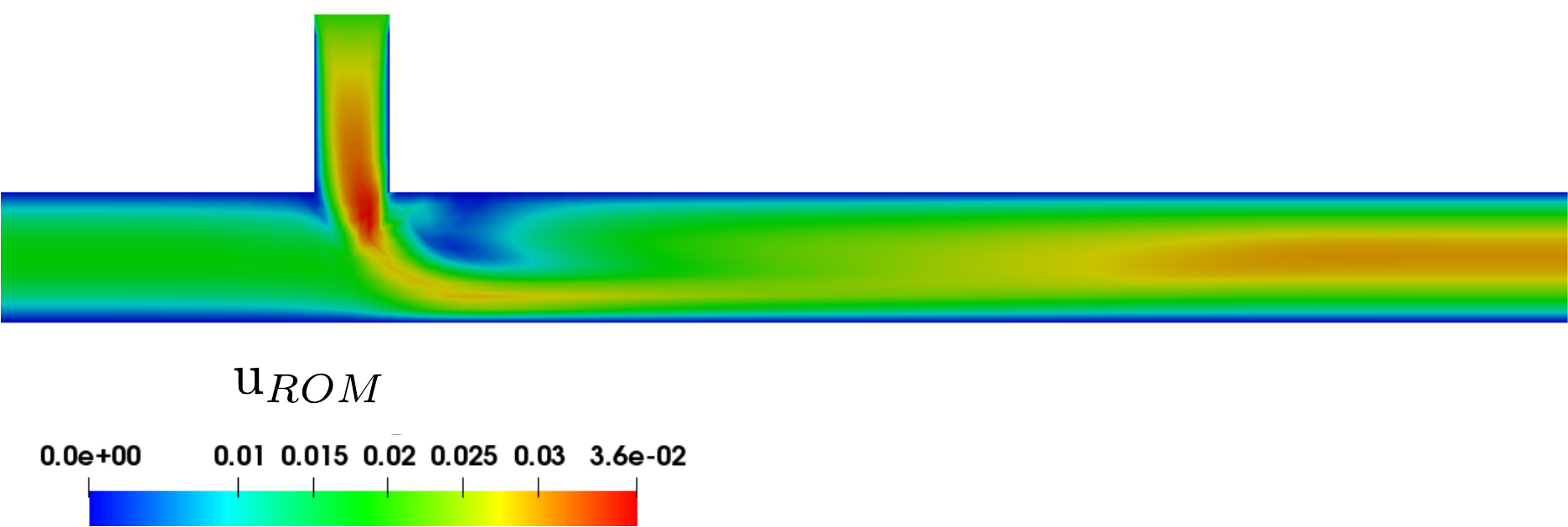}  
\includegraphics[width=0.32\textwidth]{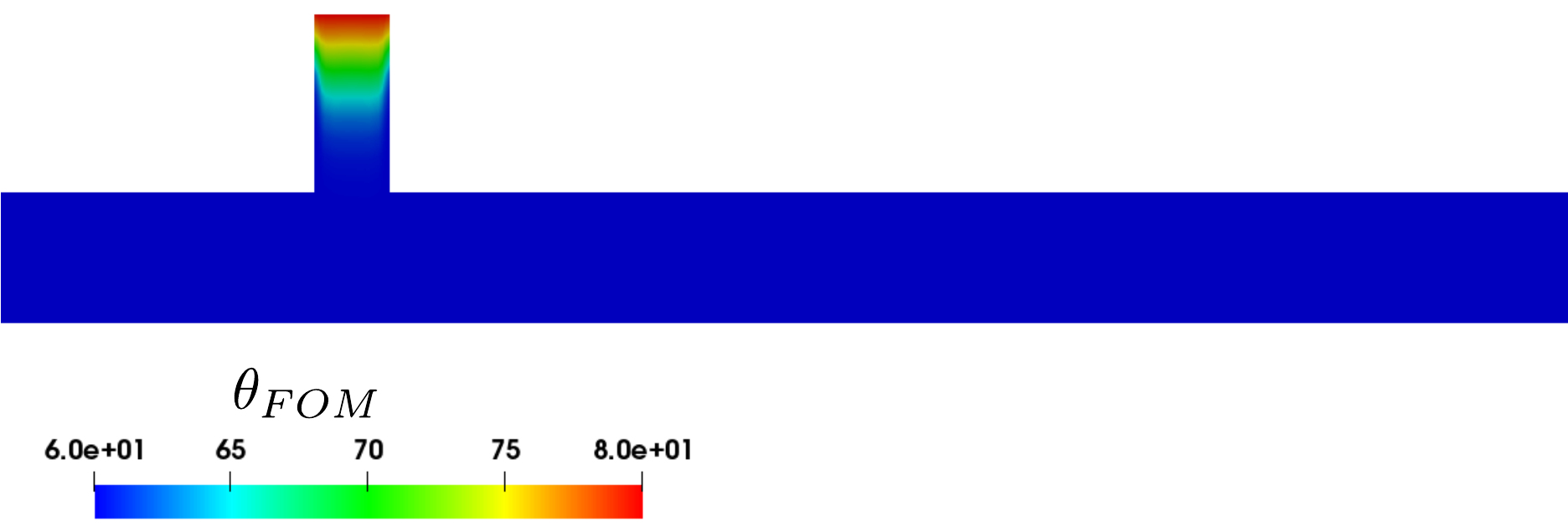}
\includegraphics[width=0.32\textwidth]{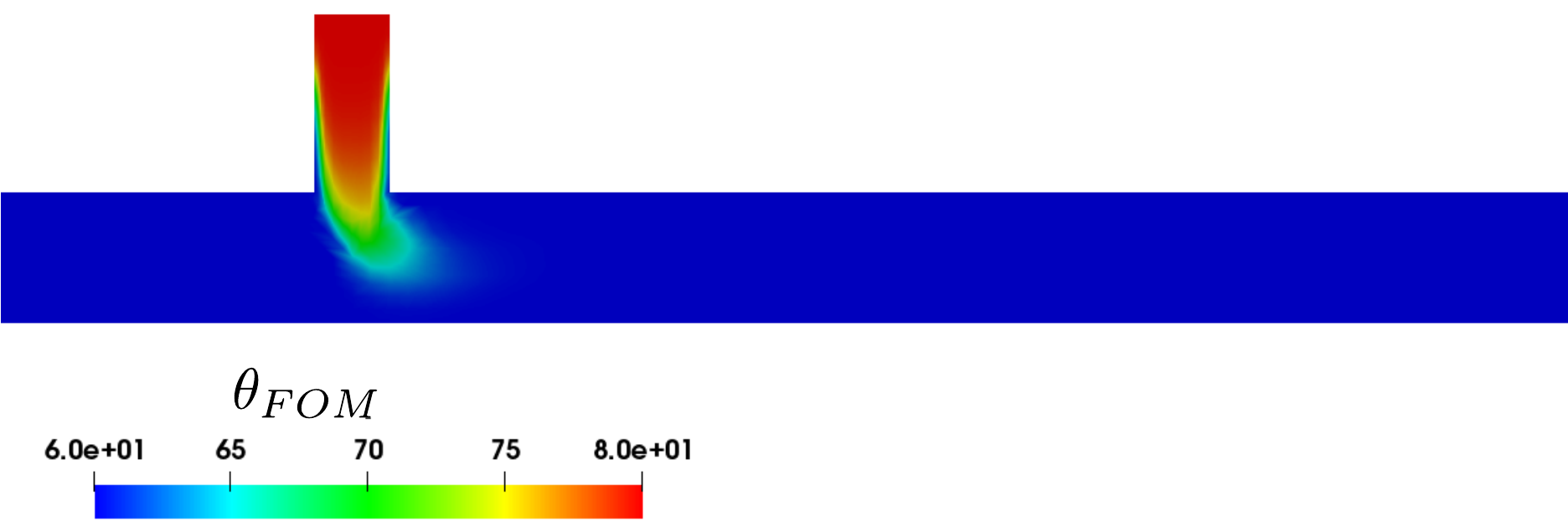}  
\includegraphics[width=0.32\textwidth]{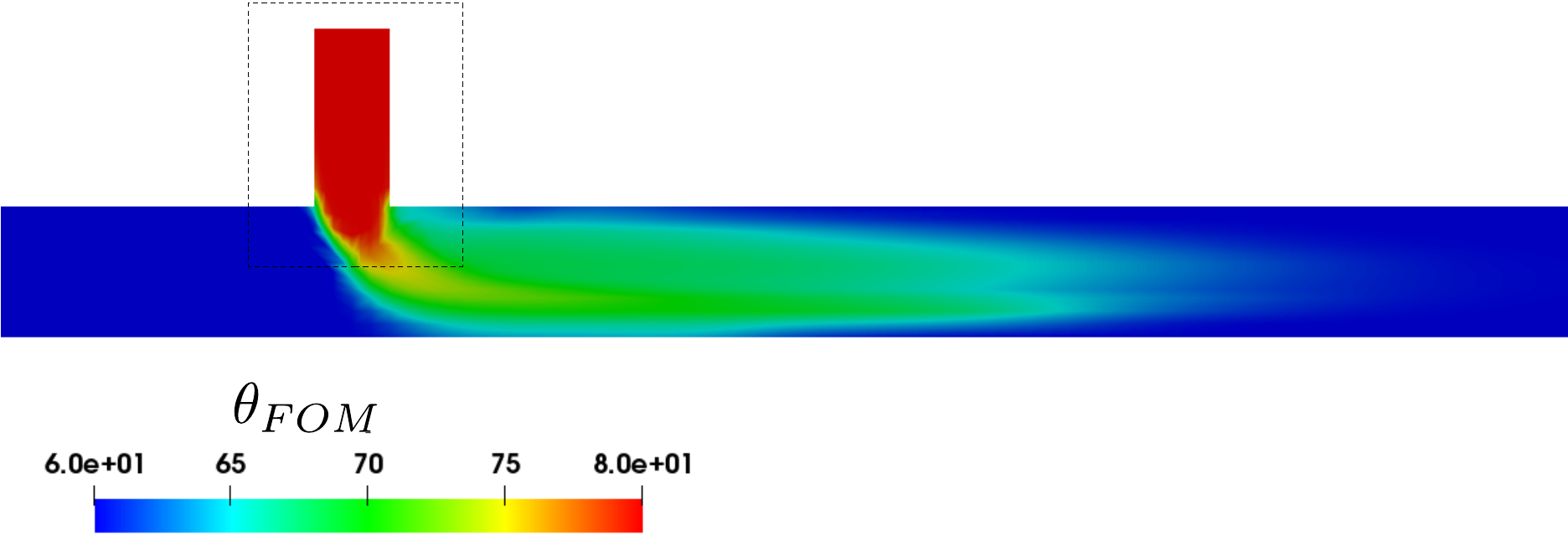}
\includegraphics[width=0.32\textwidth]{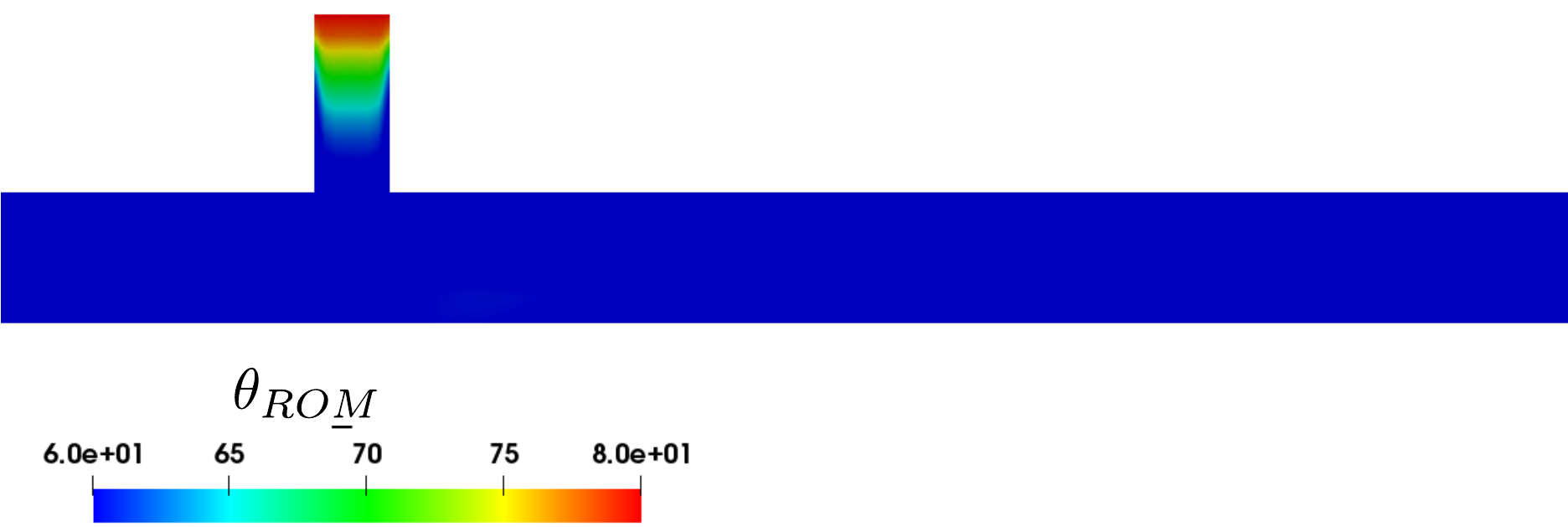}
\includegraphics[width=0.32\textwidth]{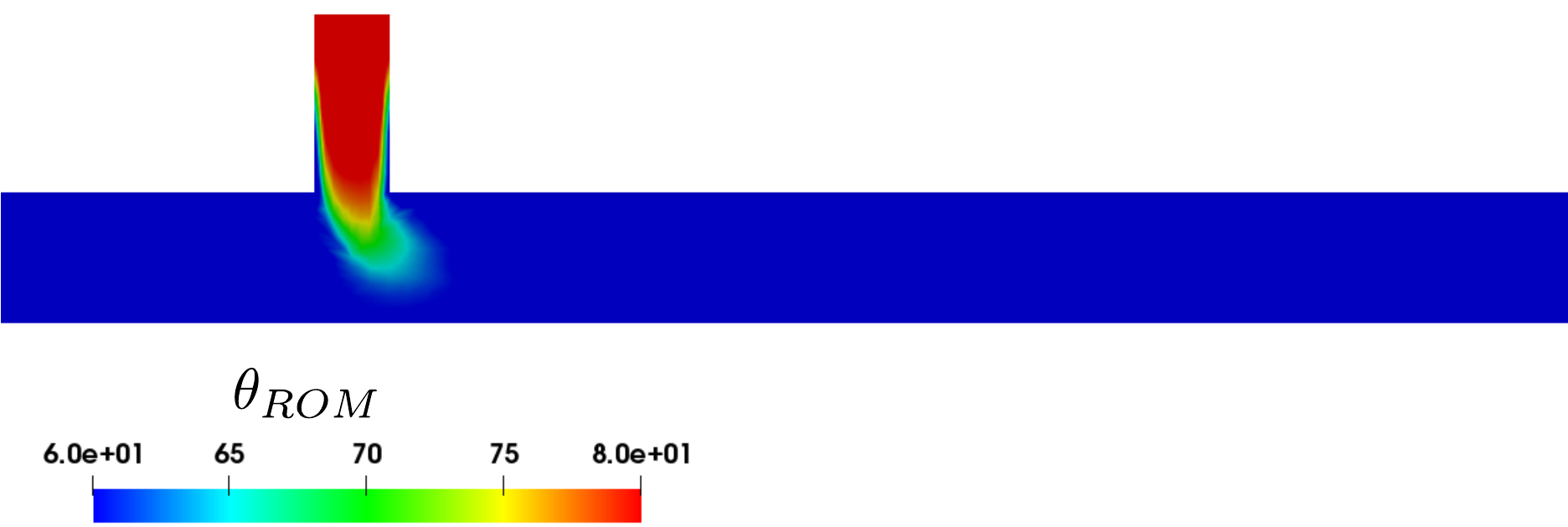}
\includegraphics[width=0.32\textwidth]{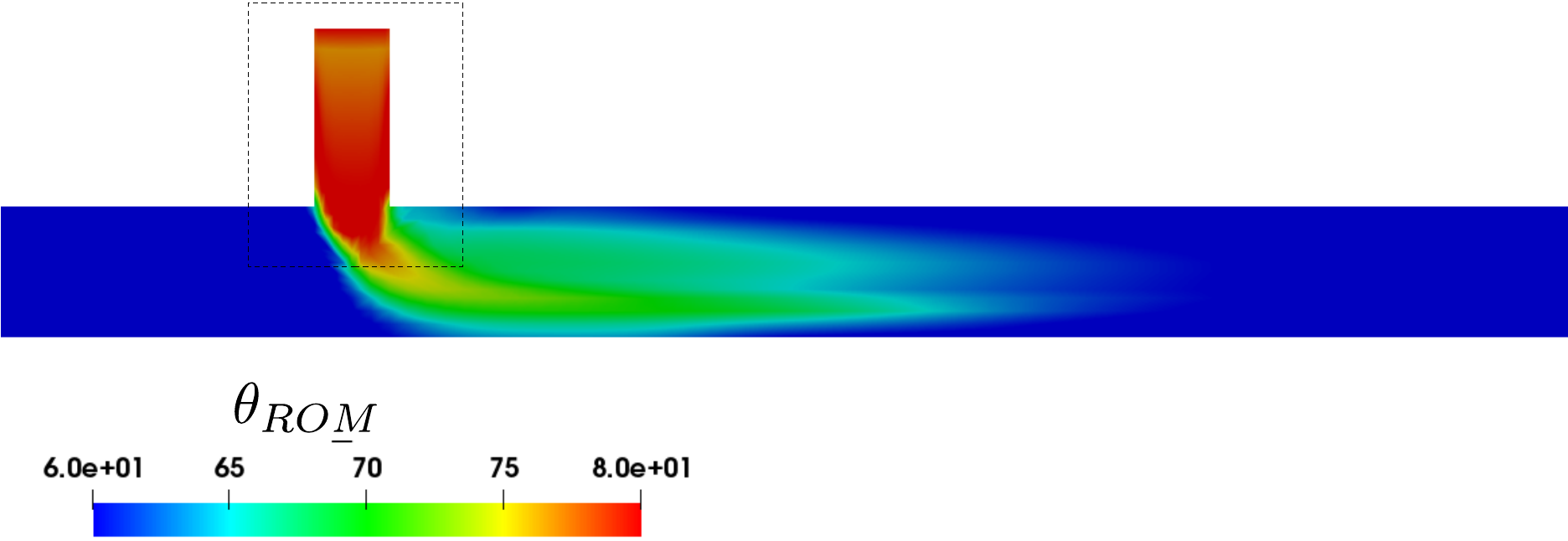}  
\includegraphics[width=0.32\textwidth]{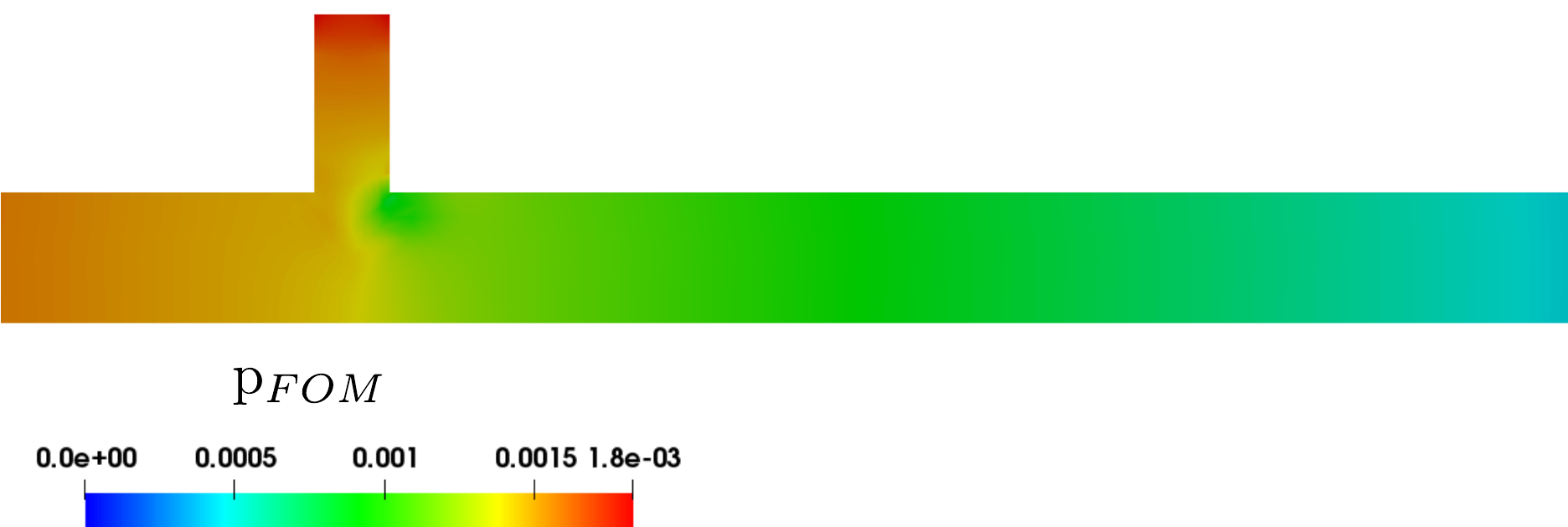}
\includegraphics[width=0.32\textwidth]{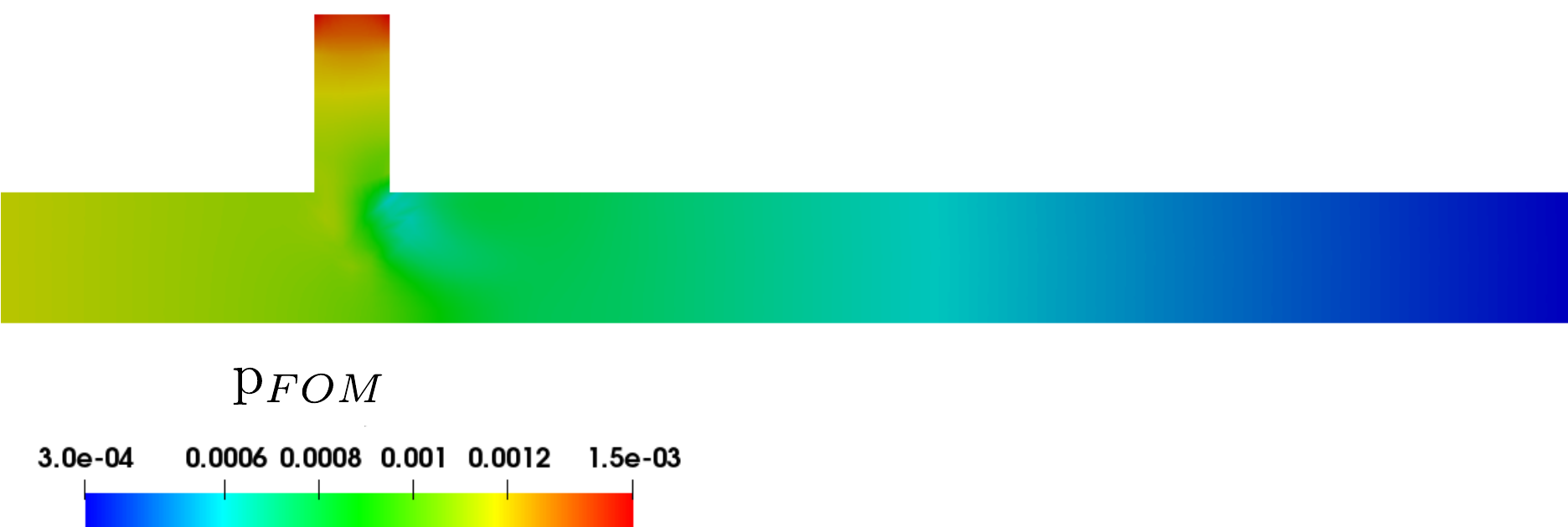}
\includegraphics[width=0.32\textwidth]{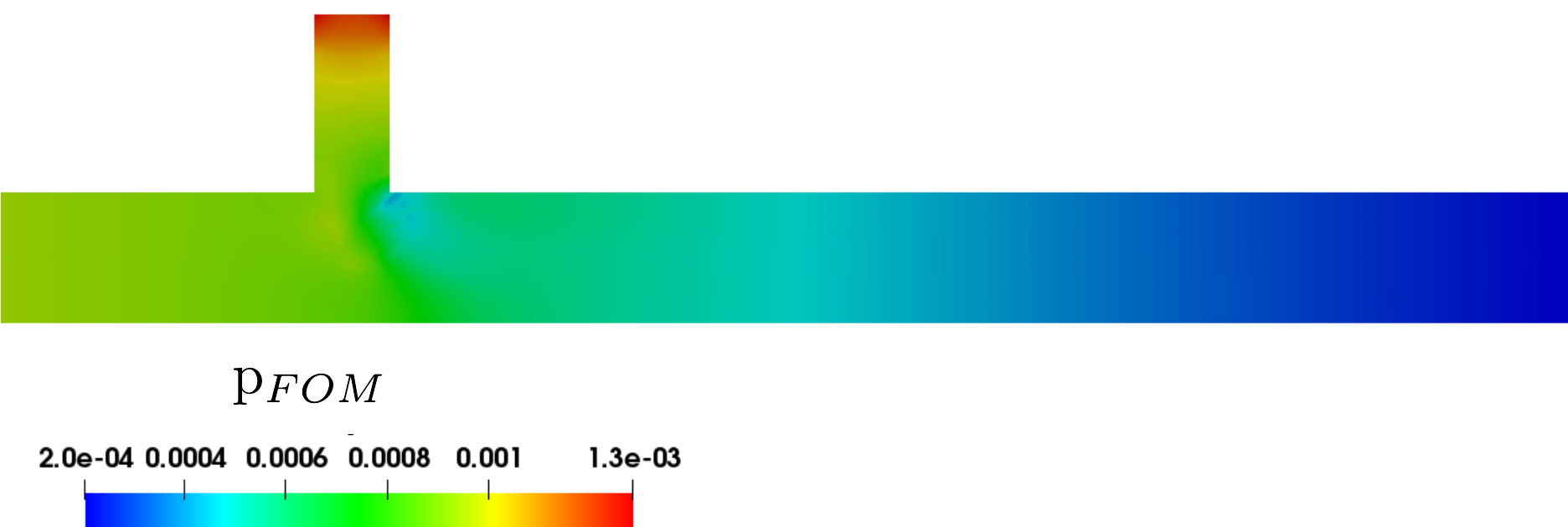}
\includegraphics[width=0.32\textwidth]{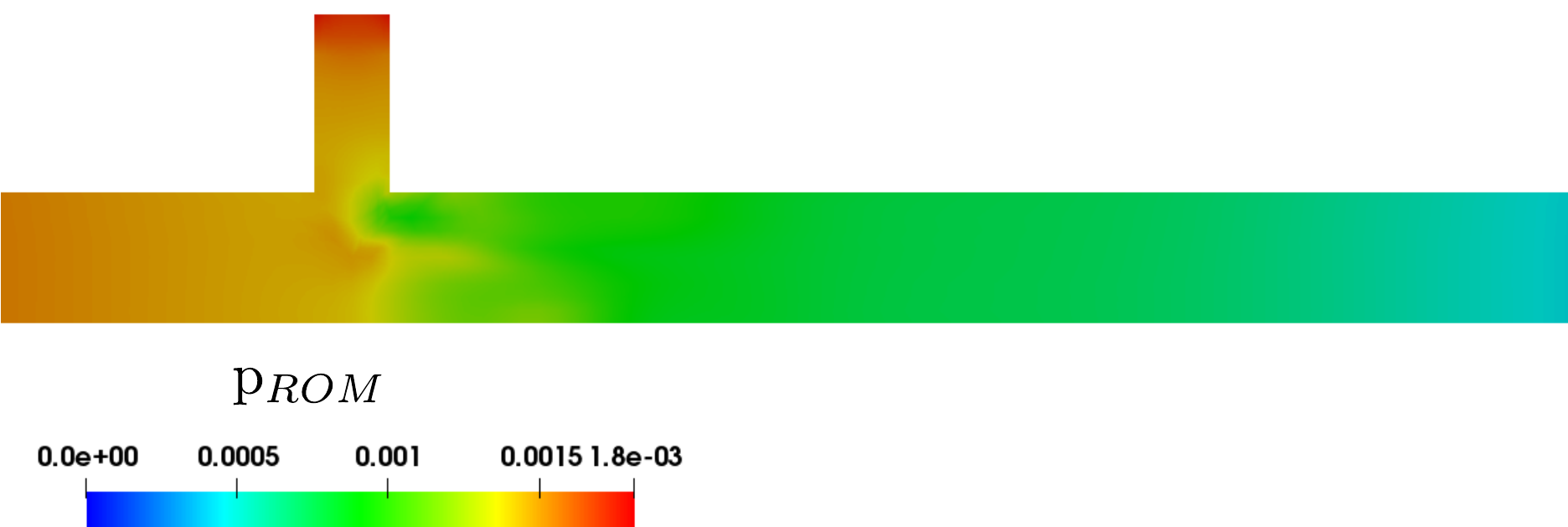}
\includegraphics[width=0.32\textwidth]{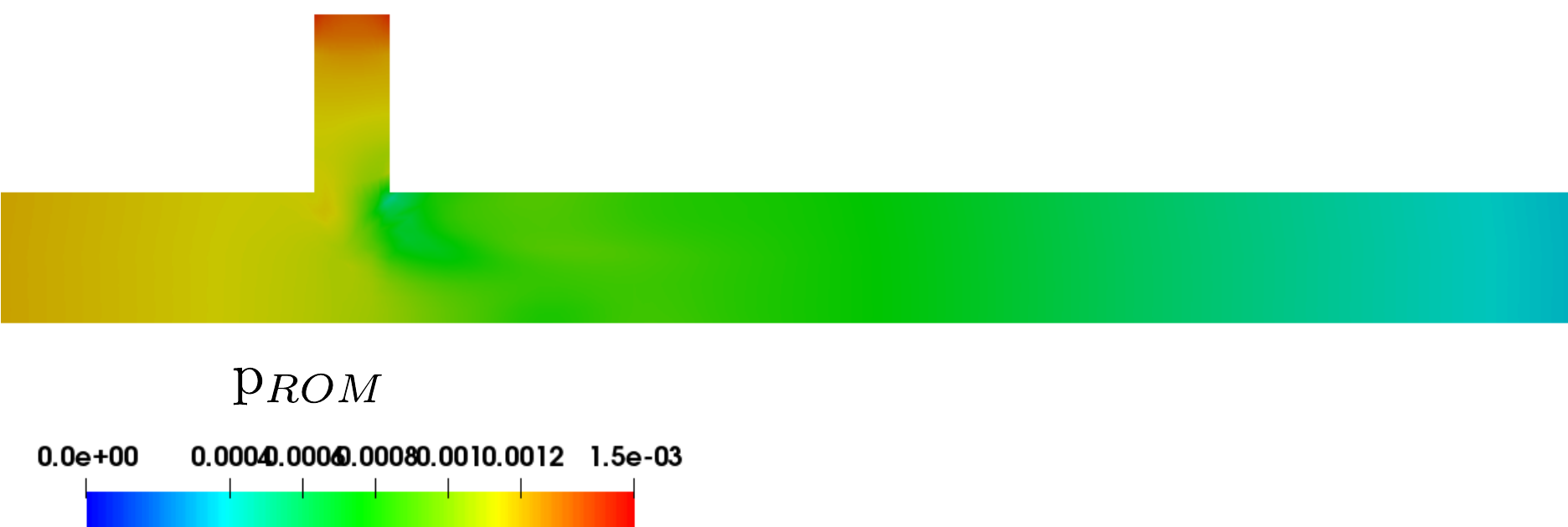}
\includegraphics[width=0.32\textwidth]{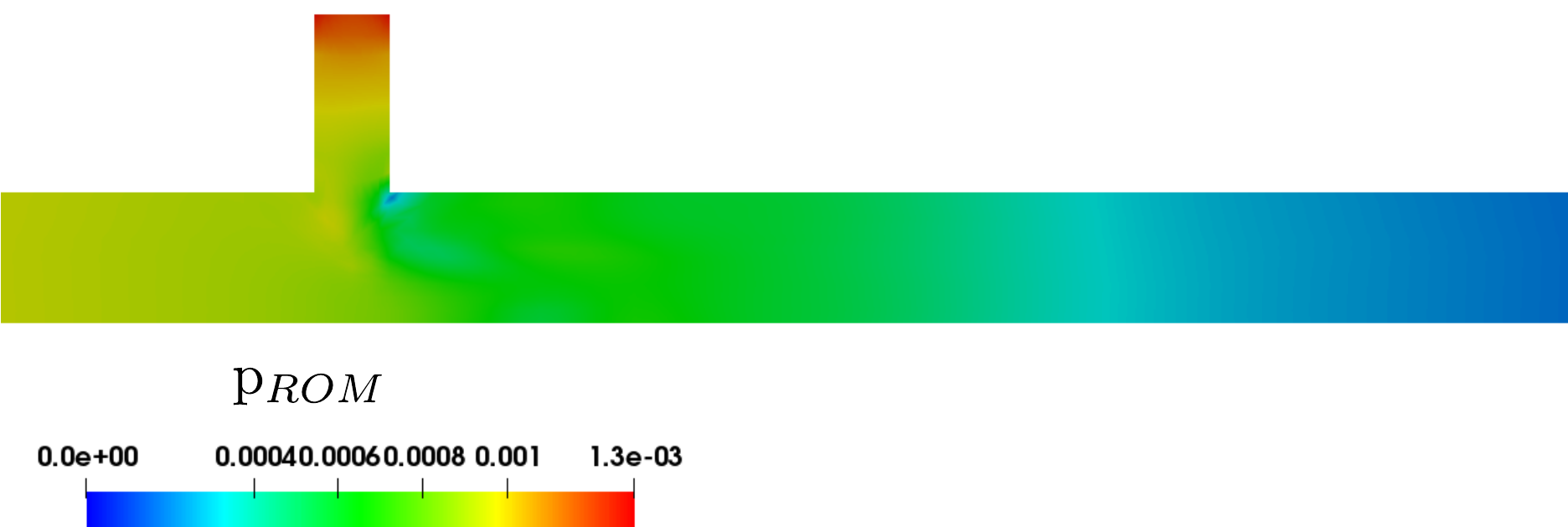}  
\   
\end{minipage}
\caption{Comparison of the velocity field for the full order (first row) and reduced order model (second row) as well as temperature full order (third row) and temperature reduced order model (4th row) and pressure full order  (5th row) with pressure reduced order model (6th row). The temperature inlets are $\theta_{m}=60^{\circ}$C and $\theta_{b}=80^{\circ}$C The fields are depicted for different time instances equal to $t=3 \si{s},10 \si{s}$ and $45 \si{s}$ and increasing from left to right. With a dotted black line we report the area zoomed in Figure~\ref{fig:marker}}.\label{fig:comparison}\label{fig:comparison_case1} 
\end{figure*}

\begin{figure*}[!tbp]  
\begin{minipage}{1\textwidth}
\centering
\includegraphics[width=0.41\textwidth]{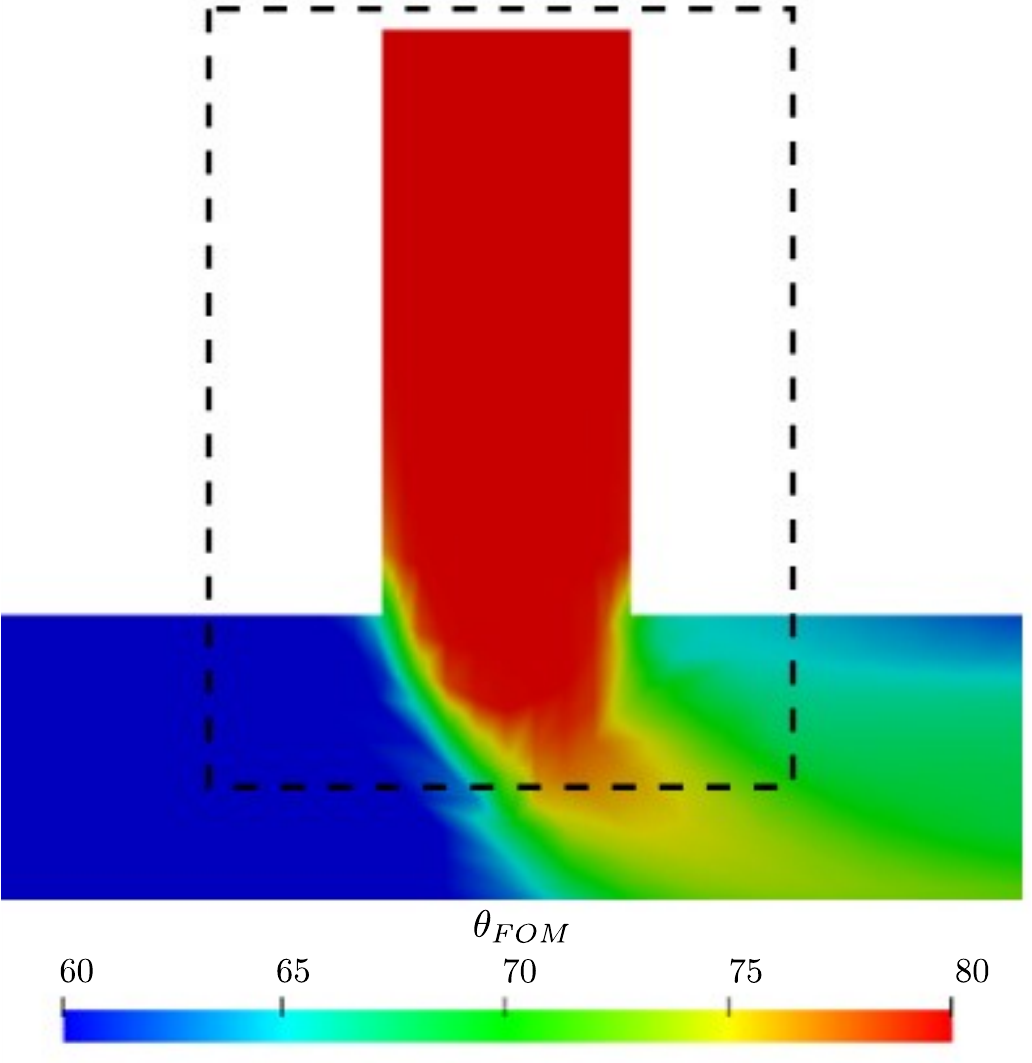}
\includegraphics[width=0.4\textwidth]{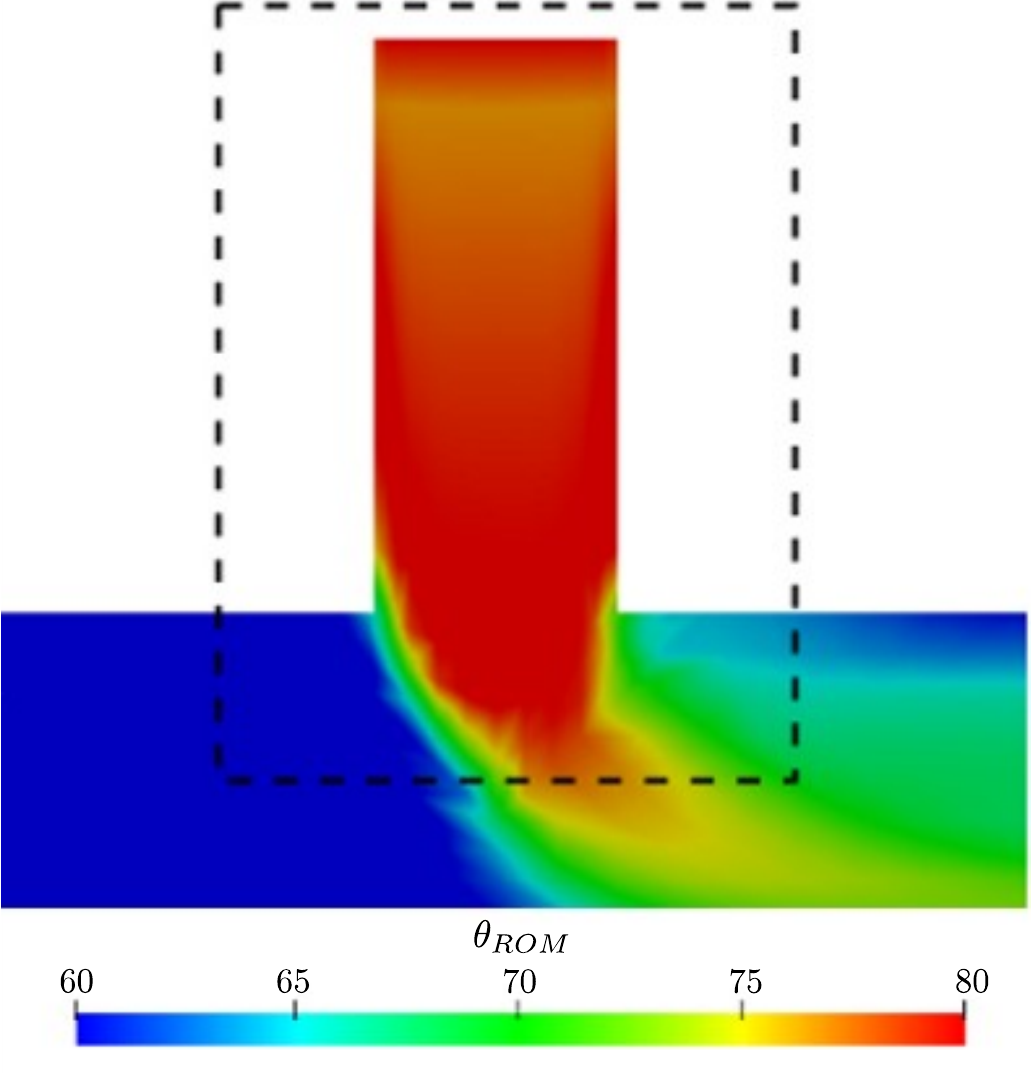}
\end{minipage} 
\caption{Zoom of the area with the biggest relative error between the FOM (left) and the ROM (right) for temperature field. The temperature inlets are $\theta_{m}=60^{\circ}$C and $\theta_{b}=80^{\circ}$C.}\label{fig:marker}  
\end{figure*}

\begin{figure*}[!tbp]
\begin{minipage}{1\textwidth}
\centering 
\includegraphics[width=0.30\textwidth]{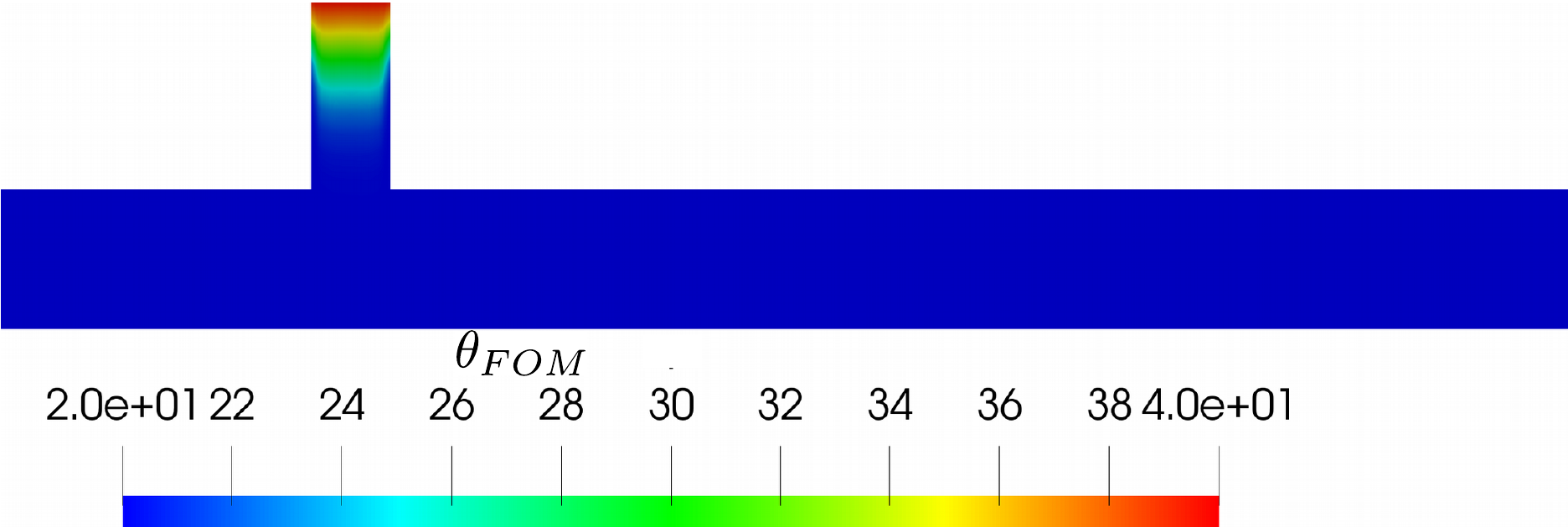}
\includegraphics[width=0.30\textwidth]{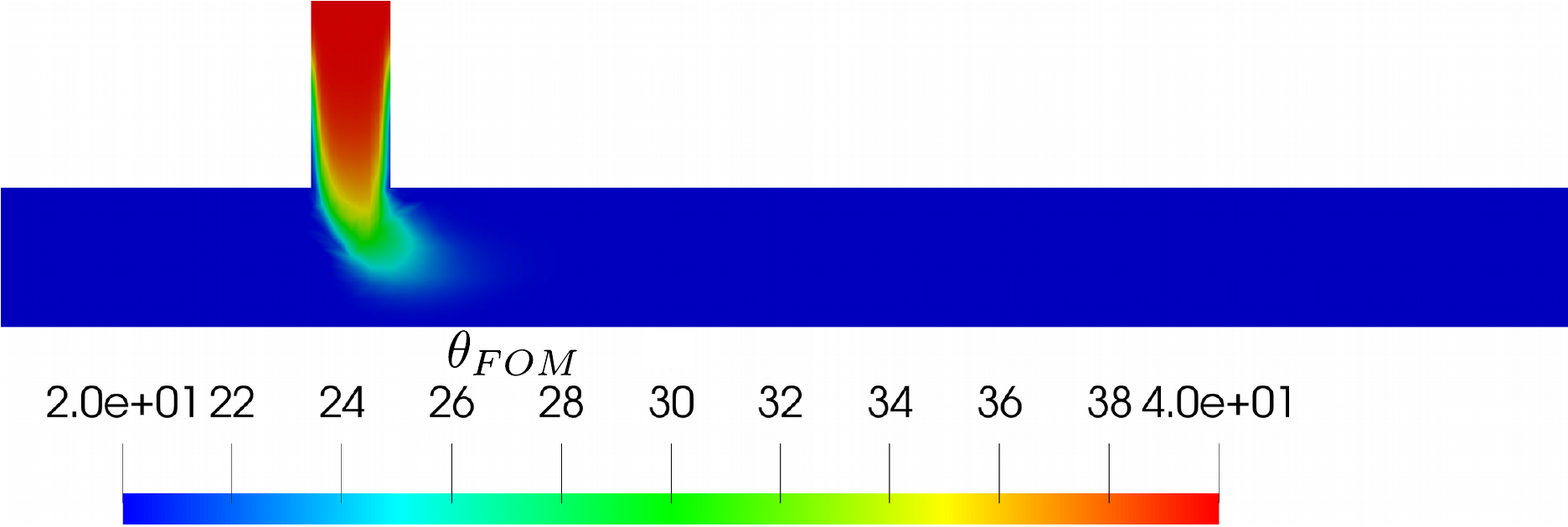}
\includegraphics[width=0.30\textwidth]{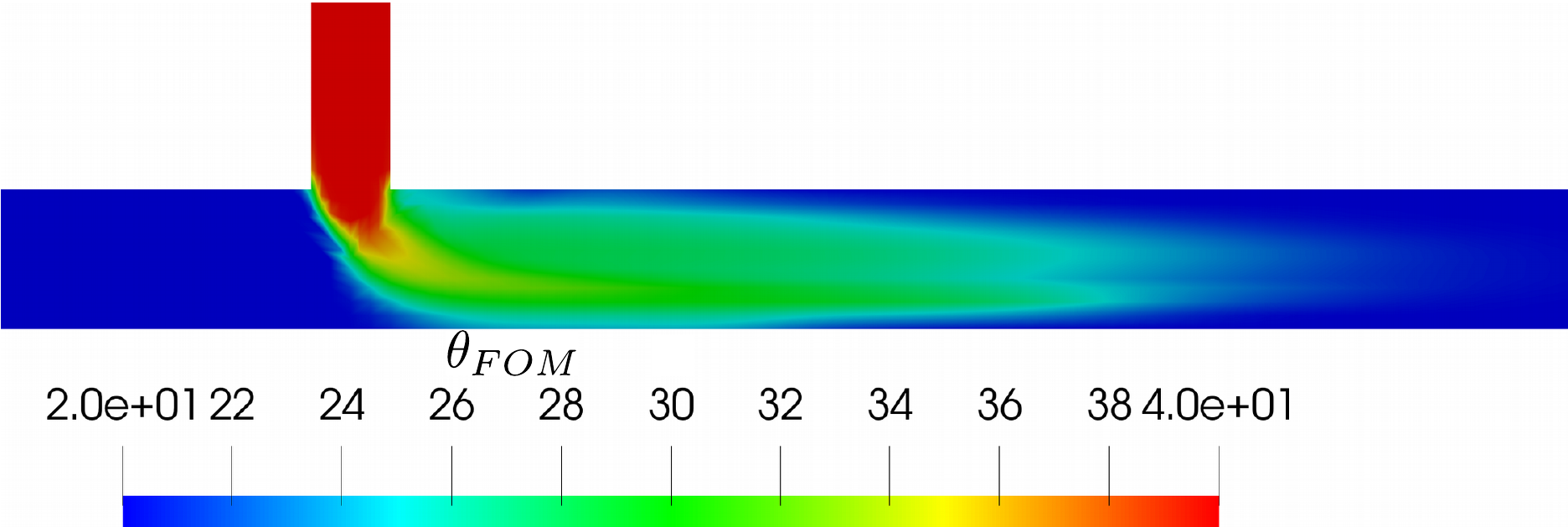}
\includegraphics[width=0.30\textwidth]{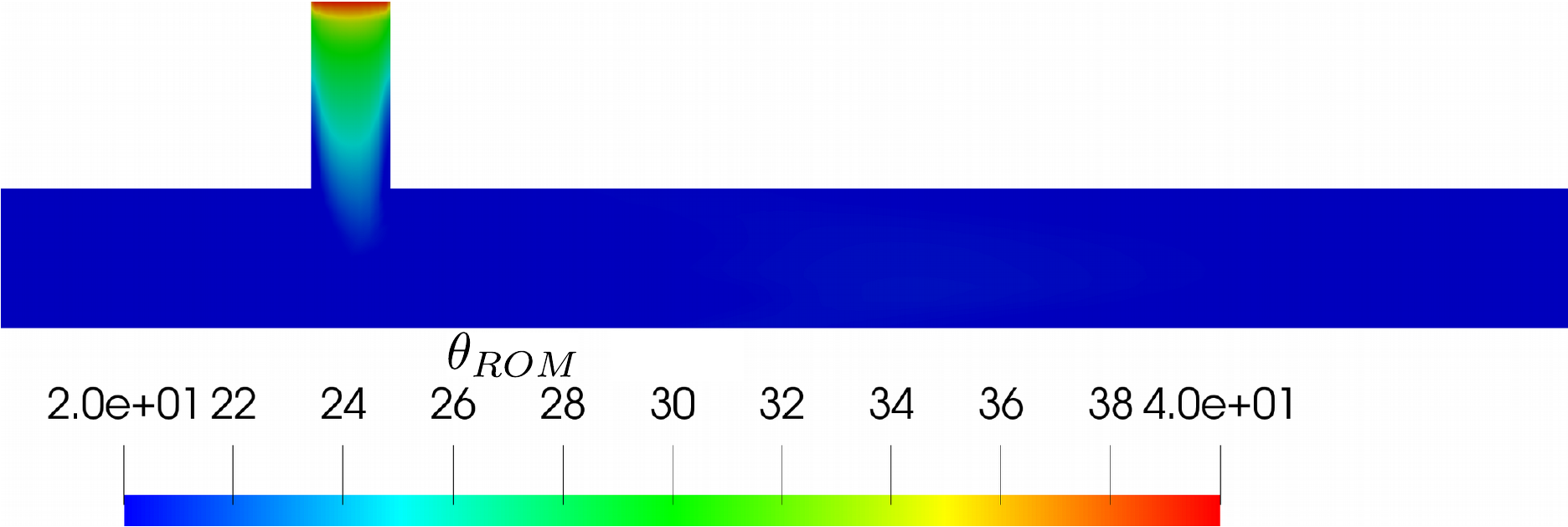}
\includegraphics[width=0.30\textwidth]{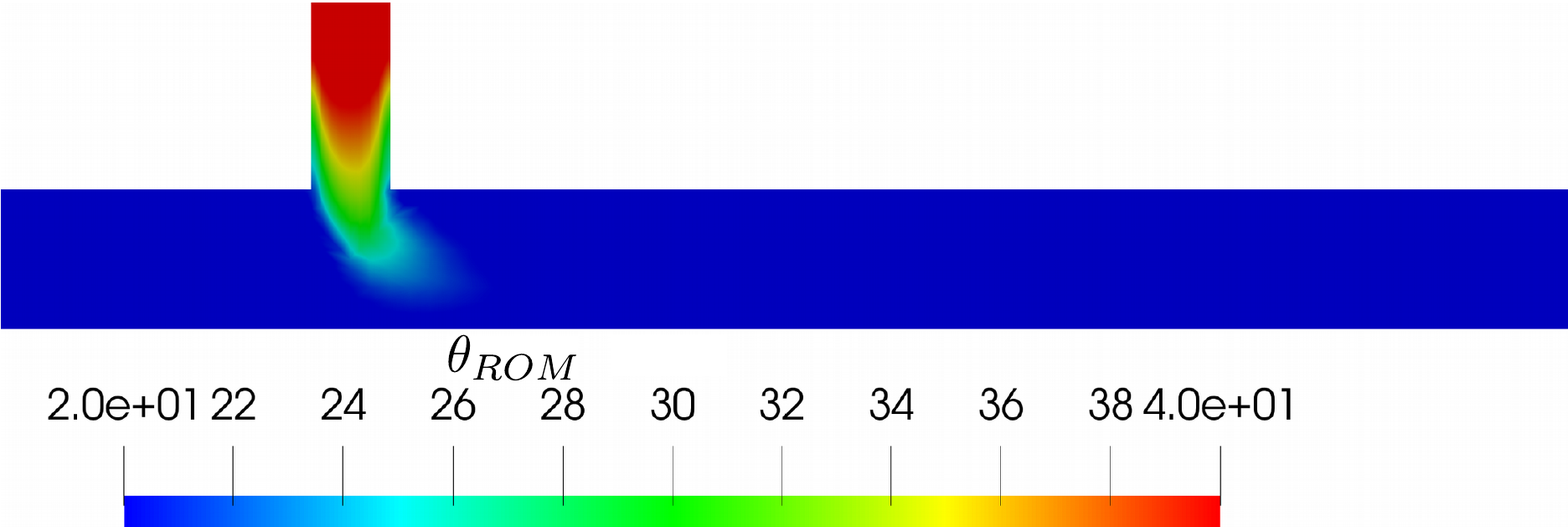}
\includegraphics[width=0.30\textwidth]{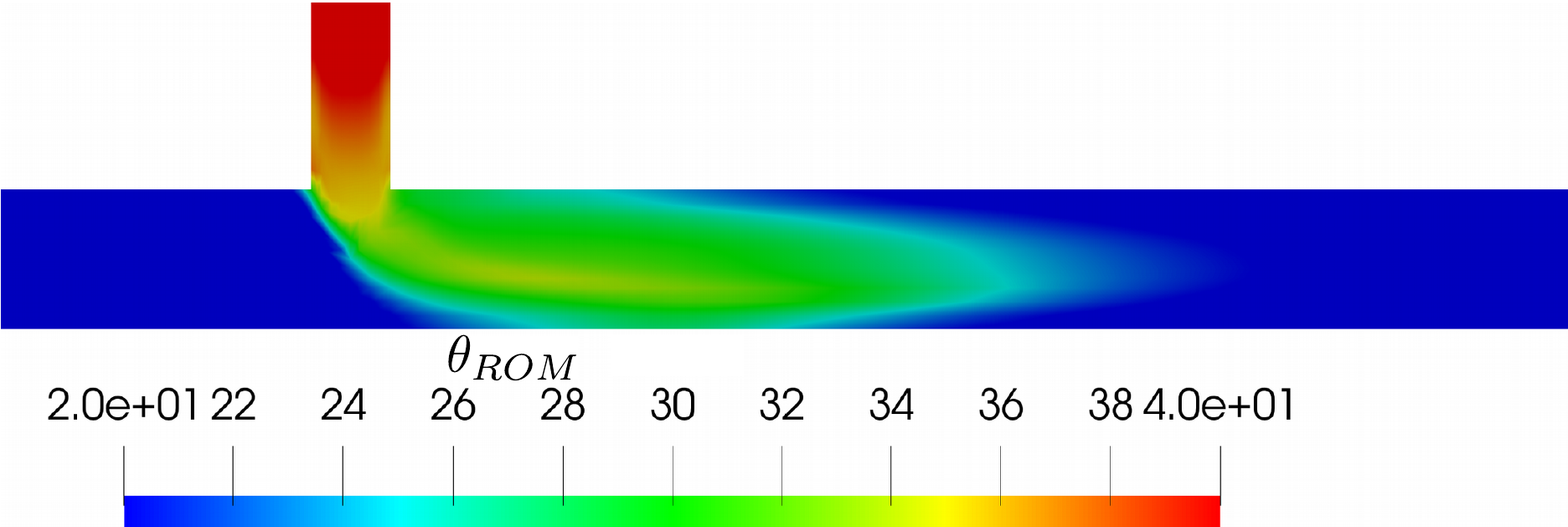}  
\end{minipage} 
\caption{Comparison of the full order field (top row) and temperature reduced order model (bottom row) for the case of temperature inlets $\theta_{m}=20^{\circ}$C and $\theta_{b}=40^{\circ}$C. The fields are depicted for different time instances equal to $t=3 \si{s}, 10 \si{s}$ and $45 \si{s}$ and increasing from left to right.}\label{fig:diff_run_20} 
\end{figure*}

The resulting velocity, temperature and pressure fields are reconstructed with $\epsilon_{L^2}$ error as shown in figure (\ref{fig:l2error}) and the minimum, maximum and average $\epsilon_{L^2}$ errors are available in the first three columns of table (\ref{fig:errorstatistics}). The error seems to be larger for velocity during the first timesteps and this could happen because of the highly transient nature of the flow. This error could be reduced by including more snapshots taken during the first timesteps. Perhaps, to enhance the results, one could also consider using a weighted-POD method \cite{navonweighted}, or a combination of a POD method in time and a greedy method in parameter \cite{eftang2011hp}. As in this case the temperature inlets are parametrized, the ROM, which is trained only on inlets $\theta_m =50^{\circ}C$ and $\theta_b = 70^{\circ}$C, has been used to simulate a set of other temperature inlets. For each case, the $\epsilon_{L^2}$ error between the FOM and the ROM is plotted and shown in figure (\ref{fig:error_dif_runs}) and the minimum, maximum and average relative errors are summarized in table (\ref{fig:temperrorstatistics}). Due to the linearity of the temperature equation, $\frac{\partial \theta}{\partial t} +(\bm{u}\cdot\nabla)\theta - \alpha_{dif}\Delta \theta$, for temperature inlet values that belong to a range close to the trained value, the ROM can reproduce the fields with good accuracy, as shown in figure (\ref{fig:error_dif_runs}), without having to sample and enrich the POD space with additional points. To compare the FOM and ROM results, a run for temperature inlet values of $\theta_m =60^{\circ}$C and $\theta_b = 80^{\circ}$C has been performed and the results are shown in figure (\ref{fig:comparison_case1}). One could observe that the biggest error is found in the area of the branch pipe, figure (\ref{fig:marker}). This error could be caused by the fact that the length of the branch pipe is not long enough, so the flow is not fully developed by the time it reaches the mixing region. Therefore, this region is characterized by large gradients. A comparison also for the case with the biggest $\epsilon_{L^2}$  error is shown in figure (\ref{fig:diff_run_20}), where the ROM is run for temperature inlets $\theta_{m}=20^{\circ}$C and $\theta_{b}=40^{\circ}$C. Even in this case,  where the inlet values are relatively far away from the ones that they were used to train the ROM, the reduced model is capable of reproducing the main flow with a good accuracy. The maximum $\epsilon_{L^2}$  for the reconstructed temperature is less than $9 \% $ (\ref{fig:temperrorstatistics}). The velocity and pressure fields are omitted in figure (\ref{fig:diff_run_20}), as the change in temperature boundary conditions does not affect the velocity and the pressure fields. Thus, they remain as in figure (\ref{fig:comparison_case1}). The CPU time of the FOM is $856.71$s whereas, for the ROM, is only $2.29$s. This corresponds to a computational speed-up factor of $\approx 374$.

\begin{table}[!tbp]
\centering
  \begin{minipage}[b]{0.6\textwidth}
  \caption{Summary of the physical parameters.}\label{fig:parameters}
  \centering
    \begin{tabular}{ l | c | c }\toprule 
    & Main Pipe & Branch Pipe \\ \hline \hline
    $u$ (m/s) & 0.01 & 0.02  \\ \hline
    $T$ ($^{\circ}$C) & 50 & 70 \\ \hline
    $D$ (mm) & 140 & 80 \\ \hline
    $Re$               & 140 & 240 \\ \bottomrule
   \end{tabular}
    \end{minipage}%
 \end{table}

\begin{table}[!tbp]
\centering
    \begin{minipage}[b]{0.6\textwidth}
      \caption{Table with the boundary conditions where $\Gamma_{m}$ refers to the main pipe inlet, $\Gamma_{b}$ to the branch pipe and $\Gamma_{0}$ is the outlet.}\label{fig:boundary_con}
      \centering 
       \begin{tabular}{ l | c | c | c | c | c  }\toprule
& $\Gamma_{m}$ & $\Gamma_{b}$ & $\Gamma_{w}$ & $\Gamma_{o}$ & $\Theta_{IF}$ \\ \hline \hline
    $\bm{u}$ & $(0.01,0,0)$ & $(0, 0, -0.02)$ & $\nabla\cdot\bm{u} = 0$ & $\nabla \bm{u} \cdot \bm{n} = 0$& $(0,0,0)$ \\ \hline
    $p$ & $\nabla p \cdot \bm{n} = 0$ & $\nabla p \cdot \bm{n} = 0$ & $\nabla p \cdot \bm{n} = 0$ & $0$& $0$\\ \hline 
    $\theta$ & $50$ & $70$ & $\nabla \theta \cdot \bm{n} = 0$ &$\nabla \theta \cdot \bm{n} = 0$& $50$ \\ \bottomrule
   \end{tabular}
    \end{minipage}%
\end{table}

\setlength\intextsep{0pt}

\begin{table}[!tbp]
\centering
      \caption{Relative $\epsilon_{L^2}(t)$ error ($\epsilon_{L^2}(t) = \frac{||X_{FOM}(t)-X_{ROM}(t)||_{L^2(\Omega)}}{||X_{FOM}(t)||_{L^2(\Omega)}}$) for velocity, temperature and pressure fields for two snapshot sampling frequencies, per $0.2$s where $N_u^s=N_{\theta}^s=N_p^s = 225$ and per $0.1s$ where $N_u^s=N_{\theta}^s=N_p^s = 450$s.}\label{fig:errorstatistics}
      \centering
       \begin{tabular}{ l | c | c | c | c | c | c  }\toprule
& $\bm{u}$ per $0.2$s & $\theta$ per $0.2$s & $p$ per $0.2$s & $\bm{u}$ per $0.1$s & $\theta$ per $0.1$s & $p$ per $0.2$s \\ \hline \hline
    Minimum $\epsilon_{L^2}(t)$ & $0.023$ & $0.004$ & $0.013$ & $0.019$ & $0.021$ & $0.010$ \\ \hline
    Maximum $\epsilon_{L^2}(t)$ & $0.031$ & $0.021$ & $0.040$ & $0.027$ & $0.020$ & $0.038$\\ \hline 
    Average $\epsilon_{L^2}(t)$ & $0.089$ & $0.035$ & $0.096$ & $0.088$ & $0.034$ & $0.034$ \\ \bottomrule
   \end{tabular}
\end{table}

\begin{table}[!tbp]
\centering 
      \caption{Relative $\epsilon_{L^2}(t)$ error ($\epsilon_{L^2}(t) = \frac{||X_{FOM}(t)-X_{ROM}(t)||_{L^2(\Omega)}}{||X_{FOM}(t)||_{L^2(\Omega)}}$) for the temperature field for five different sets of temperature inlet boundary conditions. The sets are $50,60^{\circ}$C, $40,60^{\circ}$C, $60,80^{\circ}$C, $20,40^{\circ}$C and $55,75^{\circ}$C.}\label{fig:temperrorstatistics}
      \centering
       \begin{tabular}{ l | c | c | c | c | c }\toprule
& $\theta$, $50,60^{\circ}$C & $\theta$, $40,60^{\circ}$C & $\theta$, $60,80^{\circ}$C & $\theta$, $20,40^{\circ}$C & $\theta$, $55,75^{\circ}$C \\ \hline \hline
    Minimum $\epsilon_{L^2}(t)$ & $0.004$ & $0.006$ & $0.004$ & $0.012$ & $0.004$  \\ \hline
    Maximum $\epsilon_{L^2}(t)$ & $0.033$ & $0.029$ & $0.030$ & $0.085$ & $0.033$ \\ \hline 
    Average $\epsilon_{L^2}(t)$ & $0.022$ & $0.022$ & $0.020$ & $0.057$ & $0.022$  \\ \bottomrule
   \end{tabular}
\end{table}


\subsection{Non-isothermal Mixing in T-junction - Parametrization of the Kinematic Viscosity}
\begin{figure*}[!tbp]
\centering
\includegraphics[width=0.55\textwidth]{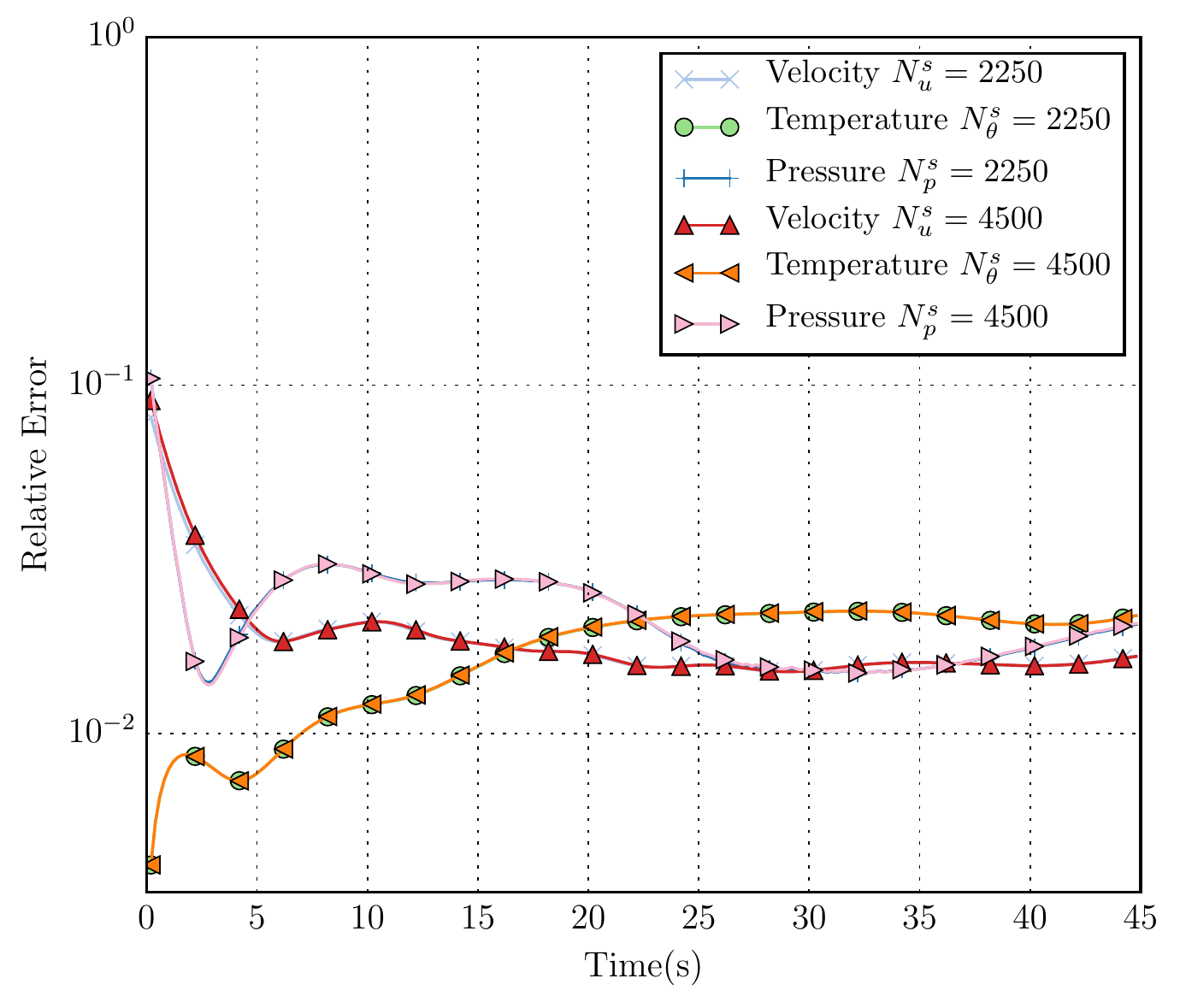}
\caption{$\epsilon_{L^2}(t)$ error ($\epsilon_{L^2}(t) = \frac{||X_{FOM}(t)-X_{ROM}(t)||_{L^2 (\Omega)}}{||X_{FOM}(t)||_{L^2 (\Omega)}}$) for two sampling spaces for the parameter (kinematic viscosity). These correspond to one with 5 sampling points for viscosity, where $N_u^s=N_{\theta}^s=N_p^s = 2250$ and one with 10 sampling points where, $N_u^s=N_{\theta}^s=N_p^s = 4500$s.  The ROM is run for $\nu=1.1e-05$.}\label{fig:convergencevisc}
\end{figure*}

The second case aims to parametrize the kinematic viscosity in the unsteady Navier-Stokes equations. Due to the non-linearity of the convective term, this case needs enrichment of the POD space with additional snapshots which are solutions of a particular range of values of the parametrized quantity. For this purpose, the same model as described in section (\ref{sec:tempinlet}) is used and the POD space is enriched with additional sampling points for the parameter of interest. Two sampling cases were considered. In the first case, 10 sampling points for the kinematic viscosity, corresponding to $\nu =  [1e-06, 2.55e-06, 4.11e-06, 5.66e-06, 7.22e-06, 8.77e-06, 1.03e-05, 1.18e-05, 1.34e-05, 1.5e-05]$ and a second one with 5 sampling points corresponding to $ \nu =  [5e-06, 7.5e-06, 1e-05, 1.25e-05, 1.5e-05]$. A convergence comparison between the two sampling spaces and the FOM is shown in figure (\ref{fig:convergencevisc}) where one could observe that the differences between the two spaces are minimal. Therefore, for computational efficiency reasons, the test case will be performed on the space with the 5 sampling points. These sampling values correspond to Reynolds numbers $\si{Re}_{\si{m}} = [280,187,140,112,93]$ for the main pipe and $\si{Re}_{\si{b}} = [320,213,160,128,107]$ for the branch. Thus, the flow remains laminar in the total pipe length. 

\begin{figure}[!tbp]
  \centering 
  \begin{minipage}[b]{0.45\textwidth}
    \includegraphics[width=\textwidth]{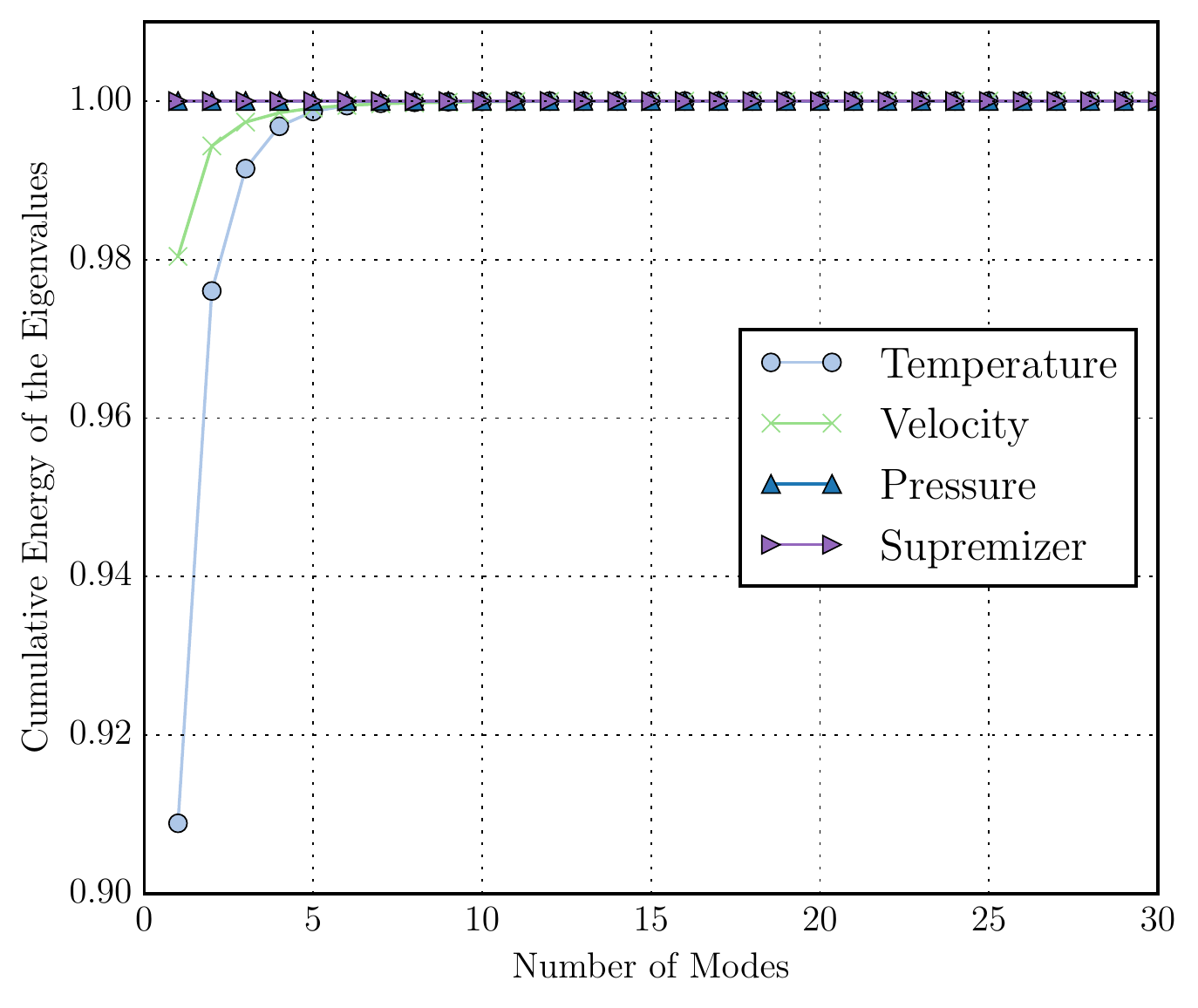}
    \caption{Cumulative energy of the eigenvalues for temperature, velocity, pressure and supremizer fields.}\label{fig:eigenvalues_case2}
  \end{minipage}
  \hfill
  \begin{minipage}[b]{0.45\textwidth}  
    \includegraphics[width=\textwidth]{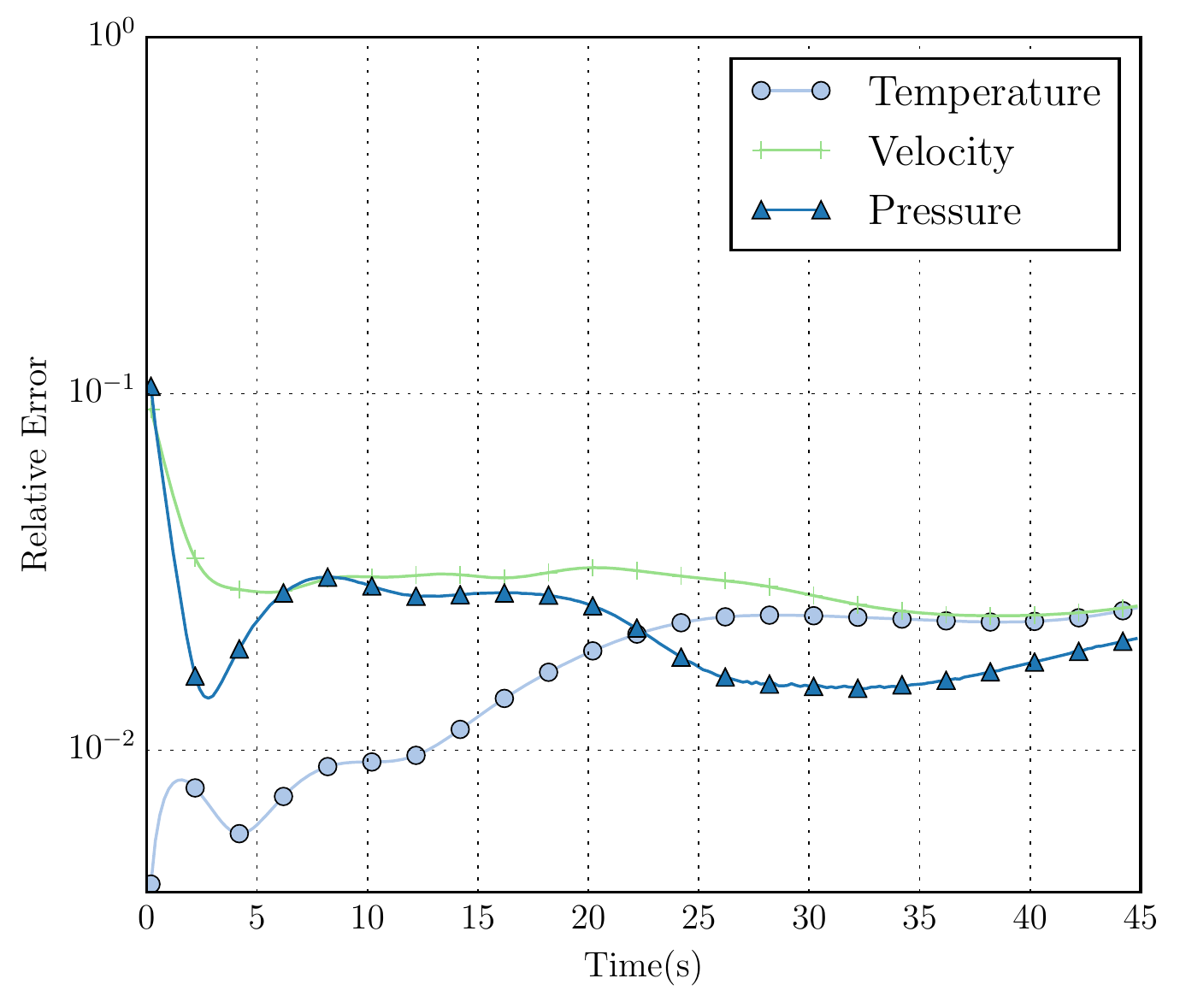}
    \caption{$\epsilon_{L^2}(t)$ error ($\epsilon_{L^2}(t) = \frac{||X_{FOM}(t)-X_{ROM}(t)||_{L^2(\Omega)}}{||X_{FOM}(t)||_{L^2(\Omega)}}$) for temperature, velocity and pressure fields for the test case with kinematic viscosity $\nu=1.1e-05$.}\label{fig:l2error_case2}
  \end{minipage}
\end{figure}

\begin{figure*}[!tbp]   
\begin{minipage}{1\textwidth}
\centering
\includegraphics[width=0.4\textwidth]{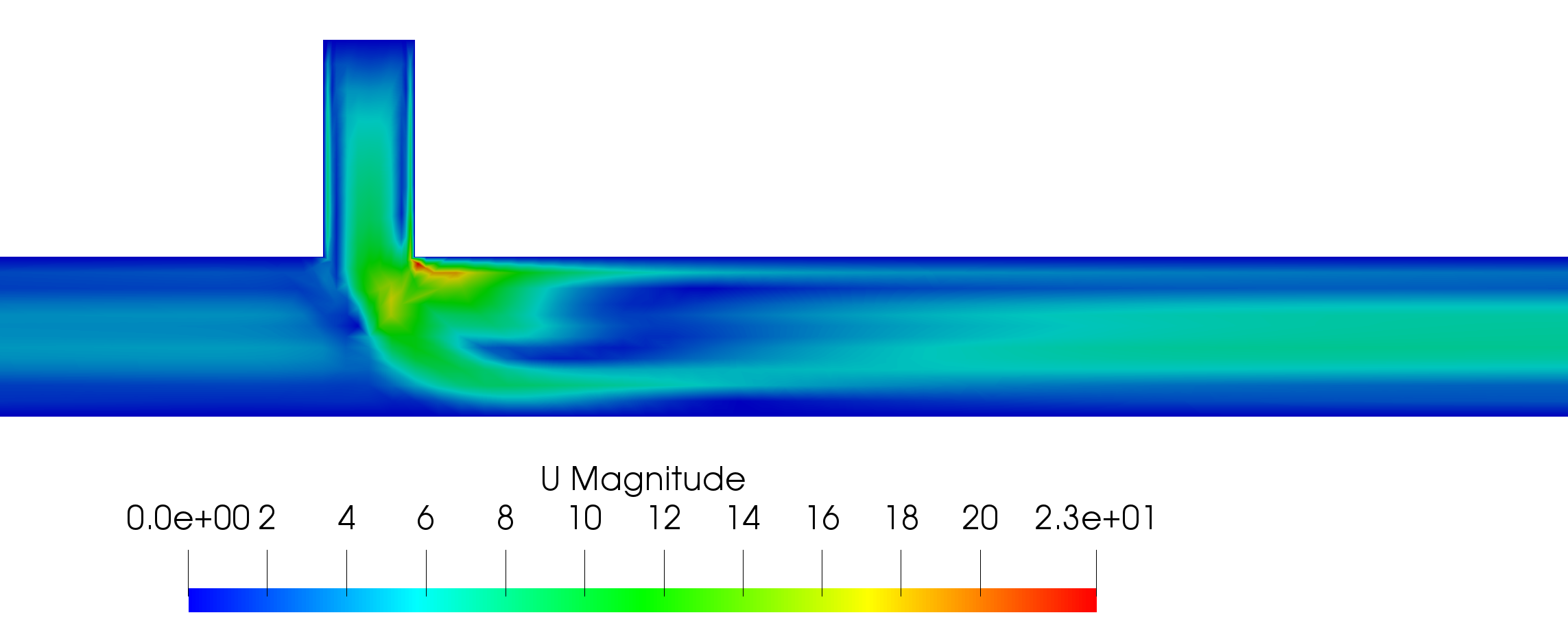}
\includegraphics[width=0.4\textwidth]{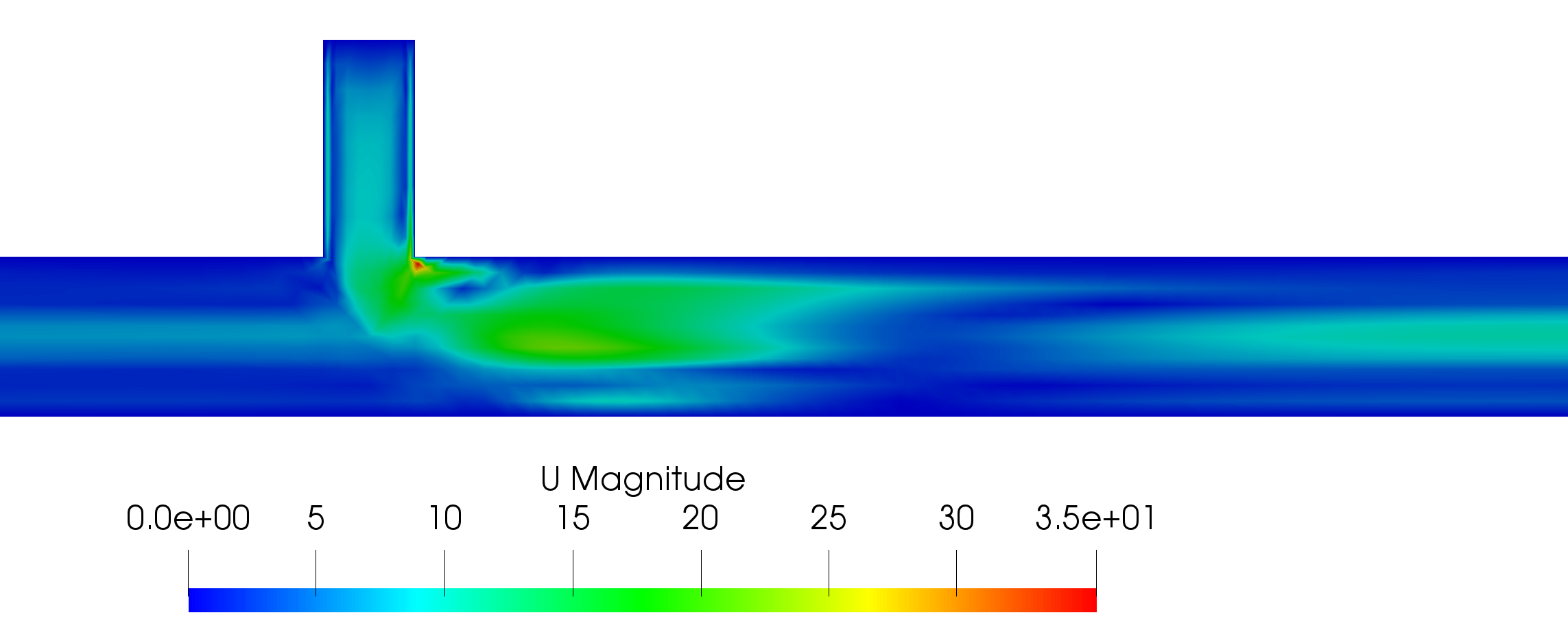}
\includegraphics[width=0.4\textwidth]{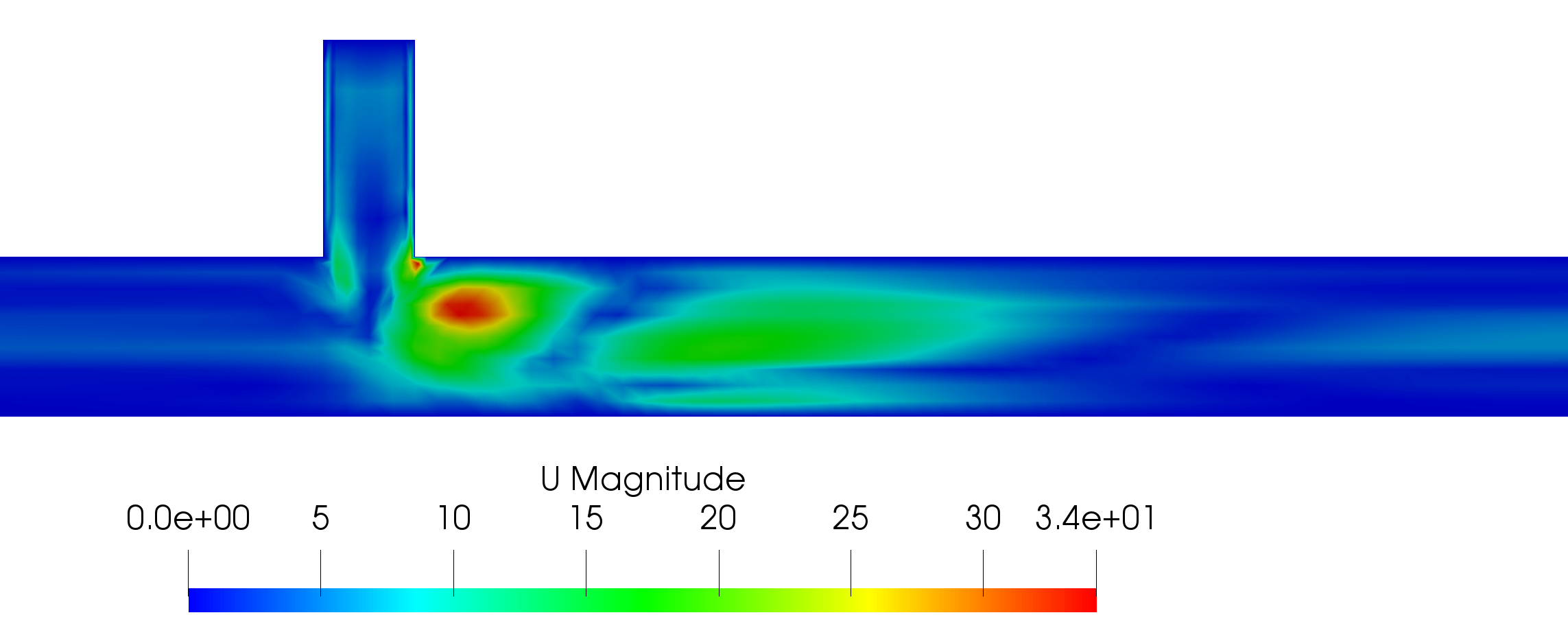}
\includegraphics[width=0.4\textwidth]{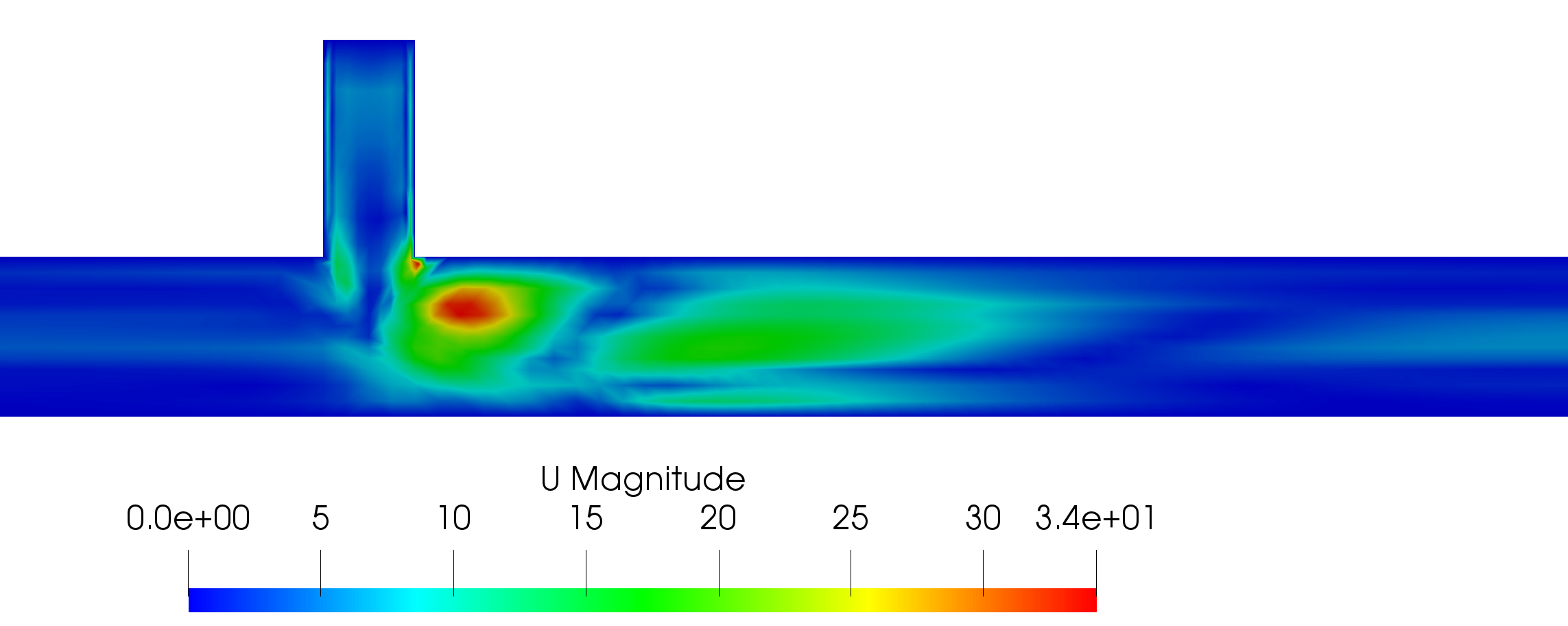}
\includegraphics[width=0.4\textwidth]{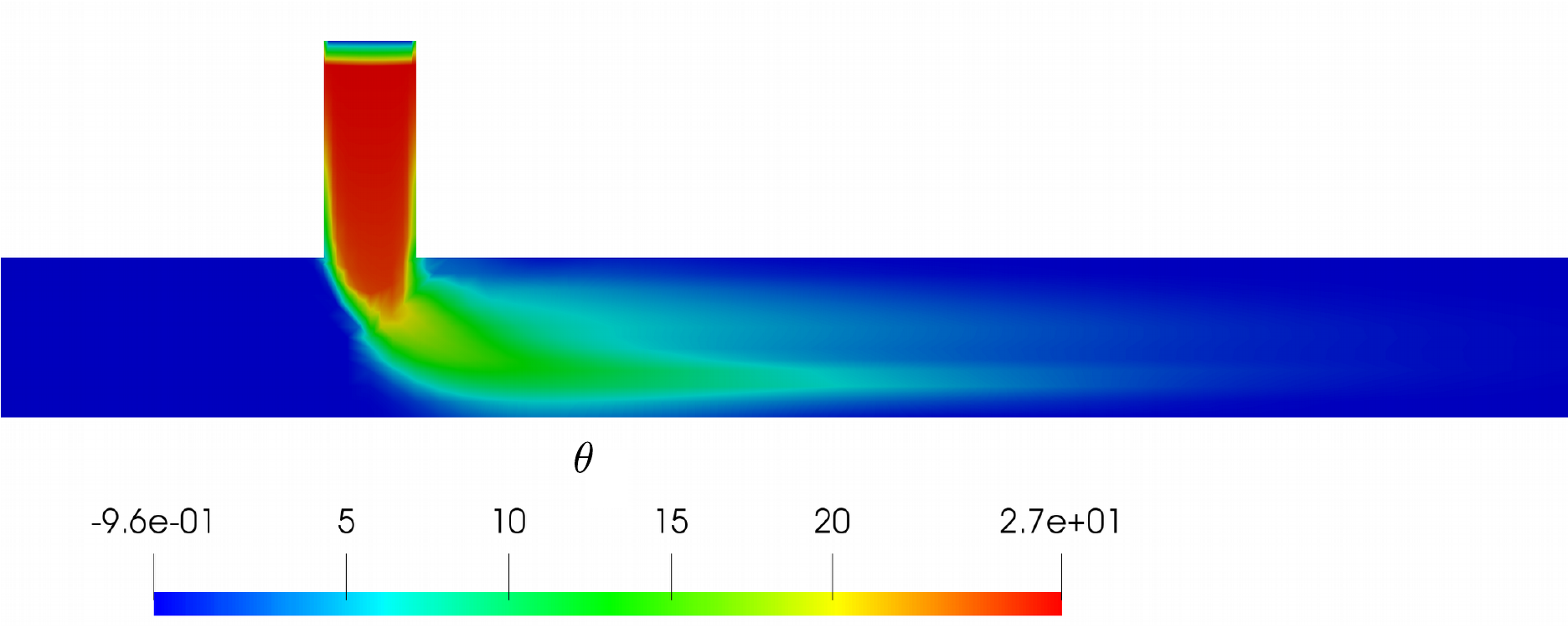}
\includegraphics[width=0.4\textwidth]{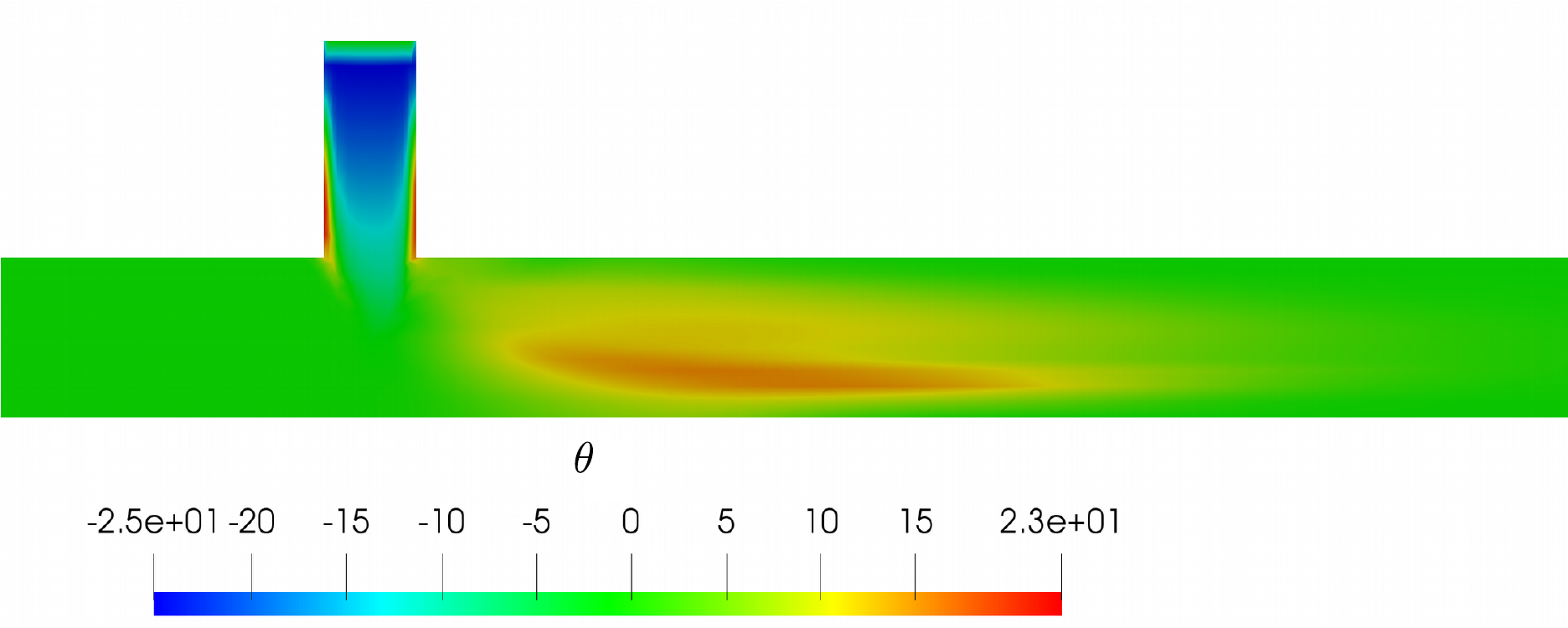}
\includegraphics[width=0.4\textwidth]{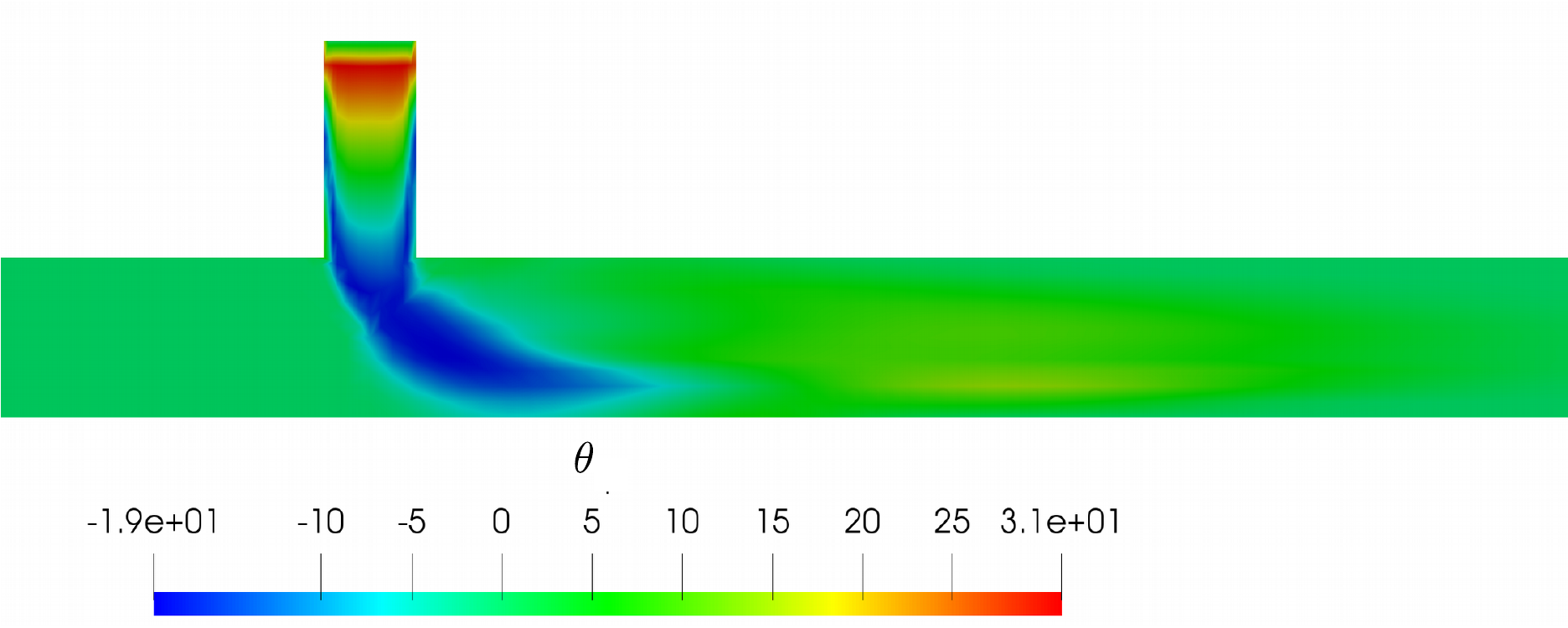}
\includegraphics[width=0.4\textwidth]{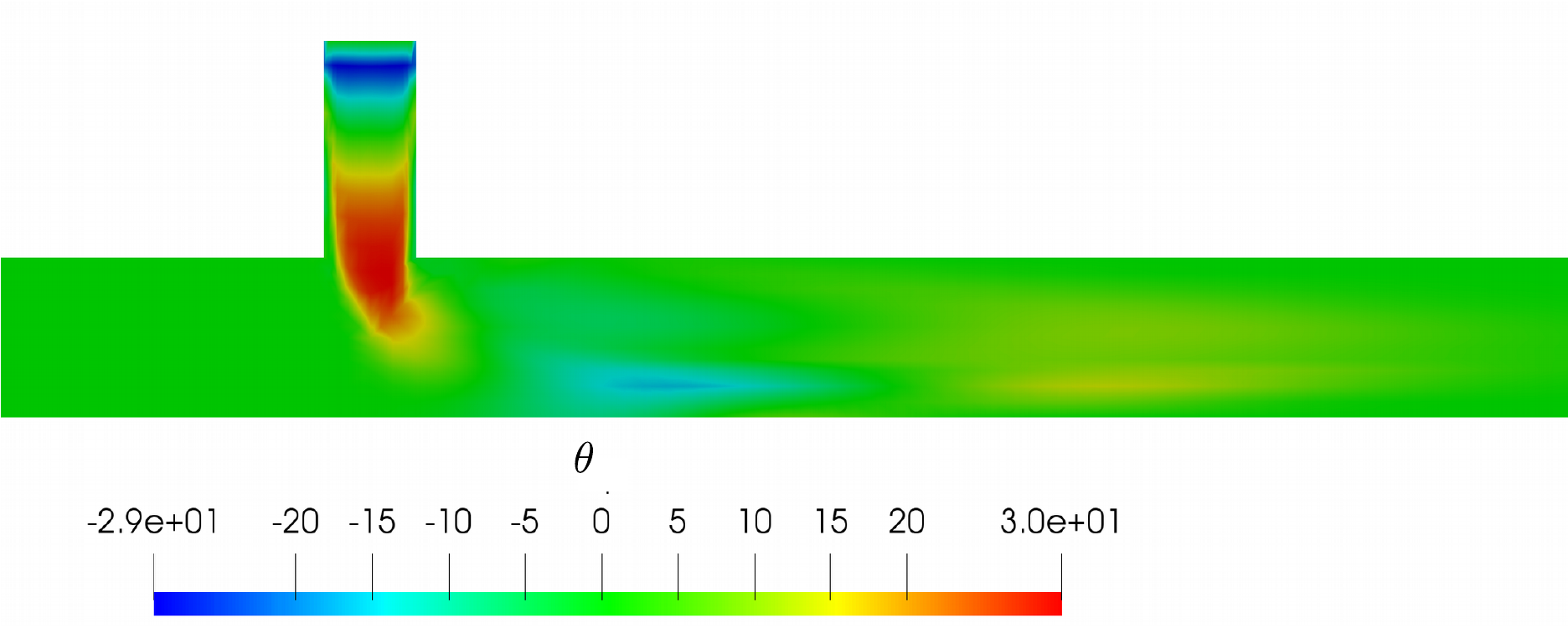}
\includegraphics[width=0.4\textwidth]{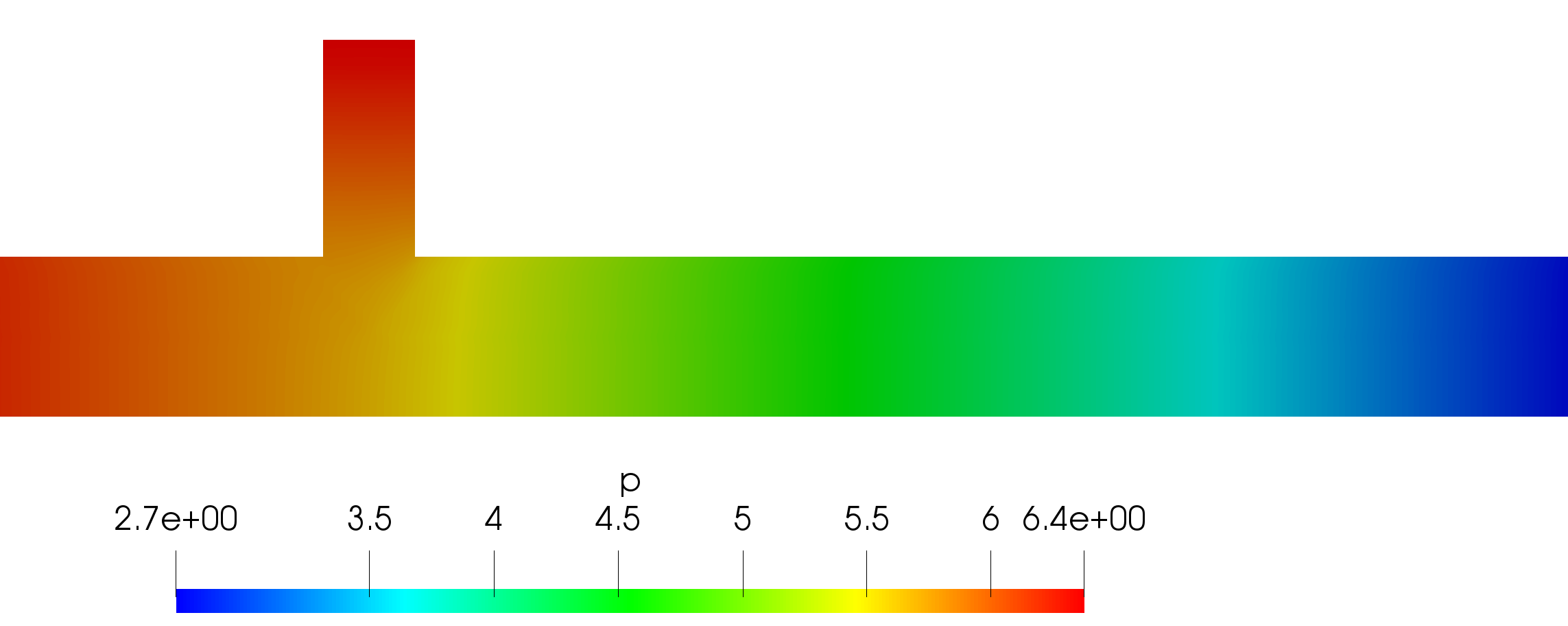}
\includegraphics[width=0.4\textwidth]{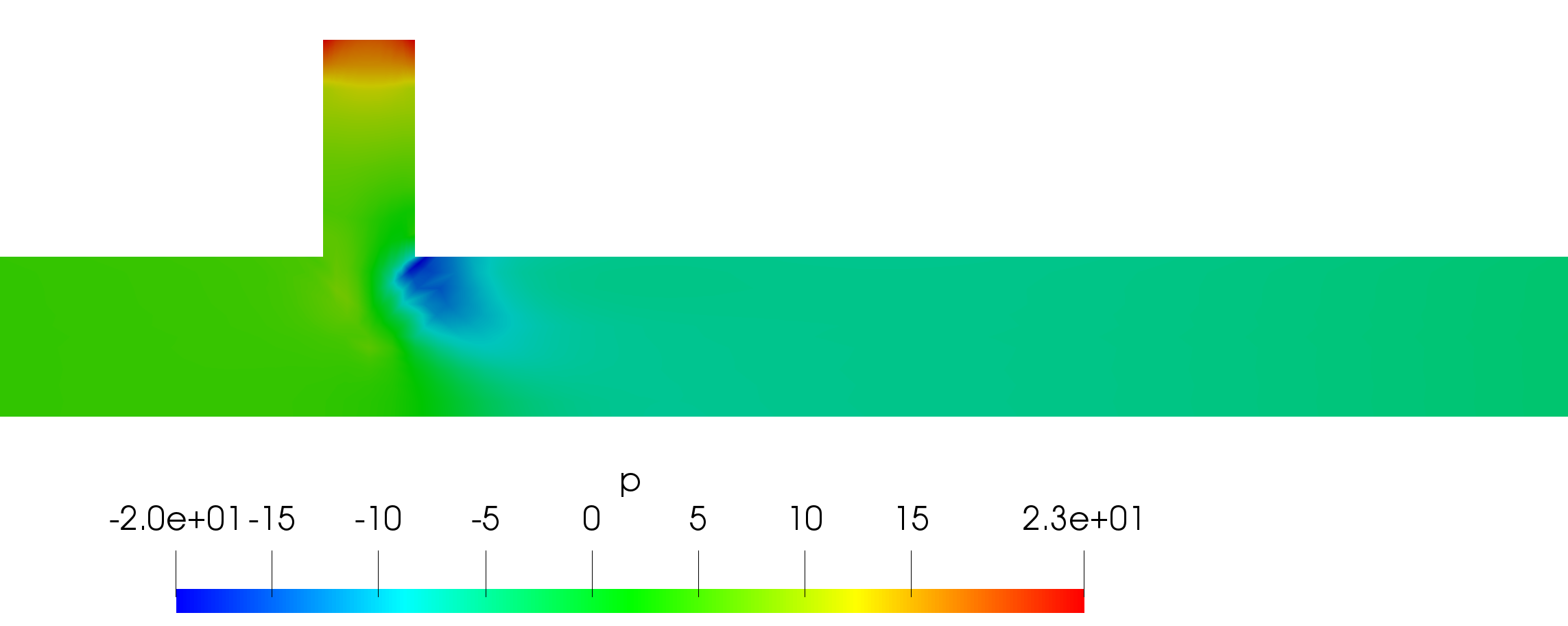}
\includegraphics[width=0.4\textwidth]{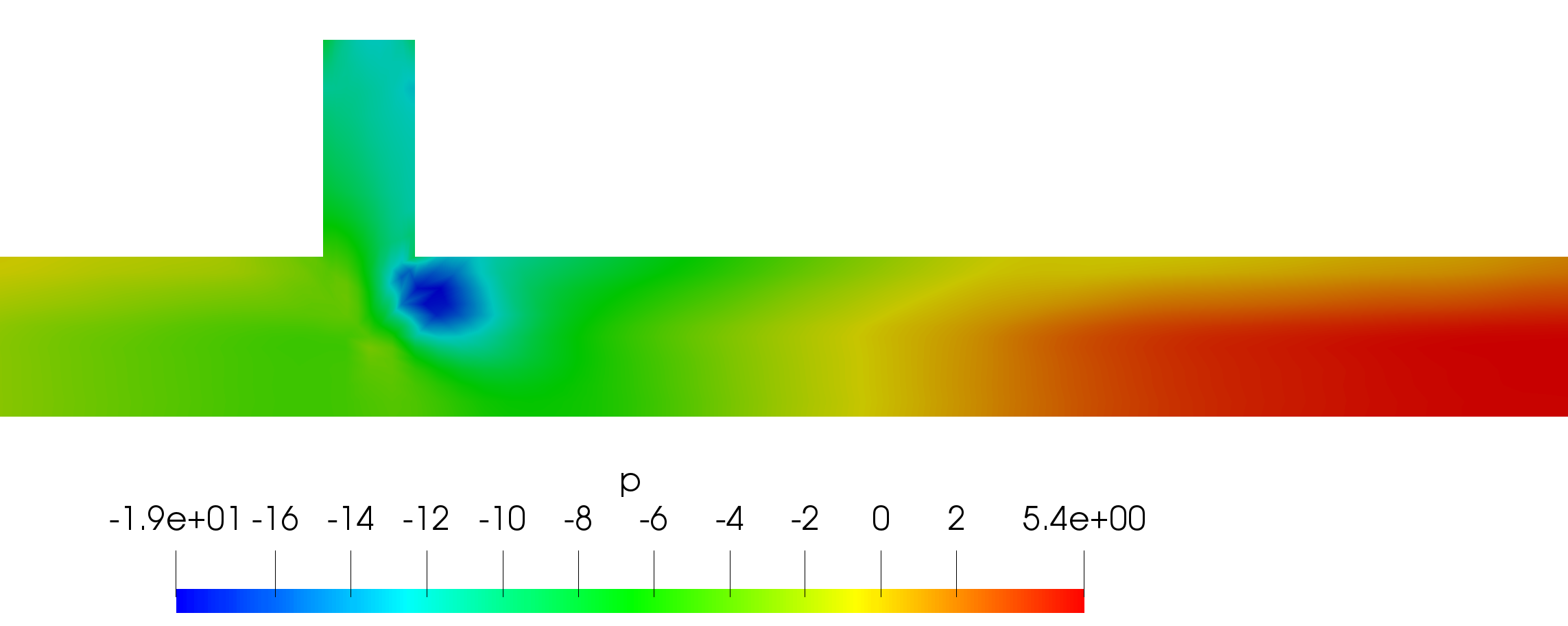}
\includegraphics[width=0.4\textwidth]{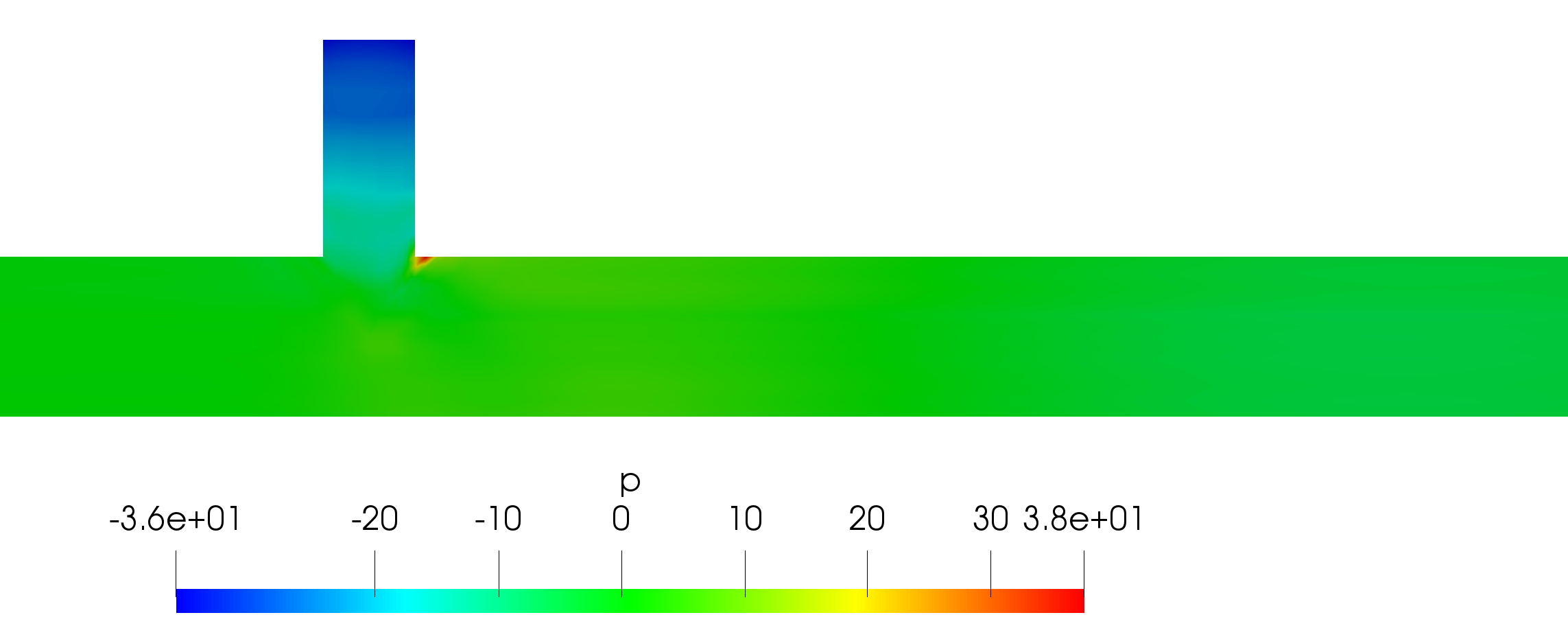}
\end{minipage} 
\caption{First four basis functions for velocity (first two rows),temperature (rows three and four) and pressure (last two rows) corresponding to $\nu = 1.1e-05$.}\label{fig:pod_modes_case2}  
\end{figure*}

\begin{figure*}[!tbp]
\begin{minipage}{1\textwidth}
\centering
\includegraphics[width=0.30\textwidth]{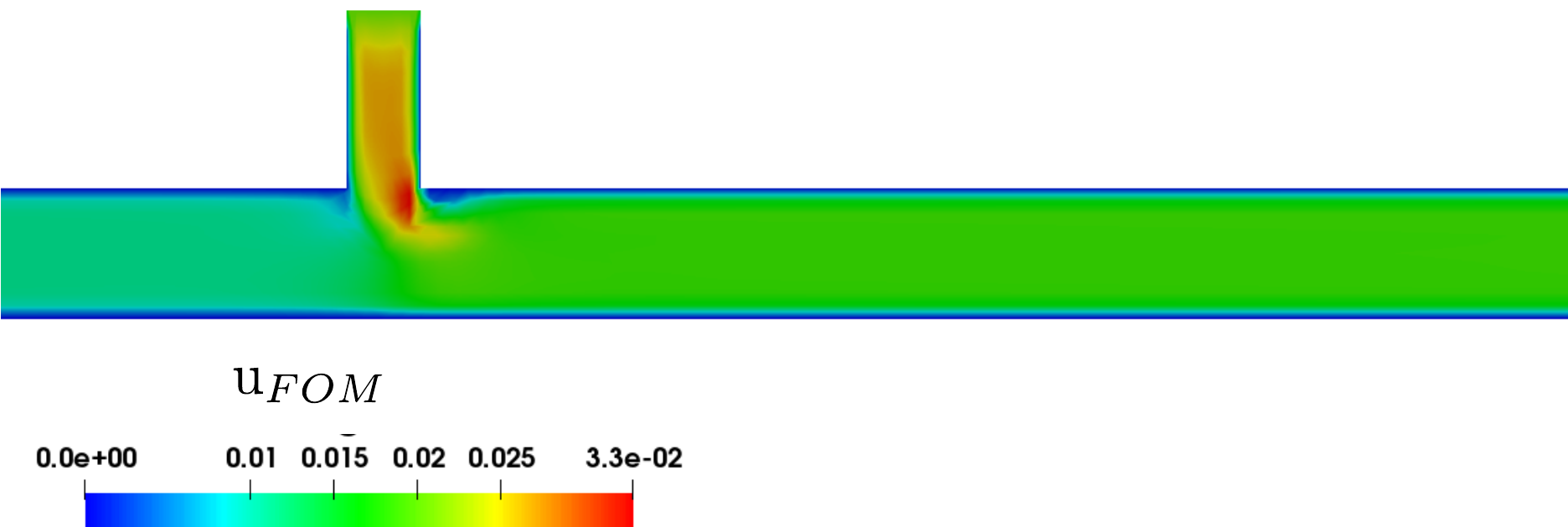}
\includegraphics[width=0.30\textwidth]{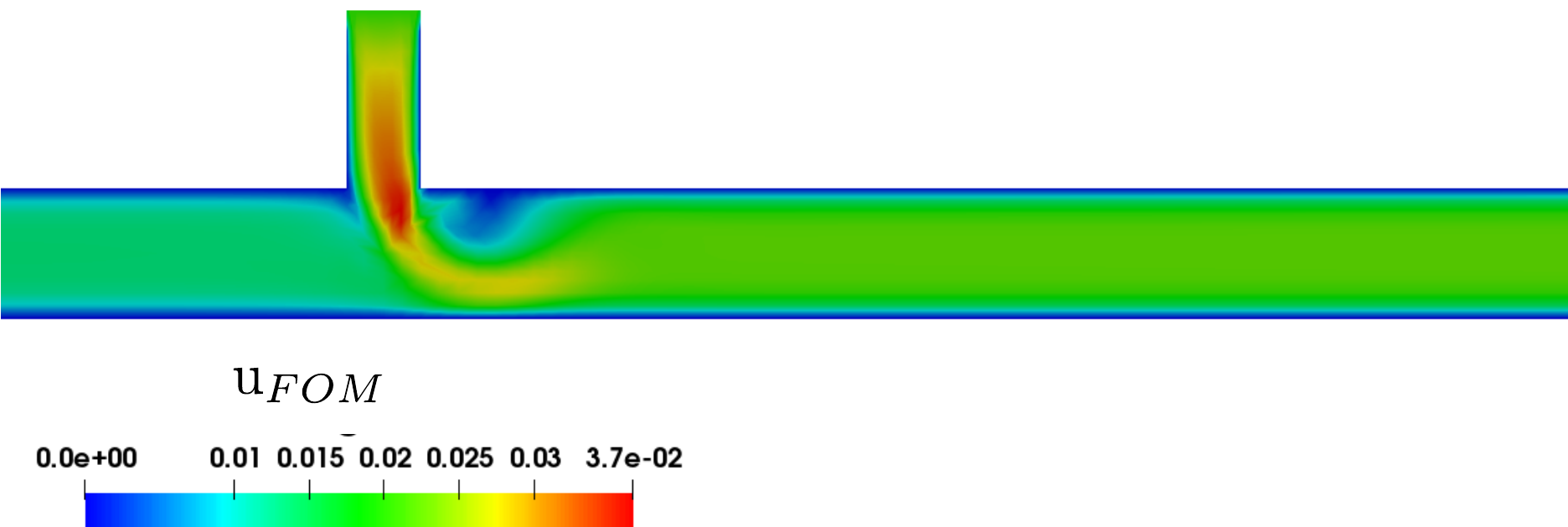}
\includegraphics[width=0.30\textwidth]{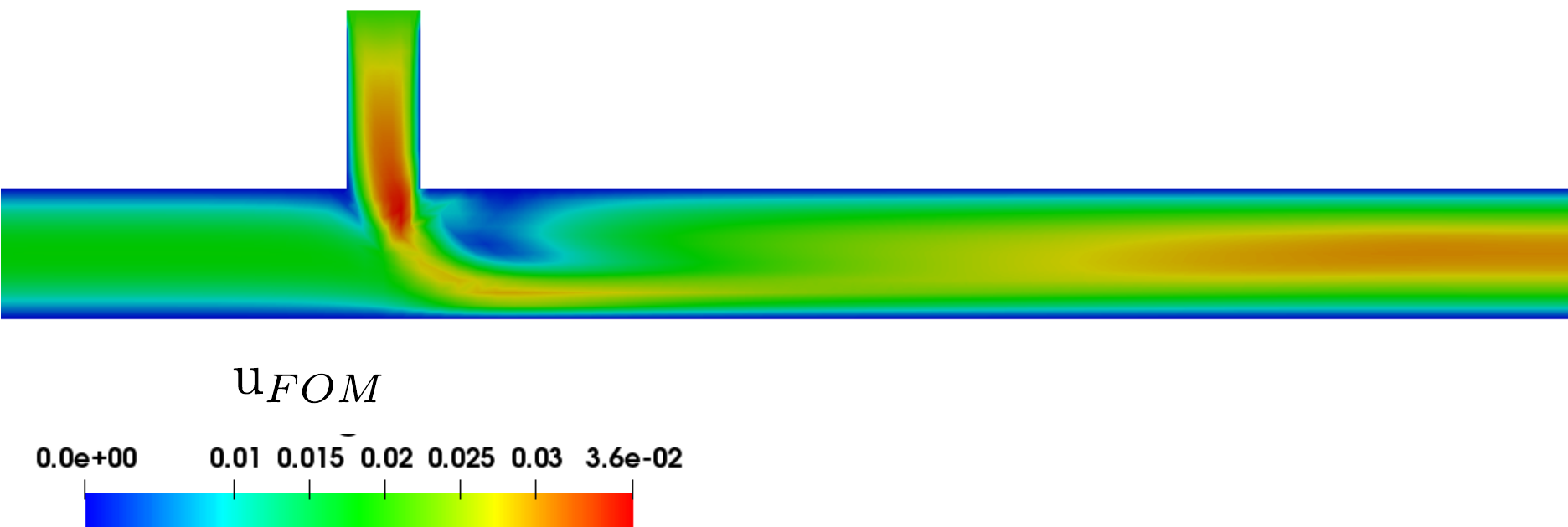}
\includegraphics[width=0.30\textwidth]{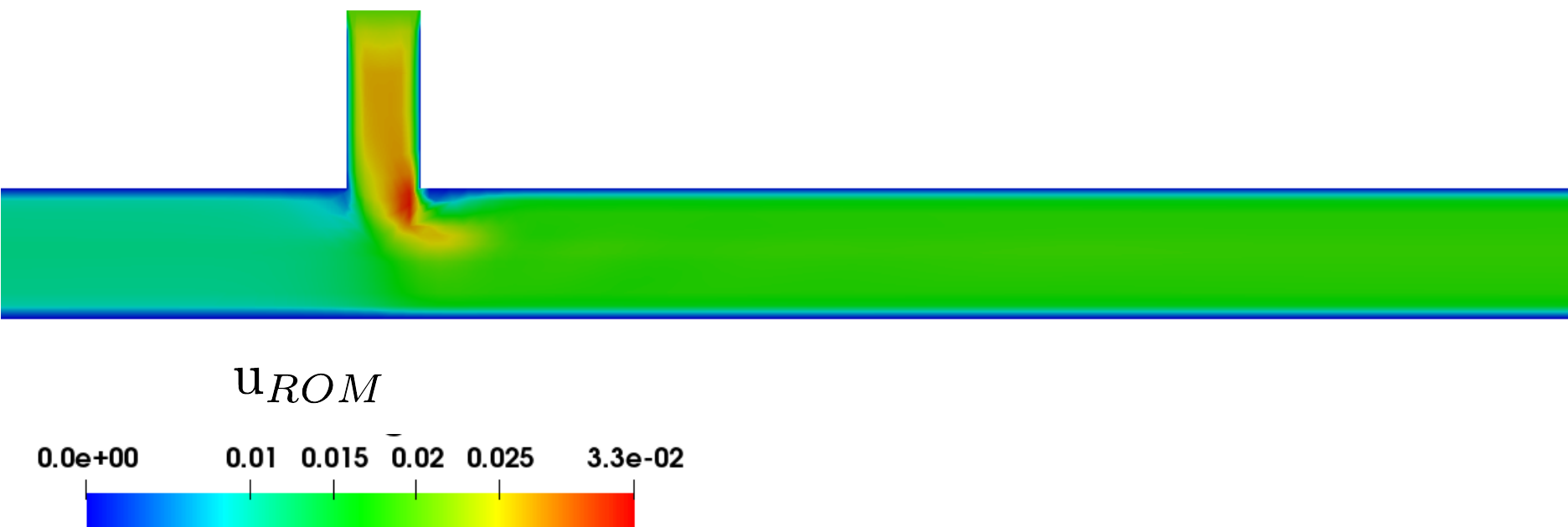}
\includegraphics[width=0.30\textwidth]{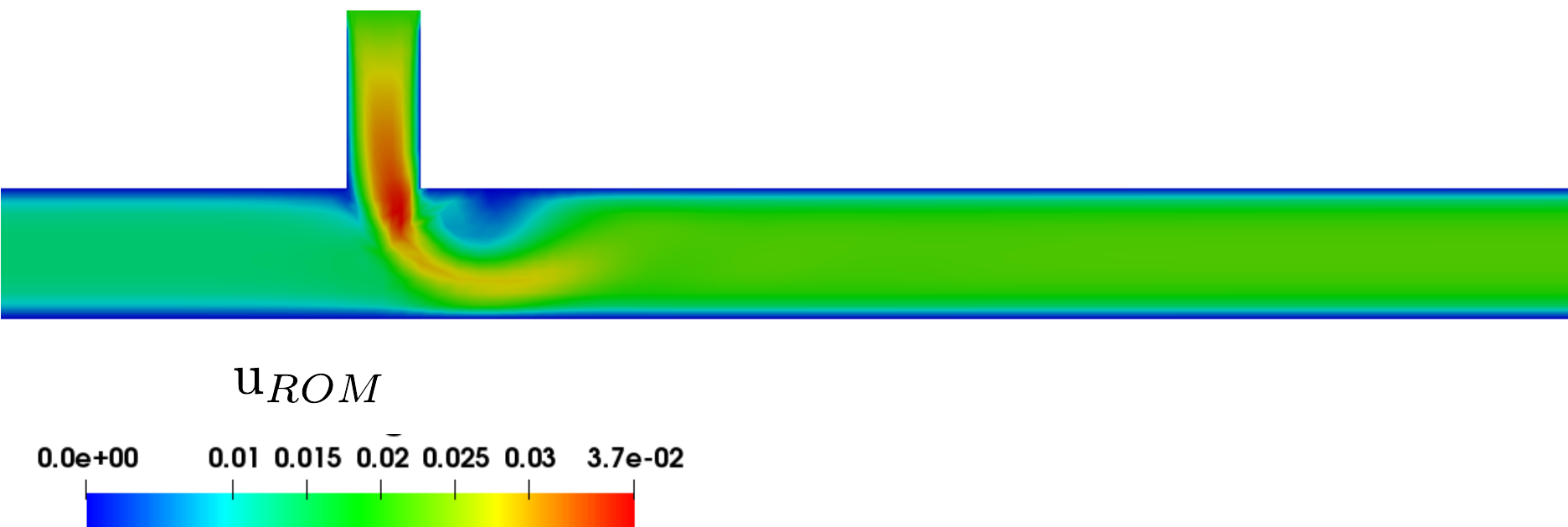}
\includegraphics[width=0.30\textwidth]{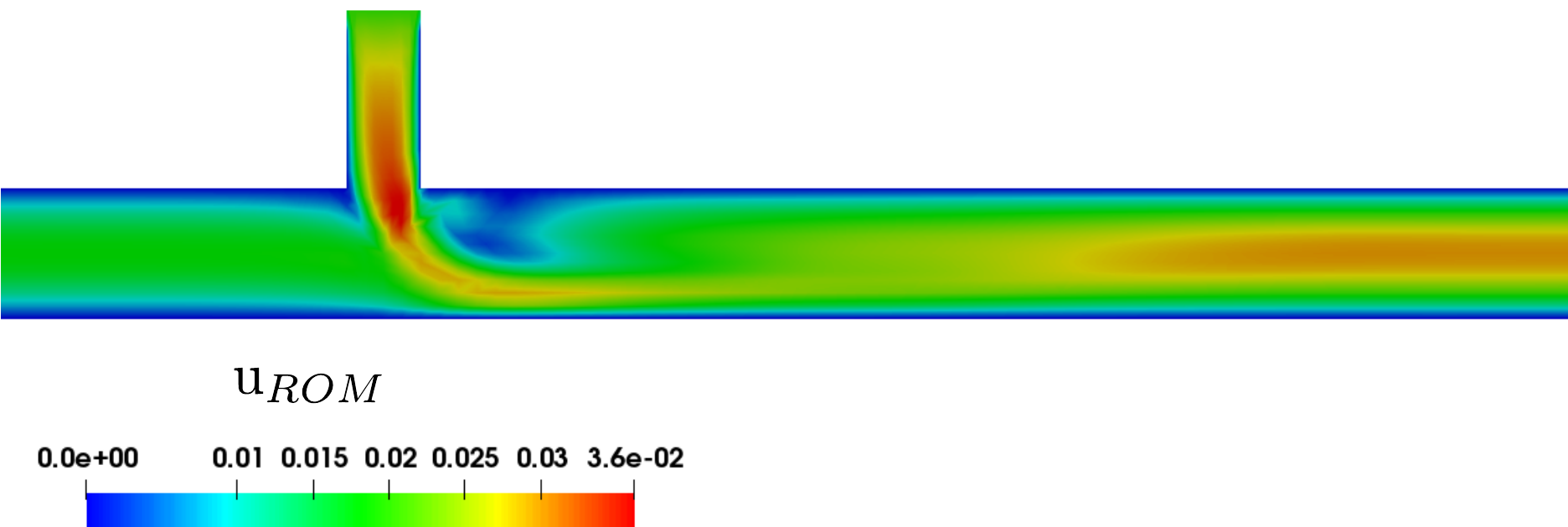}  
\includegraphics[width=0.30\textwidth]{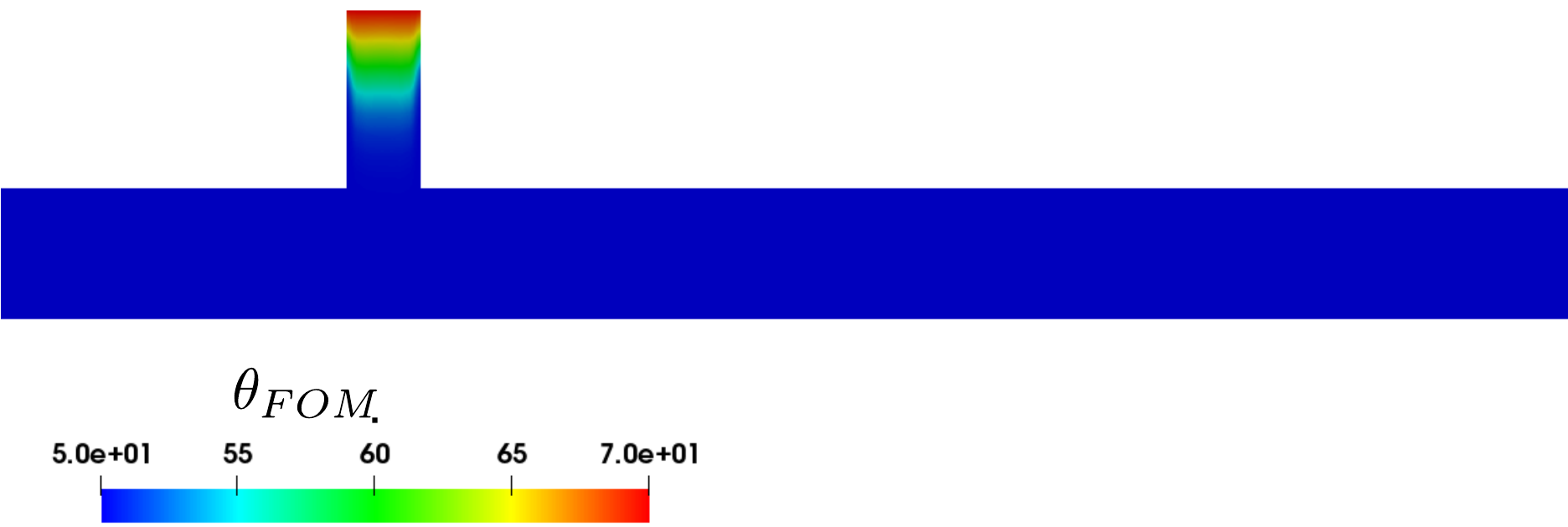}
\includegraphics[width=0.30\textwidth]{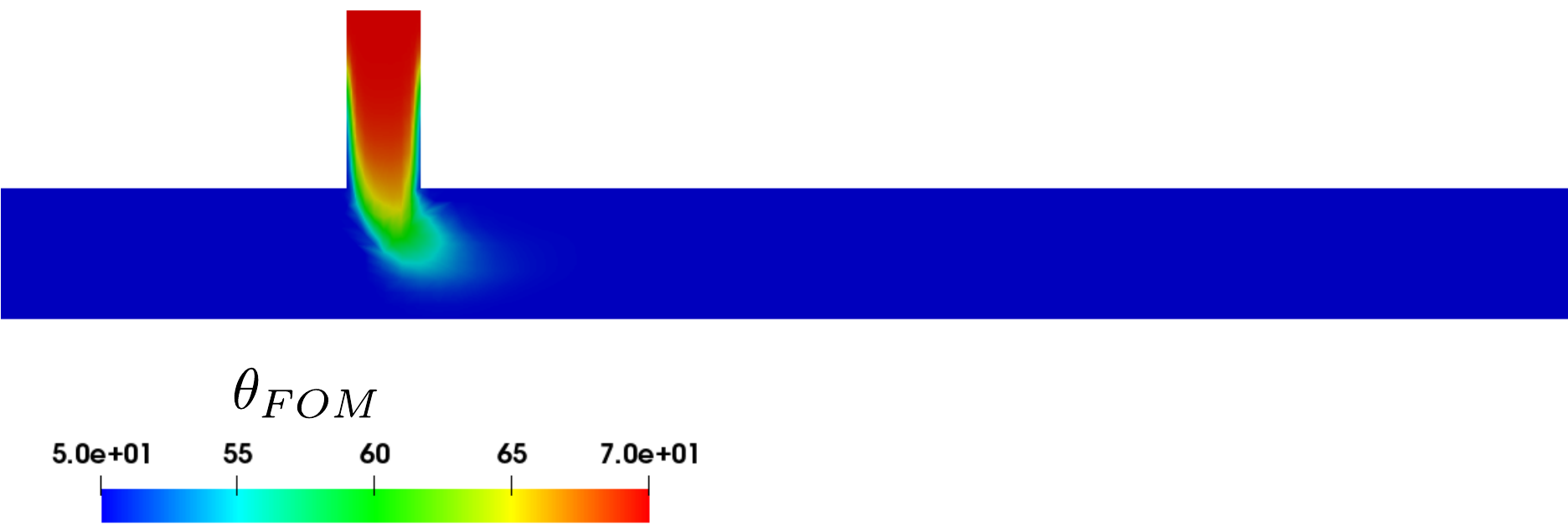}
\includegraphics[width=0.30\textwidth]{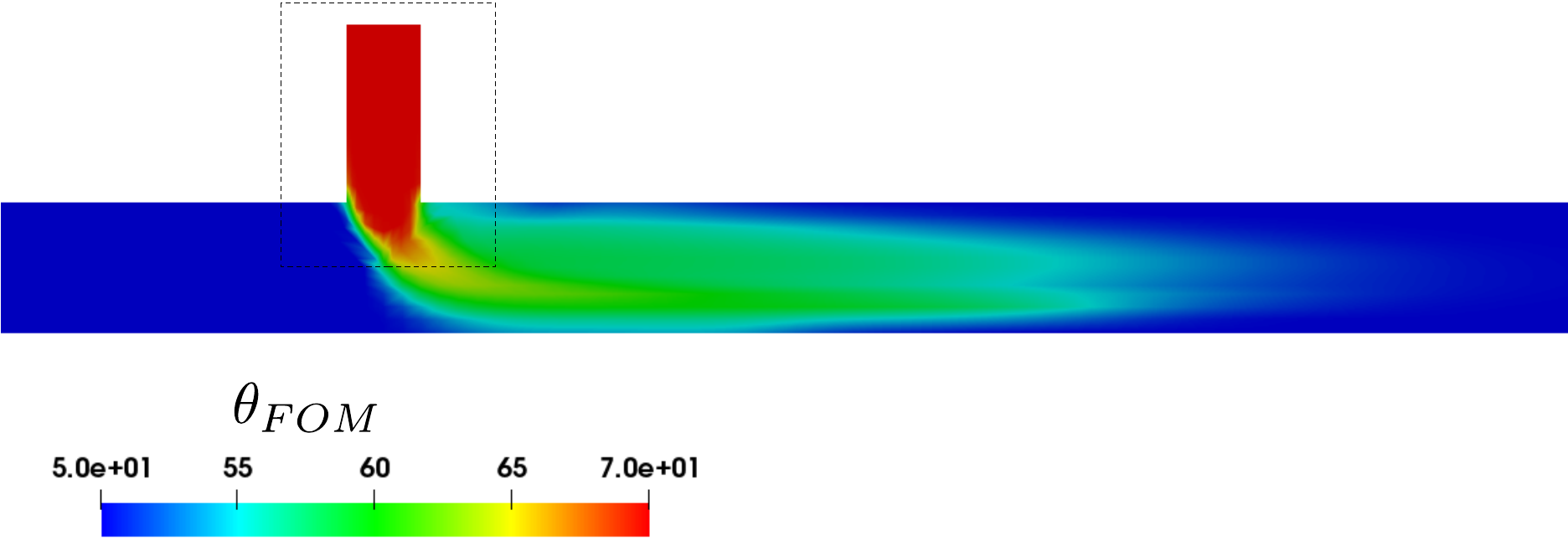}
\includegraphics[width=0.30\textwidth]{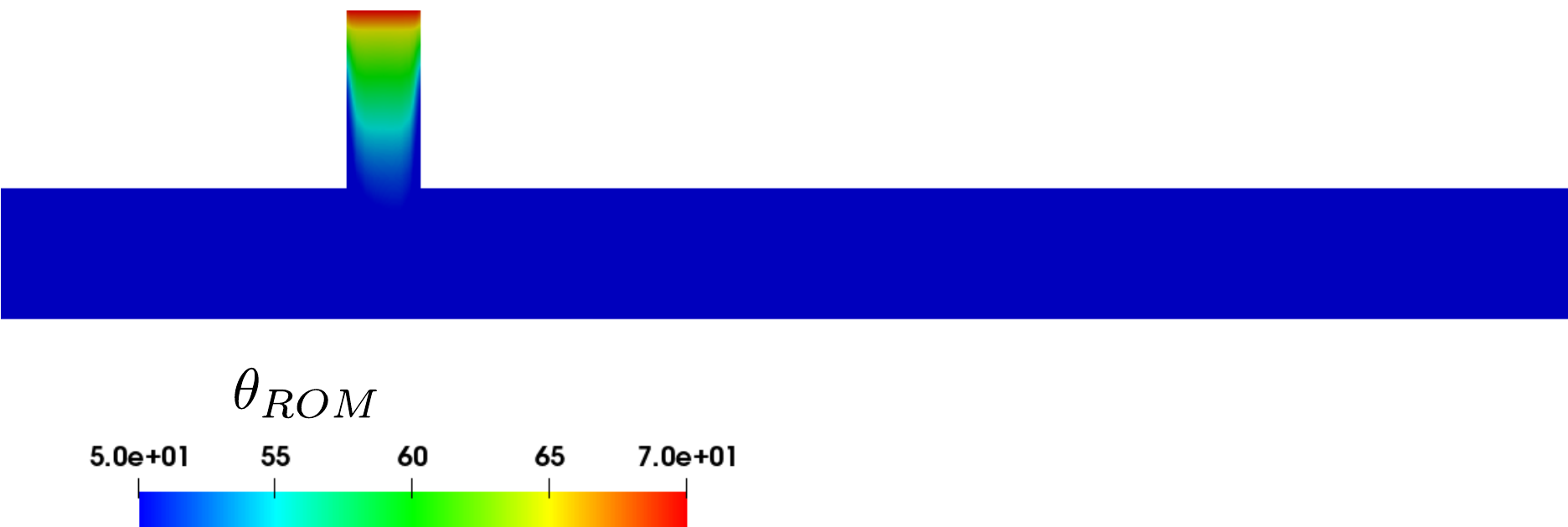}
\includegraphics[width=0.30\textwidth]{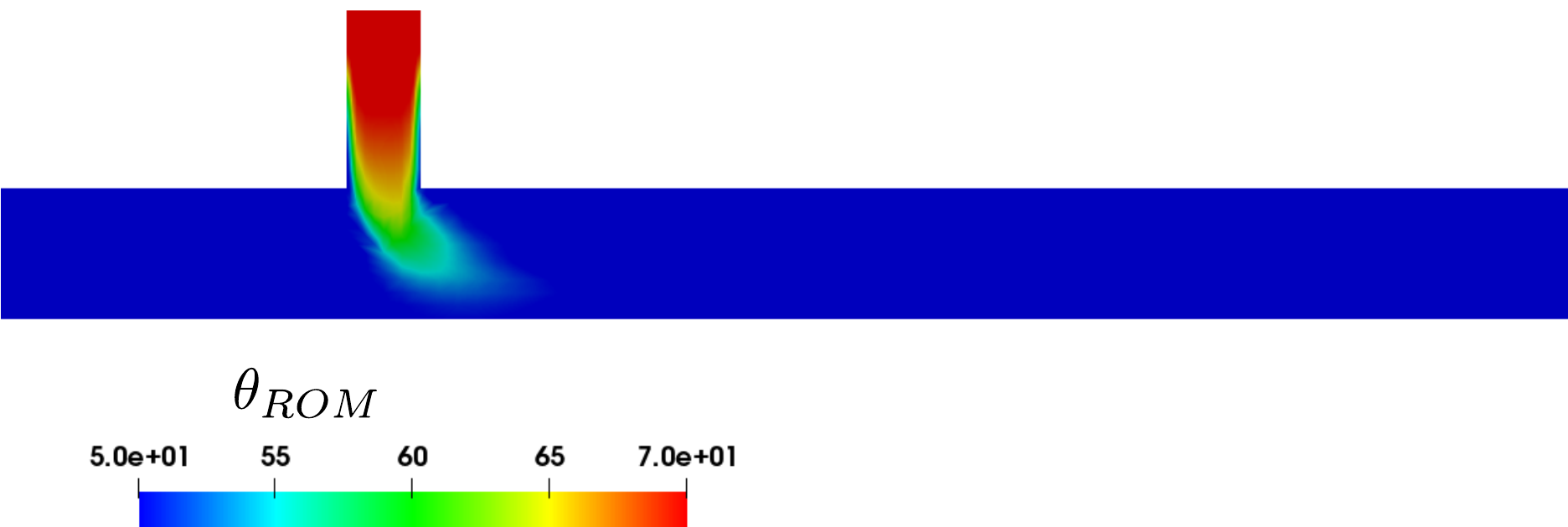}
\includegraphics[width=0.30\textwidth]{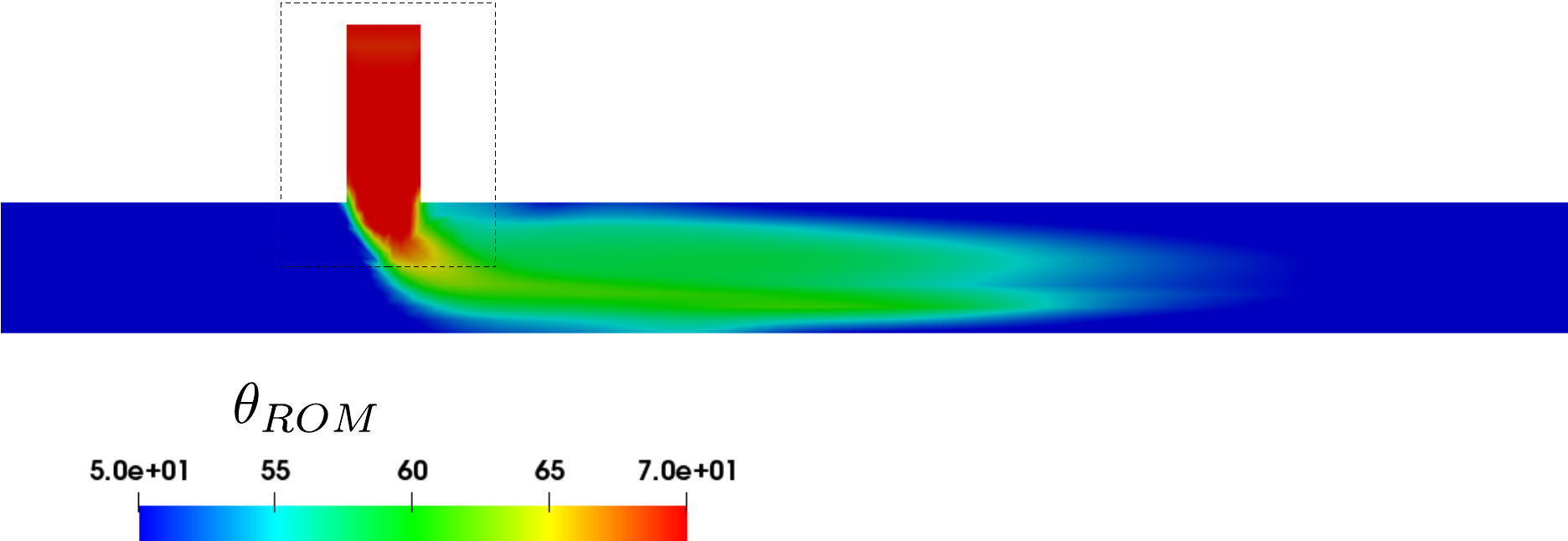}  
\includegraphics[width=0.30\textwidth]{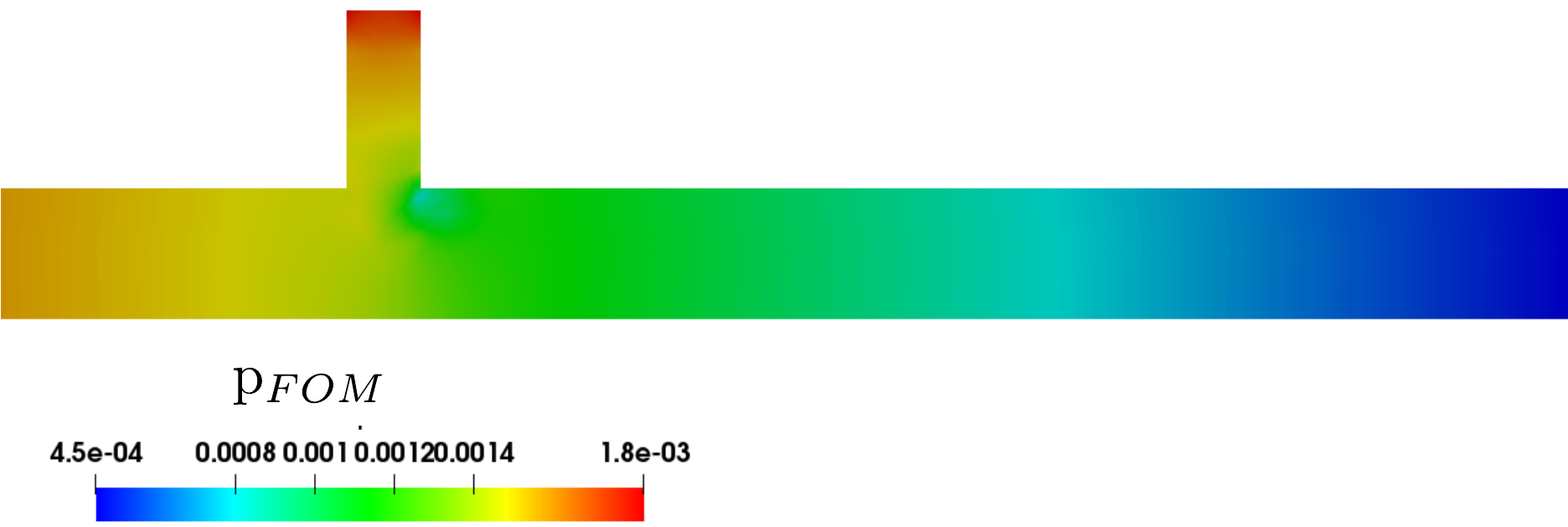}
\includegraphics[width=0.30\textwidth]{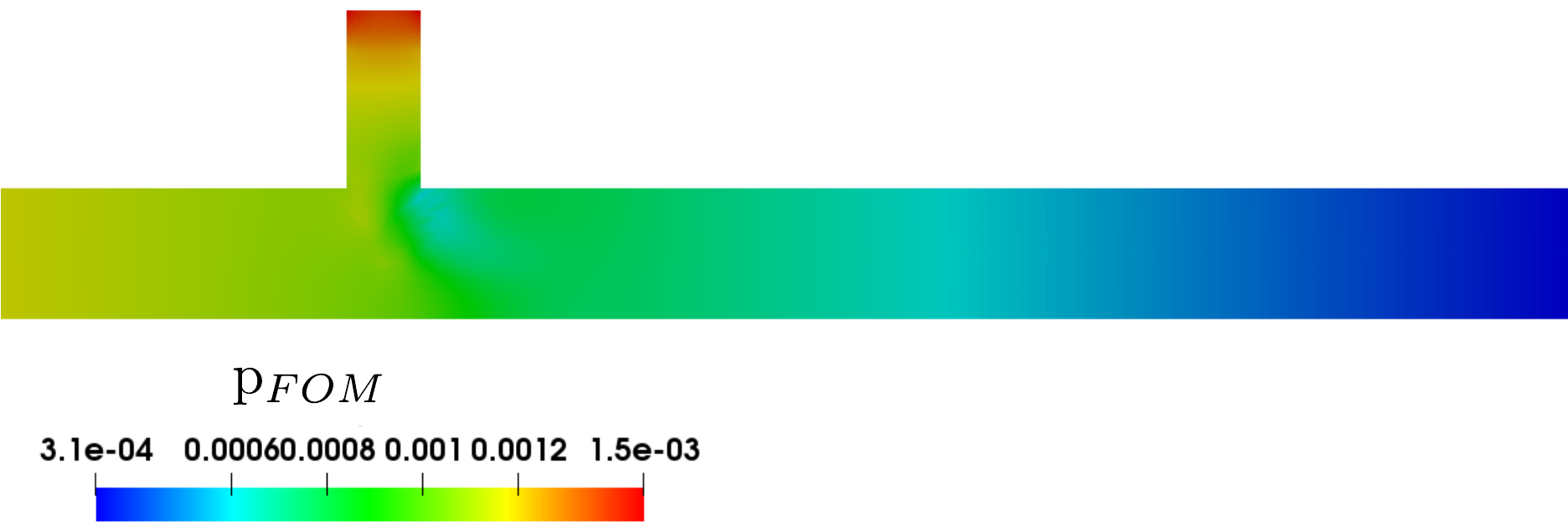}
\includegraphics[width=0.30\textwidth]{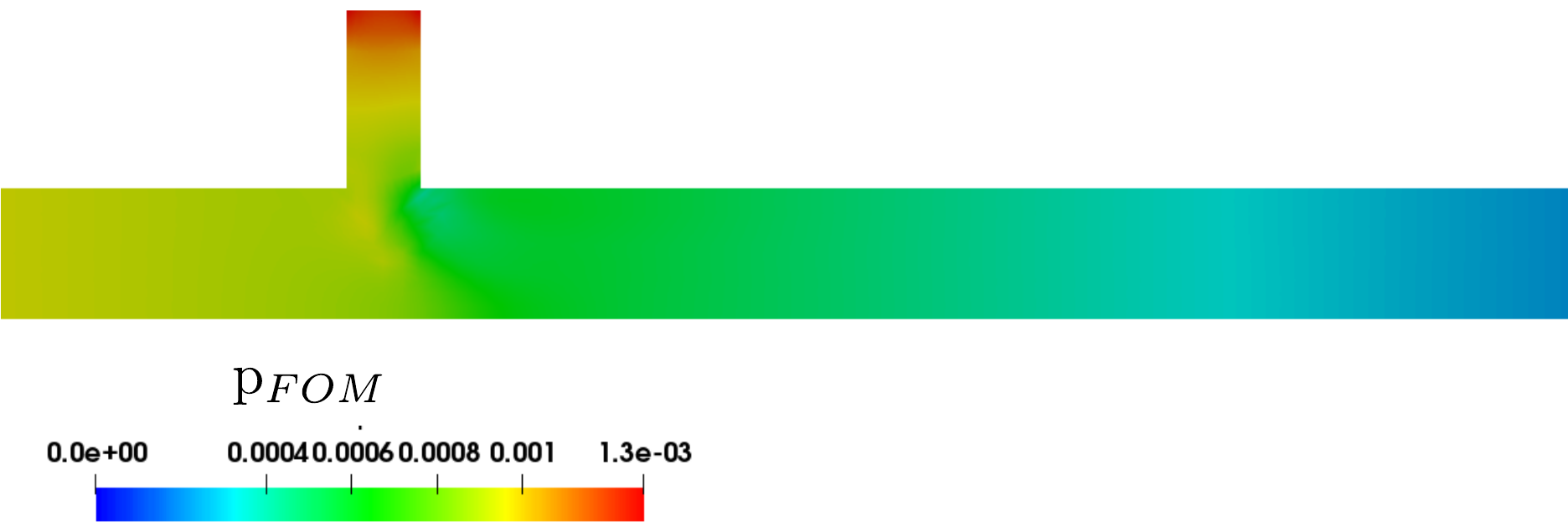}
\includegraphics[width=0.30\textwidth]{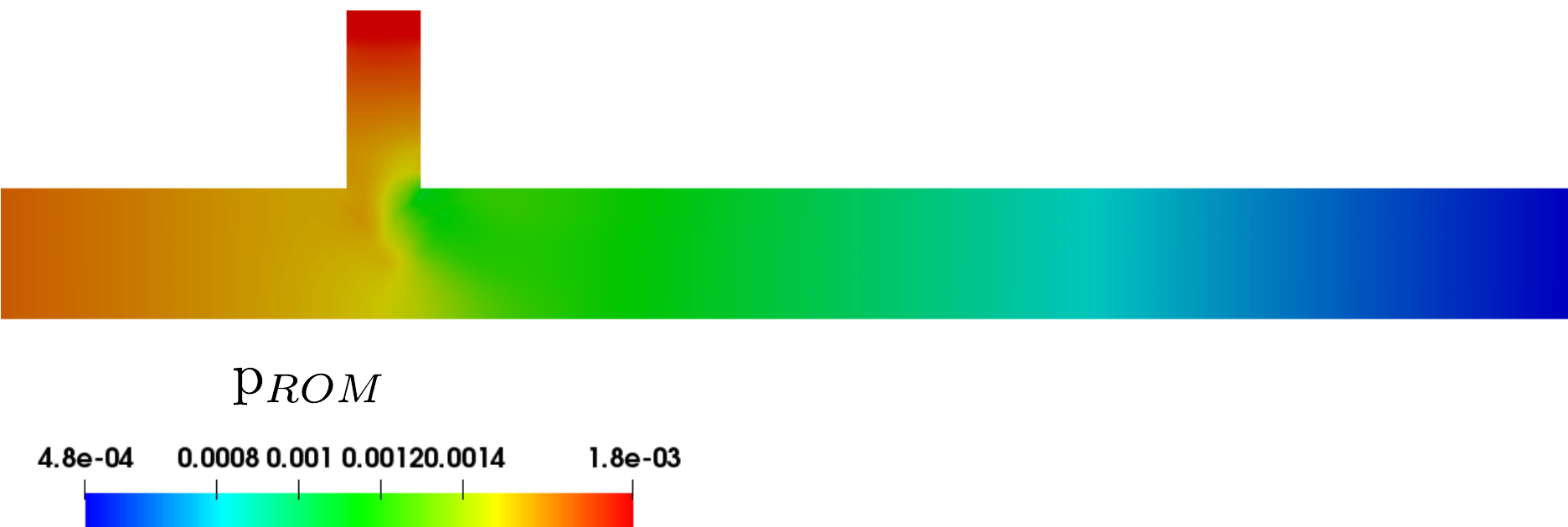}   
\includegraphics[width=0.30\textwidth]{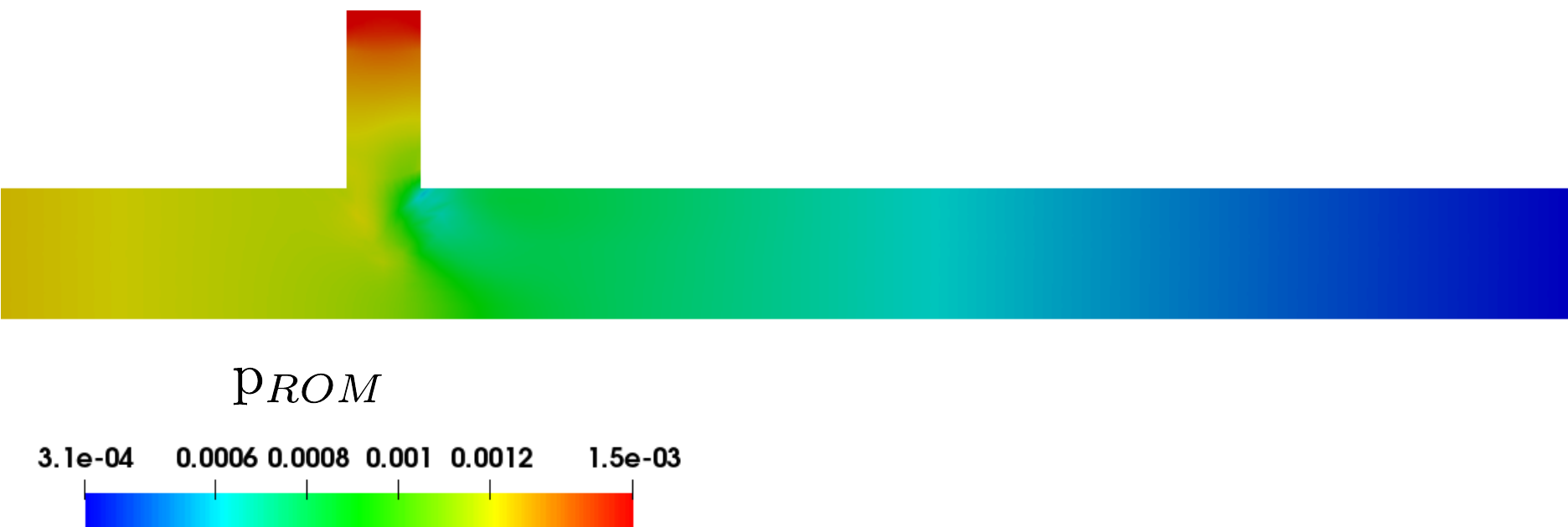}
\includegraphics[width=0.30\textwidth]{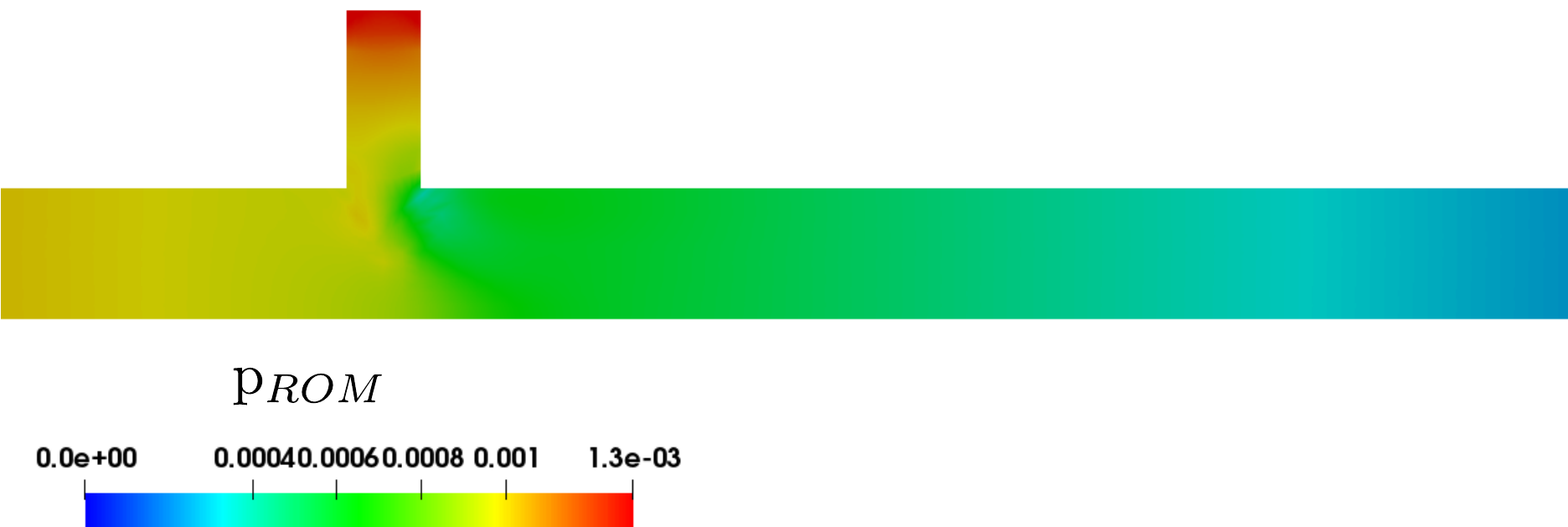}
\
\end{minipage} 
\caption{Comparison of the velocity field for the full order (first row) and reduced order model (second row) as well as temperature full order (third row) and temperature reduced order model (4th row) and pressure full order (5th row) with pressure reduced order model (6th row). The fields are depicted for different time instances equal to $t=3 \si{s}, 10\si{s}$ and $45 \si{s}$ and increasing from left to right. The viscosity is set to $\nu = 1.1e-05$.}\label{fig:comparison_case2}
\end{figure*}

\begin{figure*}[!tbp]   
\begin{minipage}{1\textwidth}
\centering
\includegraphics[width=0.4\textwidth]{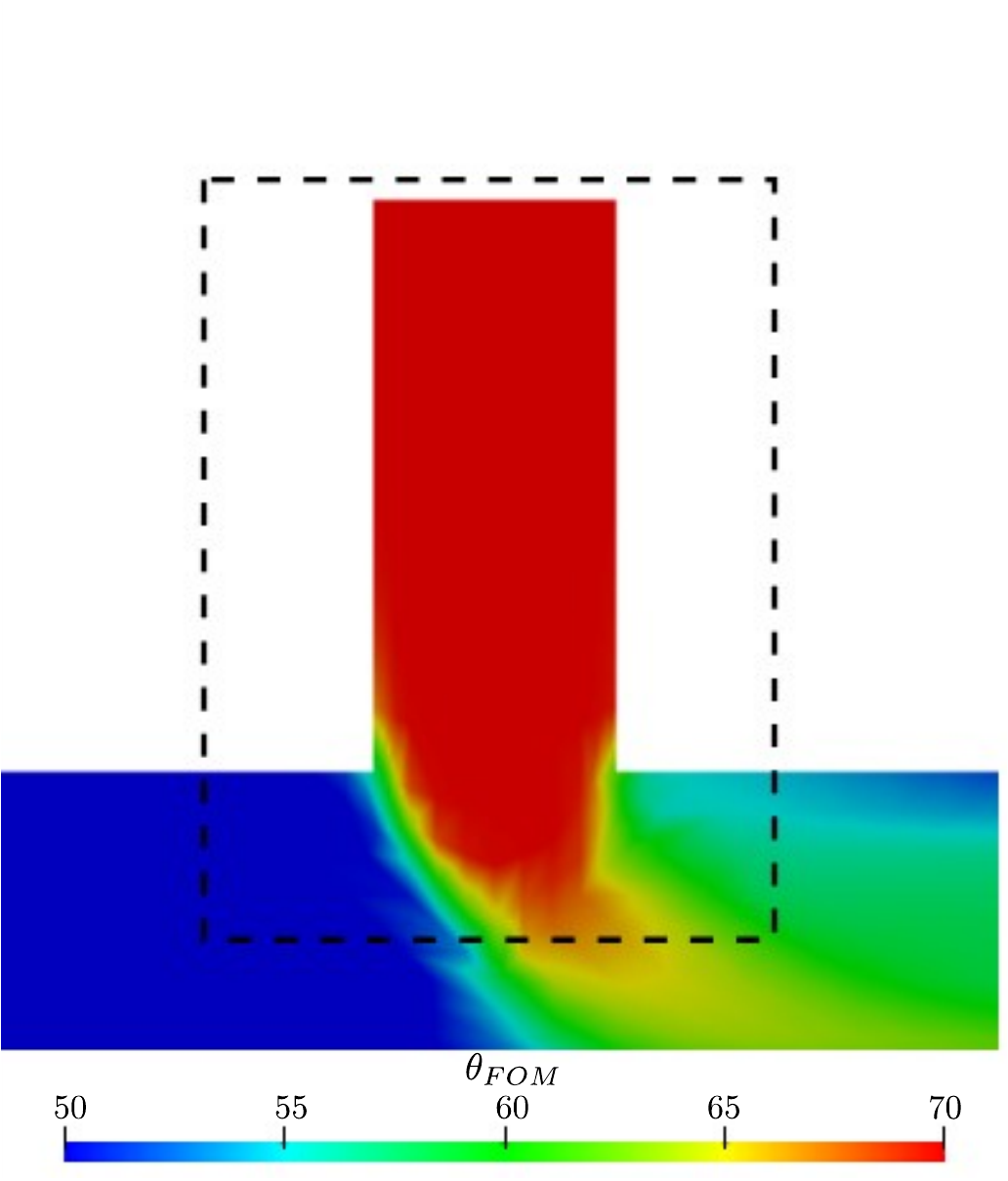}
\includegraphics[width=0.39\textwidth]{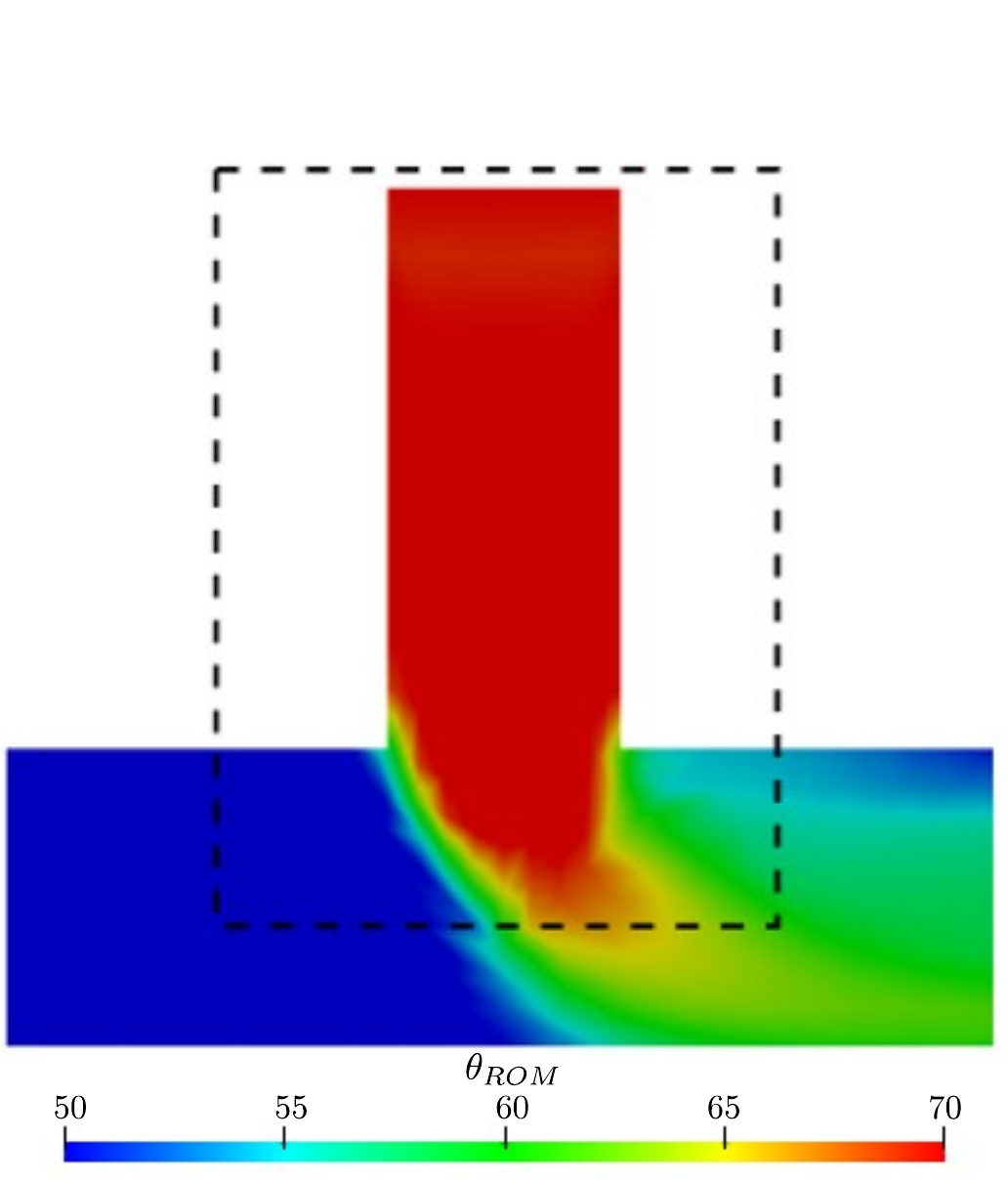} 
\end{minipage} 
\caption{Zoom of the area with the biggest relative error between the FOM (left) and the ROM (right) for temperature field, $\nu = 1.1e-05$.}\label{fig:visc_marker}     
\end{figure*}

The FOM simulation is run for each value of the kinematic viscosity in the above range, for $45$s with timestep of $\Delta T = 5 \times 10^{-3}$s. Snapshots are collected using the enhanced temporal sampling frequency according to the convergence study from the test case 1, figure (\ref{fig:convergence}). Therefore the snapshots are acquired every $0.1$s, using an equispace grid method in time and parameter, which gives a total number of $2250$ snapshots ($450$/case). A new value of the kinematic viscosity in which the ROM has not been trained but which belongs to the range of the training space, $\nu=1.1e-05$ ($\si{Re}_{\si{m}}$=$127$, $\si{Re}_{\si{b}}$ = $160$), is used to evaluate the capabilities of the parametrized ROM. To retain more than $99.9 \% $ of the system's energy, as shown in figure (\ref{fig:eigenvalues_case2}), $10$ modes for velocity, $5$ for temperature, $2$ for pressure and $3$ for the supremizer are kept The $\epsilon_{L^2}$  error between the FOM and ROM is plotted in figure (\ref{fig:l2error_case2}) which indicates that the ROM is capable of reproducing the main characteristics of the flow. Error statistics are summarized in table (\ref{fig:errorstatisticsvisc}). The first four POD modes for velocity, temperature and pressure fields are shown in figure (\ref{fig:pod_modes_case2}), in which the first mode captures most of the energy, as observed in figure (\ref{fig:eigenvalues_case2}). A comparison between the flow of the FOM and ROM models is illustrated in figure (\ref{fig:comparison_case2}), which indicates that the ROM is performing well in the reconstruction of the velocity, temperature and pressure fields. Concerning the temperature field, the area of the branch pipe, where the biggest differences were found, has been improved, figure (\ref{fig:visc_marker}), compared to the first test case (\ref{fig:marker}). The improved results could be a consequence of the enhanced sampling space which used in this test case. The error on temperature is growing as the time progresses and the two different temperature fluids start to mix in the mixing region. Taking more snapshots during the mixing period could reduce the error. In addition, to enhance the accuracy of the results, one could perform a denser sampling of the parameter space, as discussed earlier, but this increases the overall time of the offline phase and, for laminar cases, like this one, the overall improvement would be minimal (\ref{fig:convergencevisc}) . However, for more complicated cases, such as those in the turbulent range or in the transition range, enriching the POD space with additional sampling points of the kinematic viscosity would be essential. The CPU time of the FOM model is $969.23$s and the one of the ROM is $4.23$s. This corresponds to a speed-up of $\approx 211$.

\setlength\intextsep{0pt}    
 
\begin{table}[!tbp]
\centering
      \caption{Relative $\epsilon_{L^2}(t)$ error ($\epsilon_{L^2}(t) = \frac{||X_{FOM}(t)-X_{ROM}(t)||_{L^2(\Omega)}}{||X_{FOM}(t)||_{L^2(\Omega)}}$) for velocity, temperature and pressure fields for snapshot collection per $0.2$s.}\label{fig:errorstatisticsvisc}
      \centering
       \begin{tabular}{ l | c | c | c   }\toprule
& $\bm{u}$ per $0.1$s & $\theta$ per $0.1$s & $p$ per $0.1$s  \\ \hline \hline
    Minimum $\epsilon_{L^2}(t)$ & $0.024$ & $0.004$ & $0.013$ \\ \hline
    Maximum $\epsilon_{L^2}(t)$ & $0.090$ & $0.025$ & $0.104$ \\ \hline 
    Average $\epsilon_{L^2}(t)$ & $0.030$ & $0.017$ & $0.022$ \\ \bottomrule
   \end{tabular}
\end{table}


\section{Conclusions and perspectives}\label{sec:conclusions}
In this work a parametrized ROM using POD-Galerkin method is presented for applications in the study of thermal mixing in pipes. Apart from the 3D incompressible Navier-Stokes equations, a third transport equation corresponding to temperature is also considered which contains both convective and diffusive terms. Our interest is in the reconstruction of velocity, pressure and temperature fields. The proposed ROM is tested to simulate thermal mixing in a T-junction pipe, a common set-up found in nuclear power reactor cooling systems. Two different parametric cases are considering, one where the parametrization is on the temperature inlets and is considered a linear problem, and, one where a non-linear parametrization of the kinematic viscosity is concerned. In both cases, the ROM is capable of reproducing the results when run under the same conditions as in the FOM model, as well as, to predict the results on different parameters given a suitable training. In both cases a considerable computational speed up has been achieved, corresponding to a factor of approximately $374$ and $211$ respectively.
As in nuclear thermal hydraulics, the thermal mixing is studied usually in the turbulent range of Reynolds numbers, a parametric turbulent ROM for the Navier-Stokes and the temperature equation is of interest. Considering the methodology developed in the recent work of Hijazi \textit{et al.}, \cite{HijaziAliStabileBallarinRozza2018}, on modeling the turbulent parametric Navier-Stokes equations using POD-Galerkin with radial basis functions for the eddy viscosity term, a turbulent POD-Galerkin model for the unsteady Navier-Stokes and heat transport equation could be derived \cite{turbulentmixingnew}. Another future insight will be the construction of a ROM for buoyant driven flows. These will of course introduce further complexities, such as the need for additional terms in the ROM, but it will approximate much better real industrial problems \cite{kelbij}. Another challenging aspect is the computation of a-posteriori error bounds on the quantities of interest and the adaptation of the snapshots and/or parameter sampling accordingly. In regard to the FOM, even though a abundance of a-posteriori error estimates are available in finite element method, in the finite volume regime a few difficulties arise. These difficulties are mainly a consequence of the integral form of the equations found in finite volume discretization method. Methods that have been proposed rely on a-posteriori error estimates which require solutions on meshes with different spacing \cite{bergeradaptive} or on methods that treat the finite volume as a particular case of finite element and exploit the weak formulation \cite{jasak2003element}. In regard to the reduced order level, efficient and reliable a-posteriori error bounds are required. Althought a-posteriori error bounds have been proposed for elliptic PDEs \cite{hesthaven2015certified}, their determination for the weakly coupled Navier-Stokes and heat transport equations is not trivial and a further study would be of great interest.                 
 
\newpage

\section*{Appendix A. List of abbreviations and symbols}
\printnomenclature

\section{Acknowledgements} 
We acknowledge the financial support of Rolls-Royce, EPSRC, the European Research Council Executive Agency by means of the H2020 ERC Consolidator Grant project AROMA-CFD ``Advanced Reduced Order Methods  with  Applications  in  Computational  Fluid  Dynamics'' - GA  681447, (PI: Prof. G. Rozza) and INdAM-GNCS 2018.

\bibliographystyle{myIEEEtran.bst}
\bibliography{bibfile}  

\end{document}